\documentclass[12pt,preprint]{aastex}
\usepackage{graphicx}
\begin{document}

\title{Estimating the Jet Power of Mrk\,231 During the 2017-2018 Flare}
\author{Cormac Reynolds\altaffilmark{1}, Brian Punsly\altaffilmark{2}, Giovanni Miniutti\altaffilmark{3}, Christopher P. O'Dea\altaffilmark{4,5}, Natasha Hurley-Walker\altaffilmark{6}}
\altaffiltext{1}{CSIRO Astronomy and Space Science, PO Box 1130, Bentley WA 6102, Australia}
 \altaffiltext{2}{1415 Granvia Altamira, Palos Verdes Estates CA, USA
90274: ICRANet, Piazza della Repubblica 10 Pescara 65100, Italy and
ICRA, Physics Department, University La Sapienza, Roma, Italy,
brian.punsly@cox.net}
\altaffiltext{3}{Centro de Astrobiologia (CAB) ESA - European Space
Astronomy Center (ESAC)} \altaffiltext{4}{Department of Physics and
Astronomy, University of Manitoba, Winnipeg, MB R3T 2N2
Canada}\altaffiltext{5}{School of Physics \& Astronomy, Rochester
Institute of Technology, Rochester, NY 14623, USA} \altaffiltext{6}
{ICRAR-Curtin University, GPO Box U1987, Perth, Western Australia,
6102, Australia}

\begin{abstract}
Long-term 17.6~GHz radio monitoring of the broad absorption line quasar,
Mrk\,231, detected a strong flare in late 2017. This triggered four epochs of Very Long Baseline Array (VLBA)
observations from 8.4~GHz to 43~GHz over a 10-week period as well as an X-ray observation with NuSTAR. This was the third campaign of VLBA monitoring that we have obtained. The 43~GHz VLBA was degraded in all epochs with only 7 of 10 antennas available in three epochs and 8 in the first epoch. However, useful results were obtained due to a fortuitous capturing of a complete short 100~mJy flare at 17.6~GHz: growth and decay. This provided useful constraints on the physical model of the ejected plasma that were not available in previous campaigns. We consider four classes of models, discrete ejections (both protonic and positronic) and jetted (protonic and positronic). The most viable model is a ``dissipative bright knot" in a faint background leptonic jet with an energy flux $\sim10^{43}$ ergs/sec. Inverse Compton scattering calculations (based on these models) in the ambient quasar photon field explains the lack of a detectable increase in X-ray luminosity measured by NuSTAR. We show that the core (the bright knot) moves towards a nearby secondary at $\approx 0.97$c. The background jet is much fainter. Evidently, the high frequency VLBA core does not represent the point of origin of blazar jets, in general, and optical depth ``core shift" estimates of jet points of origin can be misleading.
\end{abstract}

\section{Introduction}
Mrk\,231 is a nearby quasar (redshift of $z = 0.042$) that has all of the extreme properties of the quasar population in one object. It has a high-luminosity accretion flow, a radio jet that can rival the power of other well-known nearby extragalactic radio sources and a broad absorption line (BAL) high velocity outflow. All of this combined in a nearby object provides an excellent laboratory to study the interplay between these phenomena. The radio jet makes this the brightest radio quiet quasar (RQQ) at high frequency. This means that we can capitalize on the $\sim 0.2$~pc resolution of the Very Large Baseline Array (VLBA) to explore the interior of this RQQ, a circumstance unique to Mrk\,231. This has motivated us to pursue a series 43~GHz observing campaigns with VLBA from 2006-2019 and almost continuous daily to weekly 17.6~GHz monitoring from 2013-2019 with AMI\footnote{The
Arcminute Microkelvin Imager consists of two radio interferometric
arrays located in the Mullard Radio Astronomical Observatory,
Cambridge, UK \citep{zwa08}. Observations occur between 13.9 and
18.2~GHz in six frequency channels. The Small Array (AMI-SA)
consists of ten 3.6~m diameter dishes with a maximum baseline of
20~m, with an angular resolution of 3\arcmin, while the Large Array
(AMI-LA) comprises eight 12.6~m diameter dishes with a maximum
baseline of 110~m, giving an angular resolution of 0\farcm5.}. The third VLBA campaign is reported in this paper. This study concentrates on estimating the jet power during what was the strongest 15--20~GHz flare to date.
\par The jet is extremely powerful for a RQQ during flare states with a kinetic luminosity crudely estimated to be $Q \sim 3\times10^{43}$ ergs~$\rm{s}^{-1}$ for previous flares \citep{rey09}. A small number of RQQs that have also exhibited episodes of relativistic jet formation \citep{bru00,blu03}. However, Mrk\,231 is more luminous at high frequency, more often, than other RQQs and can provide unique clues as to why some quasars are radio loud and some are radio quiet. Consequently, we
have been monitoring the radio behavior at $\sim 20$~GHz since 2009. Before 2013, the monitoring was very sporadic with the Very Large Array (VLA). Since 2011, we have detected 5 large blazar-like flares (see \citet{rey13} for a discussion) with flux densities $\geq 200$~mJy at $\sim 20$~GHz. The fourth of these major flares reached 350~mJy at the end of 2017 (see Figure~1). This initiated a target of opportunity four-epoch VLBA observation and one epoch observation with the NuSTAR X-ray telescope. Our primary aim is to improve on the crude two-epoch-based estimate of $Q$ quoted above. In Section 2, we will review our previous results of VLBA monitoring. In Section 3, we will discuss the radio light curve leading to the flare and during the flare. Section 4 collects the data obtained from our VLBA observation. We also describe the core spectra in terms of a synchrotron self-absorbed (SSA) power law. The simple spectral shape lends itself to a simple physical model of a uniform sphere that has been utilized previously for energy estimates of ejections. The details are reviewed in Section 5. The next two sections describe the solution space. In Section 8, we consider inverse Compton scattering in the ejection and compare this with our NuSTAR observation. Throughout this paper, we adopt the following cosmological parameters: $H_{0}$=71 km s$^{-1}$~Mpc$^{-1}$, $\Omega_{\Lambda}=0.73$ and $\Omega_{m}=0.27$.

\begin{figure}
\begin{center}
\includegraphics[width= 0.8\textwidth]{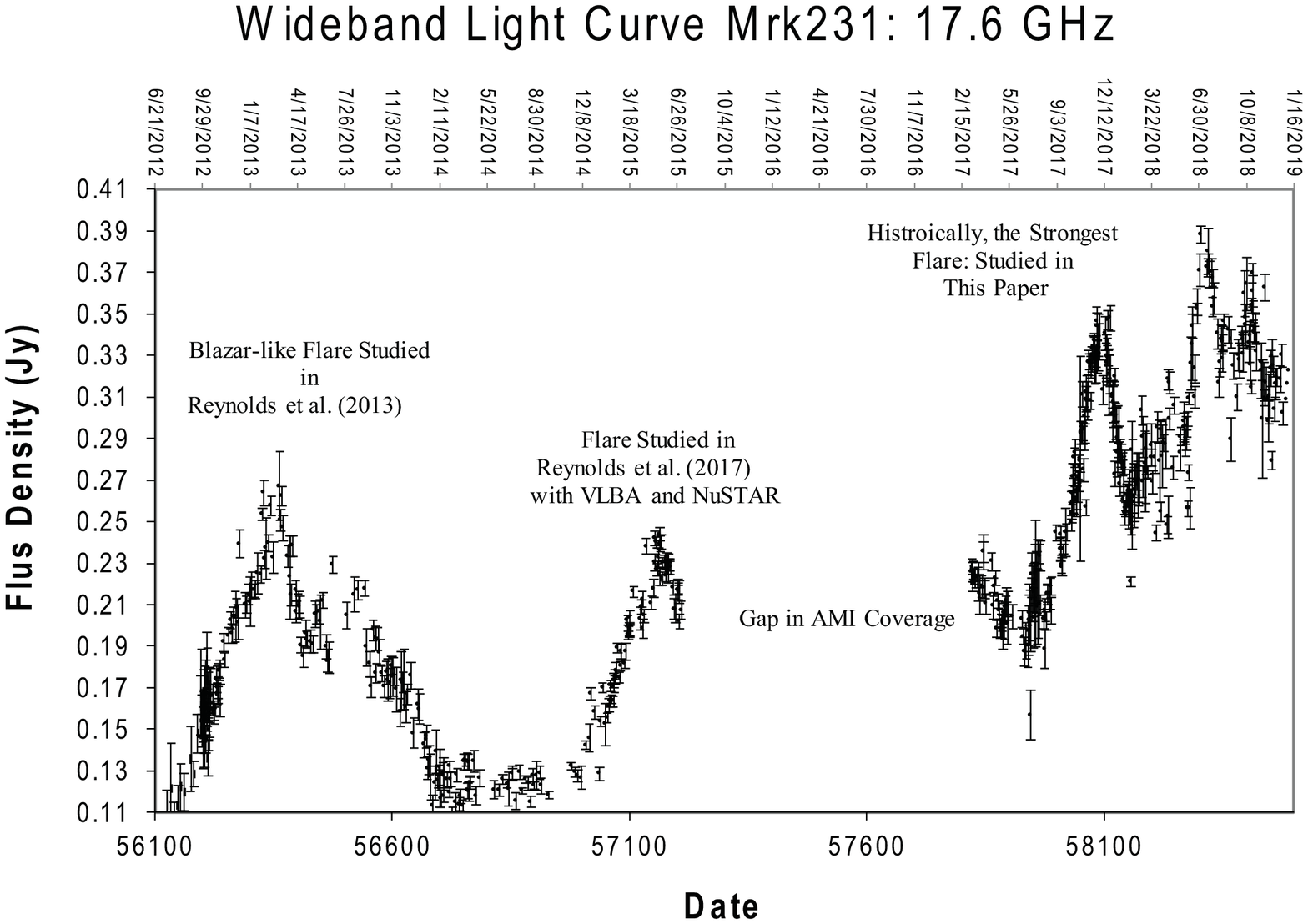}
\includegraphics[width= 0.8\textwidth]{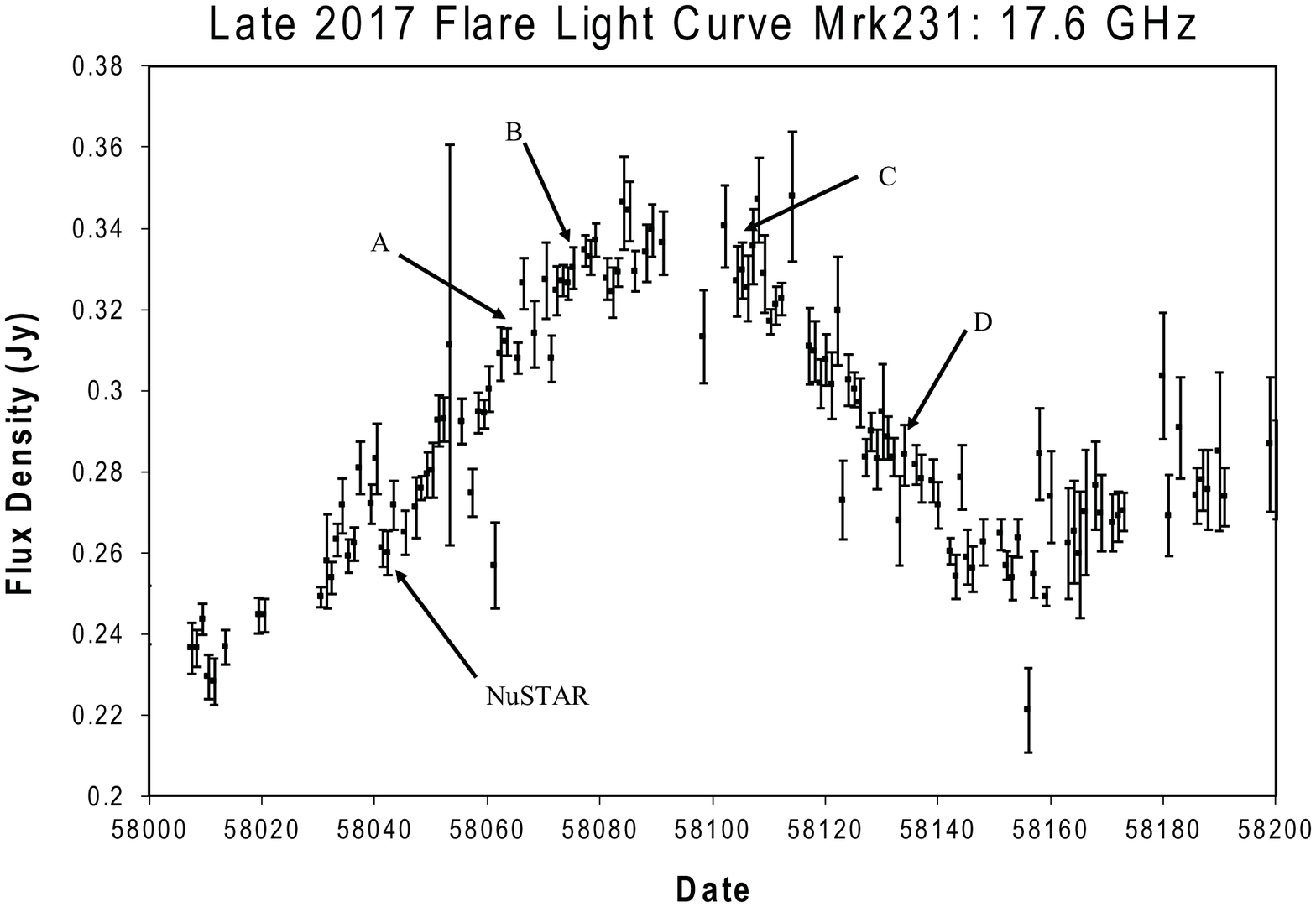}

    \caption{\label{2017.1}The top frame is the historic AMI 17.6~GHz light curve. The bottom frame is
the flare observed in this campaign with the four VLBA observation epochs A--D designated. }
\end{center}
\end{figure}

\section{Past VLBA 43~GHz Observations} The various VLBA epochs show two resolved components. There is a nuclear radio core with a highly variable spectrum. There is also a very steep spectrum, slowly variable, almost stationary secondary, K1. At 43~GHz, the core is more luminous during a flare.
\subsection{The 2006 Campaign}
The first 43~GHz observations were not triggered, but were obtained to see the benefit of higher resolution. They occurred in two epochs in 2006 separated by 3 months. In hind sight the observations were in a relatively low state for the core, $\sim 20$~mJy at 43~GHz (compare to Table~\ref{modelfits}). The results were analyzed in \citet{rey09}. Three observation frequencies were chosen: 15.3~GHz, 22.2 and 43.1~GHz. Using archival data, it was determined that time variability indicated that the line of sight, LOS, was restricted in order to avoid the inverse Compton catastrophe, $<25^{\circ}.6^{+3^{\circ}.2}_{-2^{\circ}.6}$. The core spectrum changed dramatically between the two epochs, with the 22~GHz flux density more than doubling. We were able to make a crude estimate of the jet power based on this spectral change: $Q\approx 3 \times 10^{43}$ erg/sec. It was also argued based on lower frequency VLBA in \citet{ulv99} that the spectrum of K1 is free-free absorbed near 5.4~GHz and 1.6~GHz. Based on the emission measure, K1 was interpreted as a radio lobe that results from the jet emitted from the core being stopped by entrainment of the BAL wind.
\subsection{The 2015 Campaign} It was determined that the 2006 campaign lacked the frequency coverage to properly describe the core spectrum, i.e. the SSA turnover. Thus, we proposed 8.4~GHz observations as well as 15.3~GHz, 22.2~GHz and 43.1~GHz observations. Furthermore, the observations were triggered by a strong flare at 17.6~GHz. Our primary goal was to detect a discrete ejection. We did find 1 clear and 1 marginal superluminal ejections at 43~GHz, but they were very faint, $\sim 1-2$~mJy and faded very quickly \citep{rey17}. Thus, they were not detectable in subsequent epochs 2--4 weeks later. Our three epochs of NuSTAR observations did not detect any X-ray evolution during the flare. The VLBA coverage stopped during the flare peak. There was not enough information to constrain the jet power. The spectra appeared to be SSA steep spectrum power laws (see Figure~4 for examples of SSA power laws in the current campaign), except in the last epoch. Curiously, in the last epoch the core spectrum changed to a flat spectrum power law with no evidence of SSA at frequencies lower than 8.4~GHz.
\par The clear detection of the ejection of a weak superluminal component was used in conjunction with Doppler aberration arguments to restrict the LOS to $< 23.5^{\circ}$. Similar to the upper bound found in the 2006 campaign from the time variability Doppler argument \citep{rey09}.

\section{The AMI Monitoring}
\citet{rey13} reported the methods of our radio monitoring with AMI at 13.5--18~GHz.  Historical data before AMI can be found in \citet{rey13,rey17}. Figure~1 shows the 17.6~GHz (the highest useable frequency) light curve from our more recent AMI monitoring. In November 2017, Mrk\~231 reached a historic high flux density in terms of our 17.6~GHz to 22~GHz monitoring. This triggered our 4 epoch 8.4~GHz, 15.3~GHz, 22.2~GHz and 43.1~GHz VLBA observations. The bottom frame of Figure~1 shows the 4 epochs, labeled A--D, superimposed on the flare profile. The flare amplitude is $\approx 100$~mJy above the background level, reaching 350~mJy total flux density. Epoch A is during the rise of the flare. Epochs B and C, straddle the peak of the flare and Epoch D is in the tail of the flare decay. Thus, the four epochs captured almost the entire time evolution of the flare. This is exploited in Sections 4--7. Before the flare could completely decay, another more powerful flare emerged, reaching a peak of 400~mJy.

\section{2017/18 VLBA Observations}
The VLBA project, BR214, observed 4 epochs between 2017 November 6 and 2018 January 14. The data were correlated on the VLBA DiFX correlator \citep{deller2011} and calibrated with NRAO's
Astronomical Imaging Processing System  following the standard procedures described in the AIPS Cookbook and using the updated amplitude calibration strategy described in \citet{walker2014}. All analysis was scripted in the ParselTongue interface \citep{kettenis2006}. Model fitting of the data was done in the DIFMAP package \citep{shepherd94,shepherd95}

Due to inclement weather and some equipment failures, we were unable to acquire data from all ten stations in any of the four epochs of observation. The St. Croix station was absent in all epochs due to hurricane damage and the Mauna Kea station was lost at 43~GHz in epochs B to D due to a problem with the Focus Rotation Module (FRM). These antennas provide the longest VLBA baselines and their loss significantly impacted the resolution of our observations. In addition freezing weather affected the performance of the FRM at a number of sites during all four epochs resulting in further data losses, disproportionately large at 43~GHz.

The lost stations significantly reduced the resolution and sensitivity. The resulting $u,v$ coverage was inadequate for resolving structures along the jet. We analyze the resulting data even though this circumstance made it impossible to directly monitor motion along the jet (a primary objective of the program) instead focusing on the spectral evolution of the flare.

\par The observations were phase-referenced to J1302+5748 ($1.3^{\circ}$ from Mrk\,231) at 8.4, 15, 22, and 43~GHz, following the strategies described in \citet{rey09} for projects BR295 and BP124. J1311+5513 ($\sim 80$~mJy of unresolved flux density at 43~GHz) was used as a secondary calibrator to check the quality of the phase referencing, which appeared to work well at all frequencies and epochs. For each epoch of our VLBA monitoring, an 8 hour observation with almost continuous recording at 2~Gbps (256~MHz bandwidth, dual
polarizations) was scheduled. Final on-source times and resultant image sensitivities (naturally weighted) for Mrk\,231 are summarised in Table~\ref{vlbaobs}.

\begin{table*}
\centering
\caption{\label{vlbaobs}Summary of BR214 VLBA Observations of Mrk\,231}
\begin{tabular}{ccl}
\hline \hline
Frequency~(GHz) & Sensitivity~(mJy/Beam) & Missing Antennas \\
\hline
\multicolumn{3}{c}{Epoch A -- 2017 Nov 6 -- MJD 58063} \\
\hline
8.4  & 0.35 & SC \\
15.4 & 0.33 & SC \\
22.2 & 0.13 & SC \\
43.3 & 0.19 & SC,PT \\
\hline
\multicolumn{3}{c}{Epoch B -- 2017 Nov 20 -- MJD 58077} \\
\hline
8.4 & 0.36 & SC \\
15.4 & 0.24 & SC \\
22.2 & 0.10 & SC \\
43.3 & 0.28 & BR,MK,SC \\
\hline
\multicolumn{3}{c}{Epoch C -- 2017 Dec 15 -- MJD 58102} \\
\hline
8.4 & 0.26 & SC \\
15.4 & 0.32 & SC \\
22.2 & 0.14 & SC \\
43.3 & 0.28 & BR,MK,SC \\
\hline
\multicolumn{3}{c}{Epoch D -- 2018 Jan 14 -- MJD 58132} \\
\hline
8.4 & 0.27 & SC \\
15.4 & 0.21 & SC \\
22.2 & 0.24 & SC \\
43.3 & 0.18 & BR,MK,SC \\
\hline
\end{tabular}
    \tablenotetext{}{\textbf{Note:} BR = Brewster, MK = Mauna Kea, PT = Pie Town, SC = Saint Croix}
\end{table*}

\par Table~\ref{modelfits} shows the results of our data reduction. It lists the fitted flux density of the two components in column (3) with the uncertainty in column (4). The absolute flux density uncertainties at 8.4, 15 and 22~GHz are 5\%, 5\% and 7\%, respectively \citep{hom02}. At 43~GHz, we use the amplitude stability of the phase calibrator J1302+5748 over the campaigns in 2015 and 2017 (epochs A, B and D) as the measure of the flux density uncertainty. We estimate a flux density uncertainty of 8\% at 43~GHz in this campaign, consistent with values used in our previous campaigns. The next four columns are the positional coordinates (relative to the assumed stationary component K1) and uncertainty. Columns (8)--(10) describe the Gaussian fit. The last column is the frequency.

\begin{table*}
 \centering
    \caption{\label{modelfits}Summary of Model Fits to VLBA Observations} {\scriptsize
\begin{tabular}{ccccccccccccc}
 \hline
1          & 2    &     3       & 4              & 5             & 6                &  7           & 8             & 9                        &  10   & 11   & 12  \\
 Component & MJD & Flux       & Flux           & X              & X                &  Y           & Y             & Major    & Axial & PA  & Frequency \\
           &       & Density     &  Density $\sigma$ &      & $\sigma$                 &    & $\sigma$    &  Axis                  & Ratio &       &\\
           &       &   (Jy)      & (Jy)              &  (mas)         & (mas)              &   (mas)     &    (mas)        &    (mas)               &        & (deg)       &  (GHz)\\

\hline
\hline
    \multicolumn{12}{c}{Epoch A}      \\
\hline
\hline
K1    &  58063 & 0.201 & 0.010  & 0.216 & 0.001  &-0.081 & 0.002 & 0.54 &   0.48  & 63.5 & 8.4 \\
Core  &  58063 & 0.100 & 0.005  & 1.189 & 0.002 & 0.383 & 0.003 & 0 & -    & - & 8.4 \\
K1    &  58063 & 0.081 & 0.004  & 0.154 & 0.002 & -0.145 & 0.002 & 0.32 &   0.63  & 72.0 & 15.3 \\
Core  &  58063 & 0.156 & 0.008  & 1.185 & 0.001 & 0.342 & 0.001 & 0 &  -   & - & 15.3 \\
K1    &  58063 & 0.044 & 0.003  & 0.156 & 0.001  & -0.097 & 0.001 & 0.30 &   0.86  & 66.3 & 22.2\\
Core  &  58063 & 0.131 & 0.009  & 1.236 & 0.001 & 0.389 & 0.001 & 0.10 & 0.45    & 61.7 & 22.2 \\
K1    &  58063 & 0.014 & 0.001  & 0.151 & 0.003  &0.383 & 0.003 & 0.28 & 0.67  & 83.3 & 43.1\\
Core  &  58063 & 0.082 & 0.007  & 1.22 & 0.001  &-0.137 & 0.001 & 0.14 &  0.30  & 69.1 & 43.1\\
\hline \hline
    \multicolumn{12}{c}{Epoch B} \\
\hline \hline
K1    &  58077 & 0.203 & 0.010  & 0.120 & 0.002  &-0.062 & 0.002 & 0.47 &   0.80  & 21.9 & 8.4 \\
Core  &  58077 & 0.117 & 0.006  & 1.078 & 0.003 & 0.407 & 0.003 & 0 &  -    & - & 8.4 \\
K1    &  58077 & 0.088 & 0.004  & 0.183 & 0.001  &-0.049 & 0.002 & 0.34 &   0.77  & 41.6 & 15.3 \\
Core  &  58077 & 0.178 & 0.009  & 1.192 & 0.001 & 0.432 & 0.001 & 0 &  -   & - & 15.3 \\
K1    &  58077 & 0.050 & 0.004  & 0.192 & 0.001  &-0.049 & 0.001 & 0.37 &   0.77  & 48.1 & 22.2\\
Core  &  58077 & 0.158 & 0.011  & 1.210 & 0.001 & 0.434 & 0.001 & 0.16 & 0.65    & 28.3 & 22.2 \\
K1    &  58077 & 0.015 & 0.001  & 0.145 & 0.004  &-0.003 & 0.004 & 0.28 &   0.84  & 23.5 & 43.1\\
Core  &  58077 & 0.100 & 0.008  &  1.199 & 0.001 & 0.496 & 0.001 & 0.18 & 0.42    & 25.9 & 43.1\\
\hline \hline
    \multicolumn{12}{c}{Epoch C}  \\
\hline \hline
K1    &  58102 & 0.209 & 0.010  & 0.253 & 0.002  & 0.143 & 0.002 & 0.54 &   0.75  & 87.3 & 8.4 \\
Core  &  58102 & 0.120 & 0.006  & 1.212 & 0.03 & 0.619 & 0.003 & 0 &  -    & - & 8.4 \\
K1    &  58102 & 0.086 & 0.004  & 0.264 & 0.002  & 0.081 & 0.002 & 0.38 &   0.88  & 59.9 & 15.3 \\
Core  &  58102 & 0.179 & 0.009  & 1.264 & 0.001 & 0.578 & 0.001 & 0.40 &  0.40    & 55.4 & 15.3 \\
K1    &  58102 & 0.049 & 0.003  & 0.217 & 0.002  & -0.040 & 0.002 & 0.41 &   1  & 63.2 & 22.2\\
Core  &  58102 & 0.137 & 0.010  & 1.209 & 0.001 & 0.453 & 0.001 & 0.28 & 0.67    & -33.8 & 22.2 \\
K1    &  58102 & 0.014\tablenotemark{a} & 0.002  & 0.219 & 0.007  &-0.063 & 0.008 & 0.32 &   0.27  & -74.3 & 43.1\\
Core  &  58102 & 0.077\tablenotemark{a} & 0.006  &  1.227 & 0.001 & 0.457 & 0.002 & 0.14 & 0.75    & -63.2 & 43.1\\
\hline \hline
    \multicolumn{12}{c}{Epoch D}    \\
\hline \hline
K1    &  58132 & 0.234 & 0.012  & 0.235 & 0.002  &-0.099 & 0.001 & 0.70 &   1  & 63.2 & 8.4 \\
Core  &  58132 & 0.115 & 0.006  & 1.195 & 0.004 & 0.395 & 0.003 & 0 &  -    & - & 8.4 \\
K1    &  58132 & 0.088 & 0.004  & 0.096 & 0.002  &-0.062 & 0.001 & 0.37 &   0.35  & -18.6 & 15.3 \\
Core  &  58132 & 0.148 & 0.008  & 1.084 & 0.001 & 0.416 & 0.001 & 0.366 & 0.55    & 54.0 & 15.3 \\
K1    &  58132 & 0.057 & 0.004  & 0.157 & 0.002  &-0.003 & 0.002 & 0.37 &   0.84  & 22.4 & 22.2\\
Core  &  58132 & 0.108 & 0.008  & 1.139 & 0.001 & 0.465 & 0.001 & 0.24 & 0.69    & 34.3 & 22.2 \\
K1    &  58132 & 0.016 & 0.001  & 0.154 & 0.003 & 0.001 & 0.003 & 0.27 &  0.75   & 65.5 & 43.1\\
Core  &  58132 & 0.056 & 0.005  & 1.171 & 0.001  & 0.482 & 0.001 & 0.24 &   0.39  & 54.2 & 43.1\\

\hline
\hline
\end{tabular}}
\tablenotetext{a}{Due to an absolute flux calibration issue at 43~GHz in epoch~C, the flux density at 43~GHz was scaled by the phase calibrator, J1302+5748. The flux density of J1302+5748 was scaled to its average value from epochs B and D. This produced a factor 1.37 re-scaling of the flux densities. After re-scaling, the flux density of K1 is constant at 43~GHz during epochs A--D within uncertainties as expected.}
\end{table*}

\par Our best radio images are from epoch A because we had 8 stations and the long baselines associated with Mauna Kea at 43~GHz. The other epochs B to D were missing Mauna Kea and had varying degrees of missed observing time at other stations due to freezing weather conditions (see Table 1). The 8.4~GHz to 22.2~GHz images from epoch A are shown in Figure~2. They are very similar to images from previous campaigns in \citet{rey09,rey17}, but with lower resolution.
\begin{figure*}
\begin{center}
\includegraphics[width= 0.45\textwidth,angle=0]{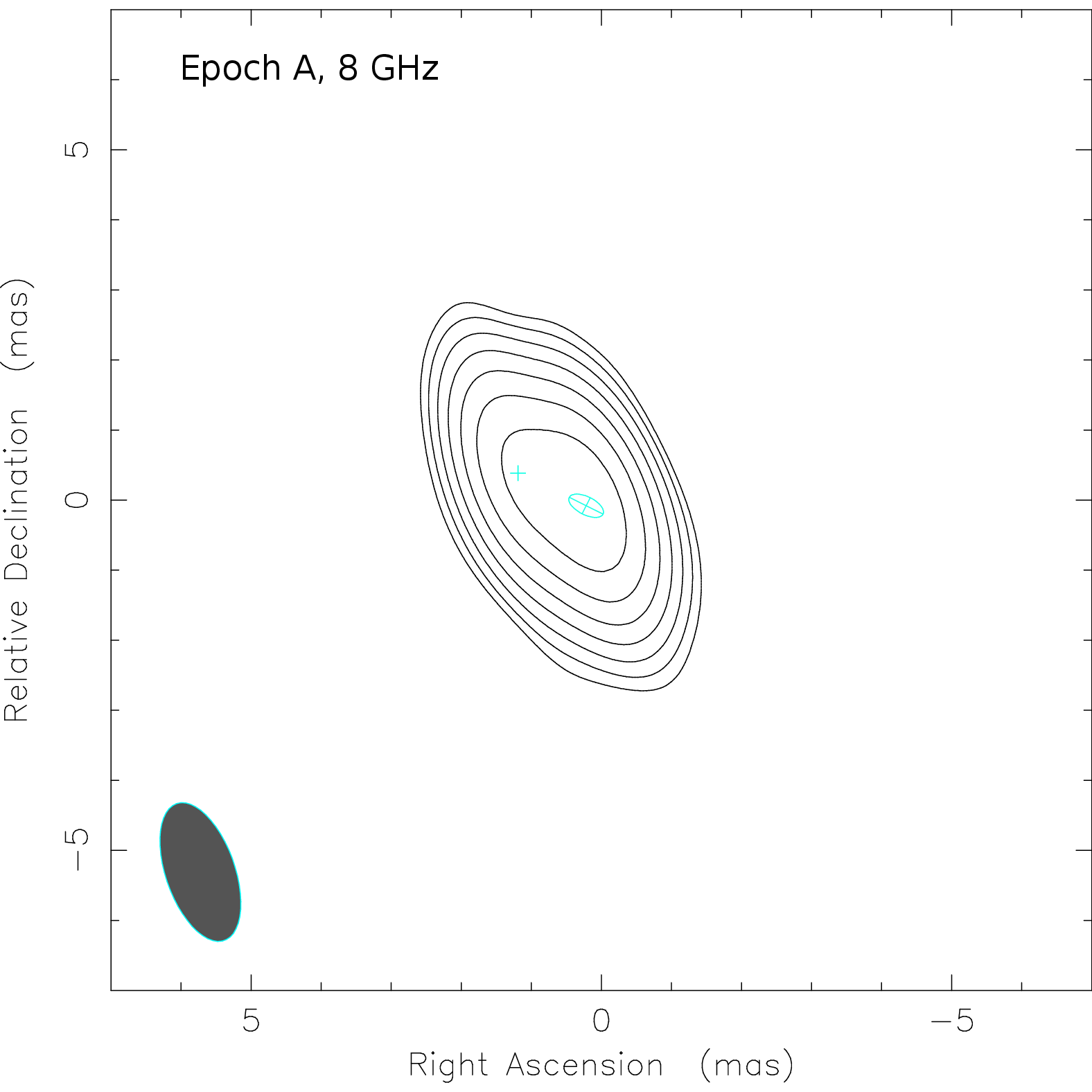}
\includegraphics[width= 0.45\textwidth,angle=0]{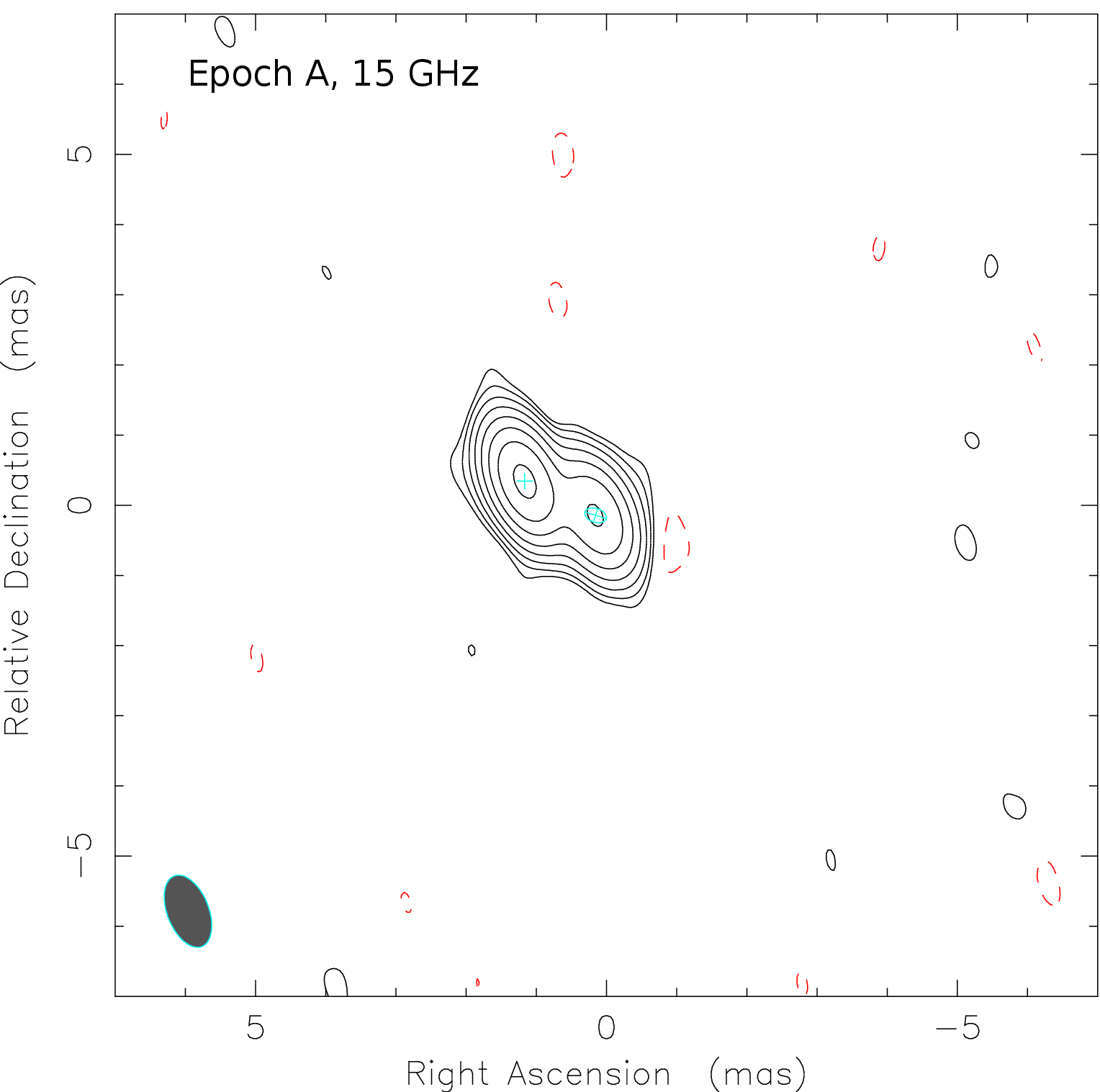}\\
\includegraphics[width= 0.45\textwidth,angle=0]{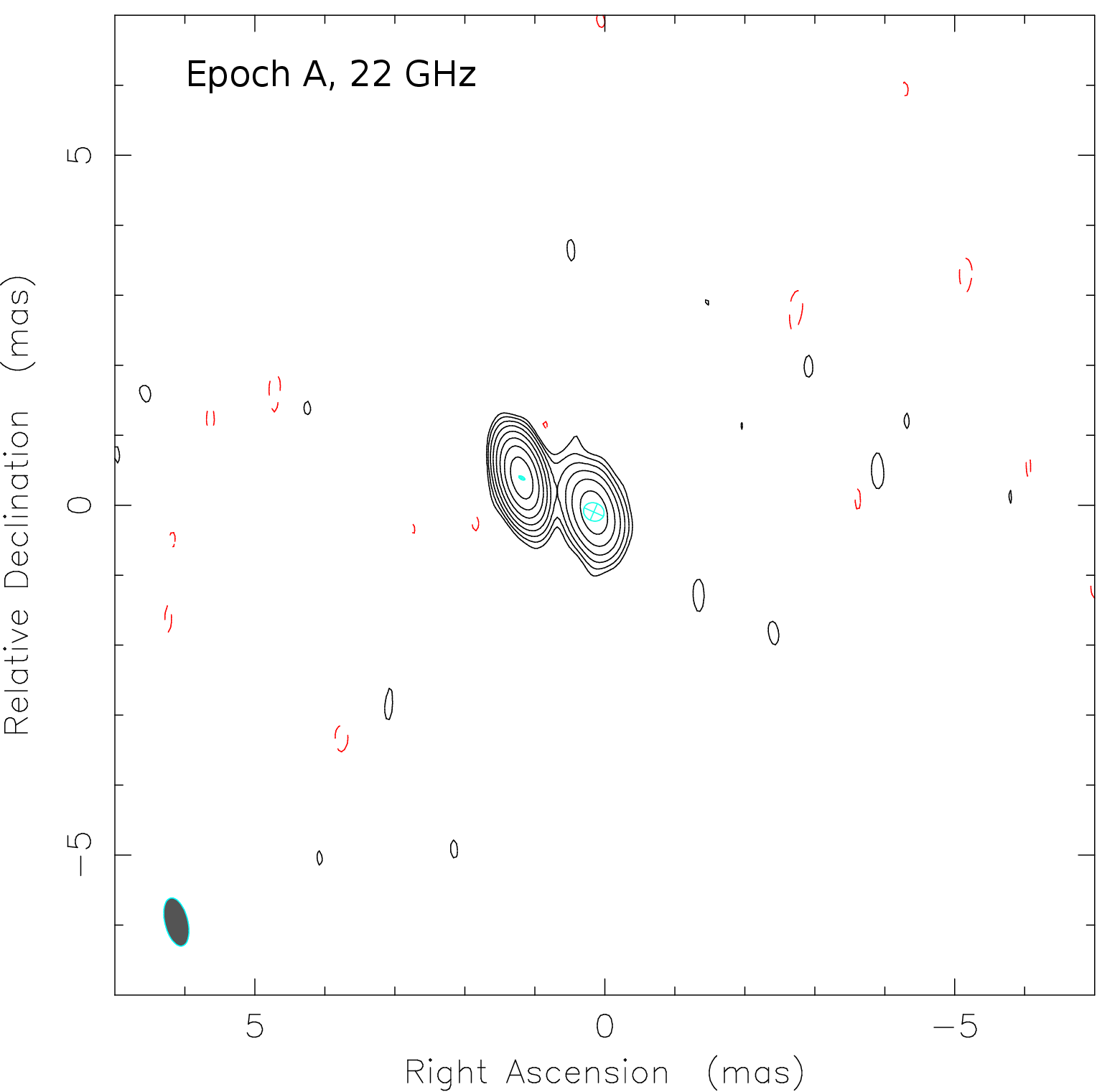}
    \caption{\label{2017A_8}Uniformly weighted VLBA images at 8.4 (top left), 15.3 (top right) and 22.2~GHz (bottom) from epoch A. Contours are factors of two from a bottom contour of $\pm$ 1.6, 1.0 and 0.6~mJy/beam respectively. Peak fluxes are 199, 156 and 127~mJy/beam respectively.  The crosses and ellipses with inscribed crosses denote the locations of point source components and the locations and dimensions of the Gaussian elliptical components in Table~\ref{modelfits}, respectively. The grey ellipse in the bottom left indicates the restoring beam.
    }
\end{center}
\end{figure*}

The 43.1~GHz images in Figure~3 have the highest resolution and the most potential for elucidating the compact nuclear structure. Due to the missing stations described in Table 1, the 43.1 GHz observations during epochs B-D were significantly compromised. Epoch C had the most issues. Even though we were able to perform a self-calibration in epoch C, we measured ``out of family" low flux densities in both the core and K1 in the following sense. K1 varies slowly and should be relatively constant from epochs B--D \citep{rey09,rey17}. Furthermore, both of our calibrator sources had reduced amplitude as well. The main suspect is an error in the opacity correction which is large at 43.1~GHz but the confounding issue has not been identified. However, we can use the quasi-stationary calibrators as amplitude calibrators.  The phase calibrator, J1302+5748, has significant flux at 43.1~GHz and is our most reliable absolute flux density calibrator. In order to re-scale the flux densities we averaged the flux density at 43~GHz for J1302+5748 in epochs B and D. This was 1.37 times larger than the flux density of J1302+5748 in epoch C. Thus, we use this factor to re-scale all the epoch~C 43.1~GHz flux densities for Mrk\~231 in Table~\ref{modelfits}, as described in the table note.
\begin{figure*}
\begin{center}
\includegraphics[width= 0.45\textwidth,angle =0]{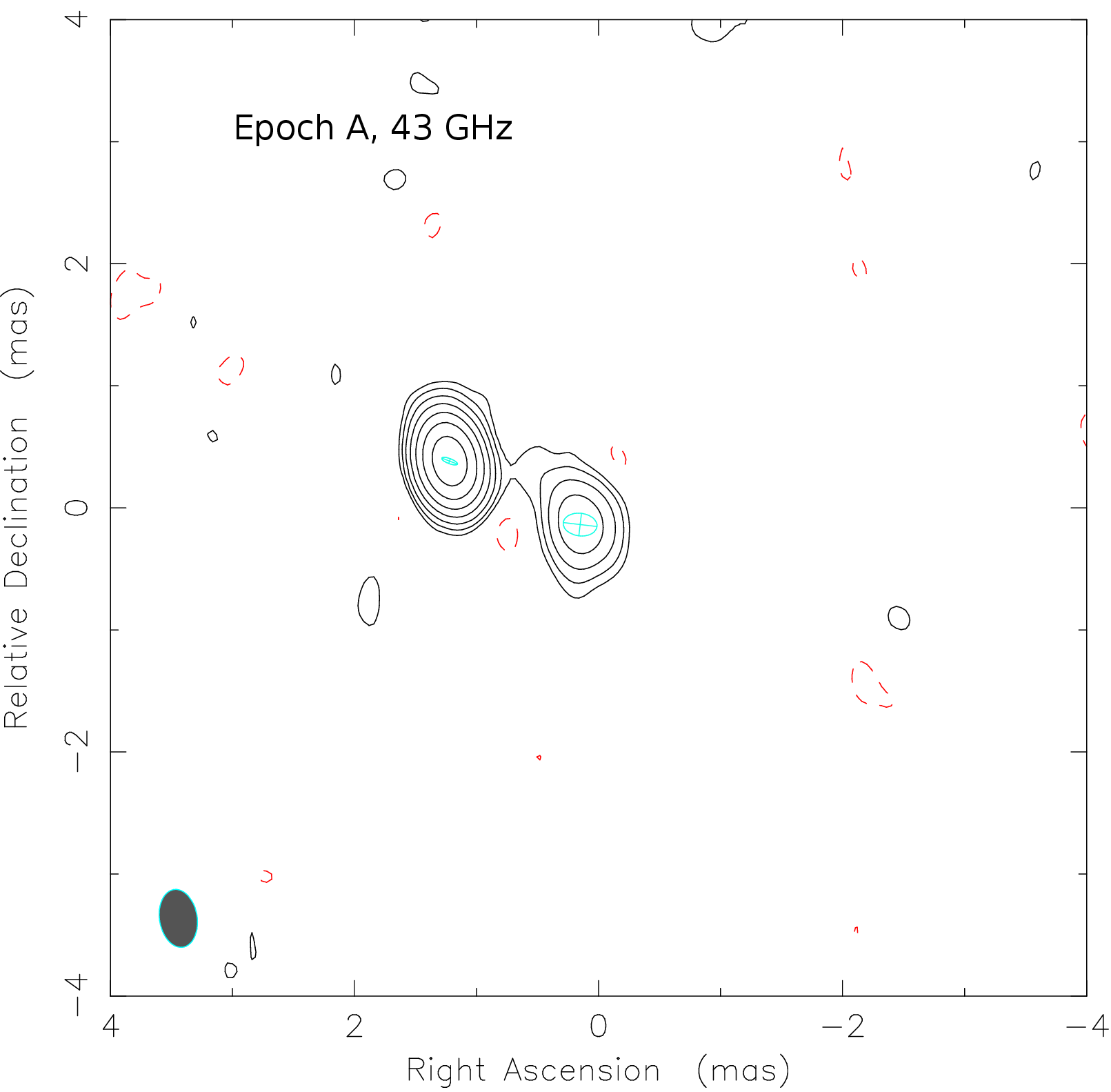}
\includegraphics[width= 0.45\textwidth,angle =0]{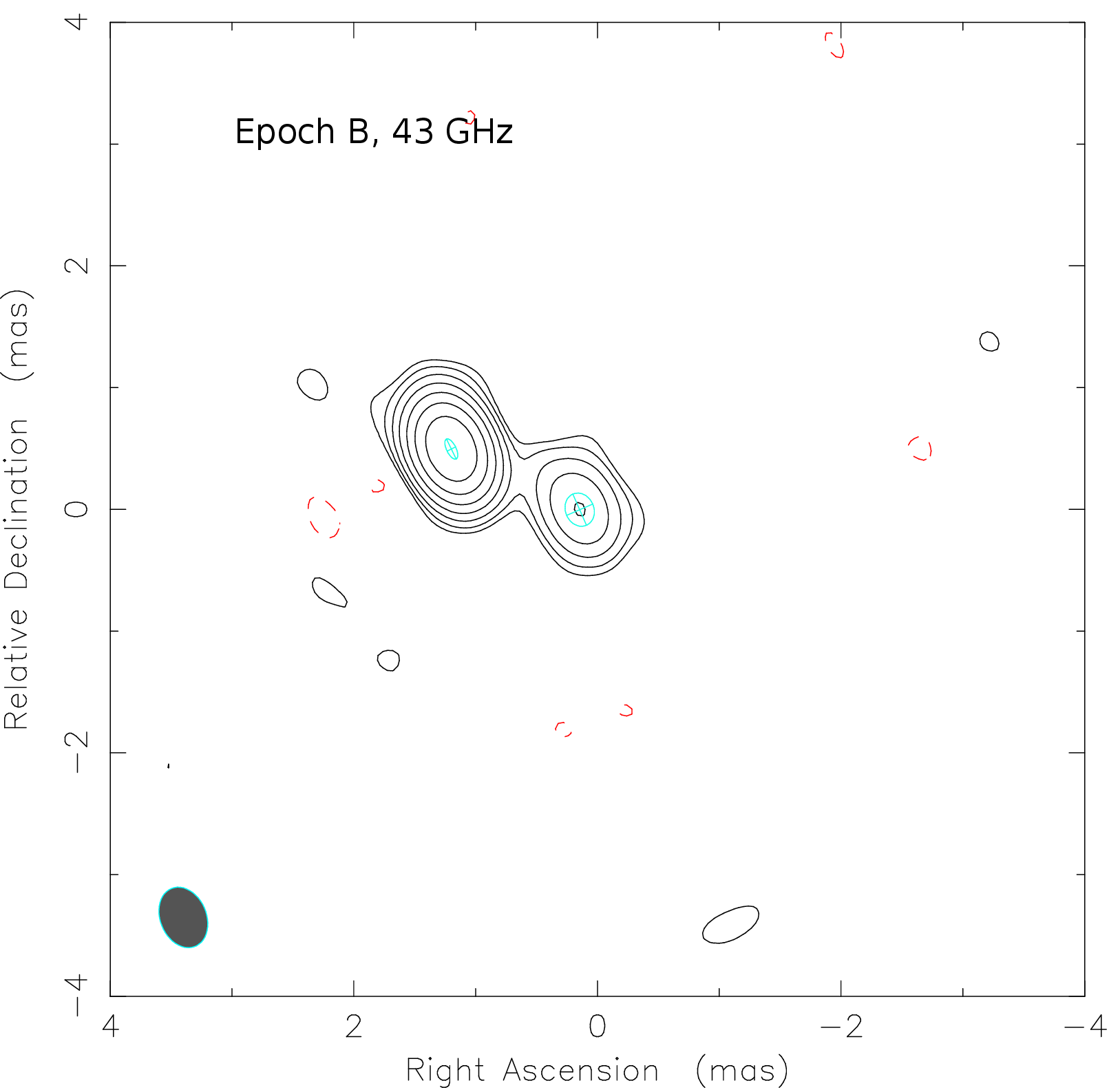}\\
\includegraphics[width= 0.45\textwidth,angle =0]{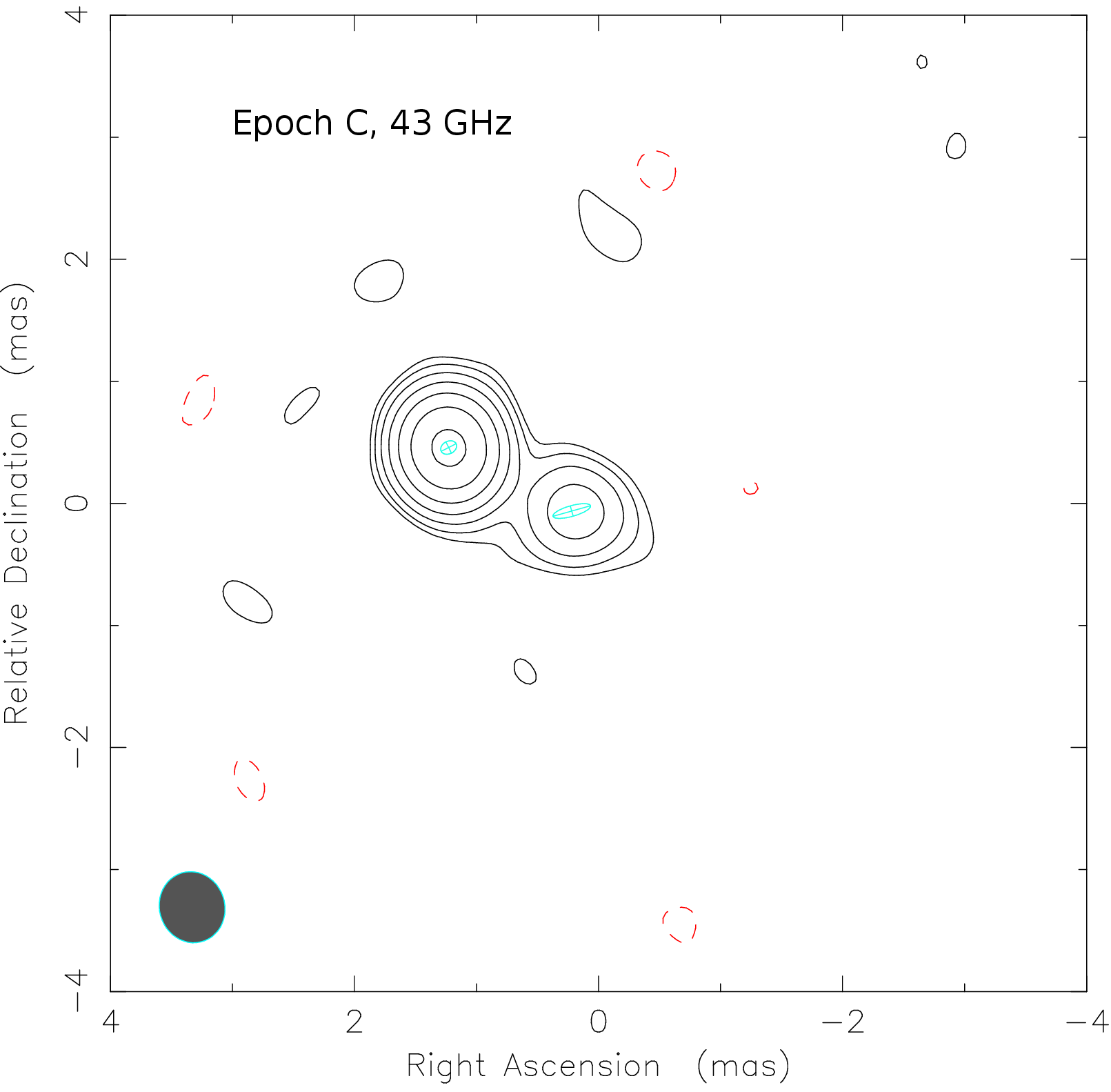}
\includegraphics[width= 0.45\textwidth,angle =0]{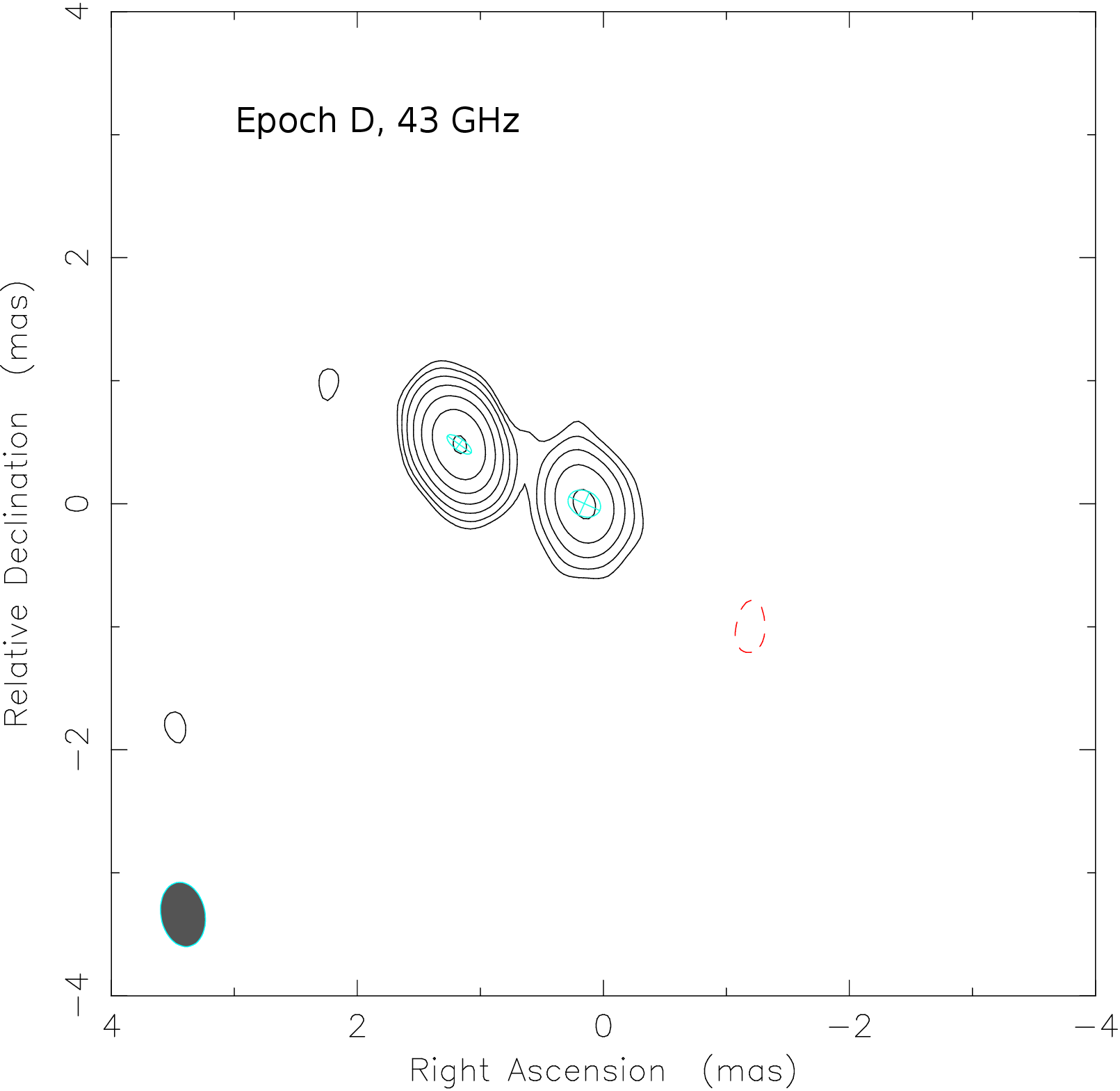}

 \caption{\label{2017A_43}Naturally weighted VLBA images at 43.1~GHz. The top row
from left to right are epochs A and B. The bottom row from left to
    right are epochs C and D. Contours are factors of two from a bottom contour of $\pm$ 0.7~mJy/beam respectively. Peak fluxes are 73, 91, 53 and 47~mJy/beam respectively. Note that epoch C was later rescaled as explained in Table~\ref{modelfits}.
The ellipses with inscribed crosses denote the locations and dimensions of the Gaussian elliptical components in Table~\ref{modelfits}. The grey ellipse in the bottom left indicates the restoring beam.}
\end{center}

\end{figure*}
\begin{figure*}
\begin{center}
\includegraphics[width= 0.9\textwidth,angle =0]{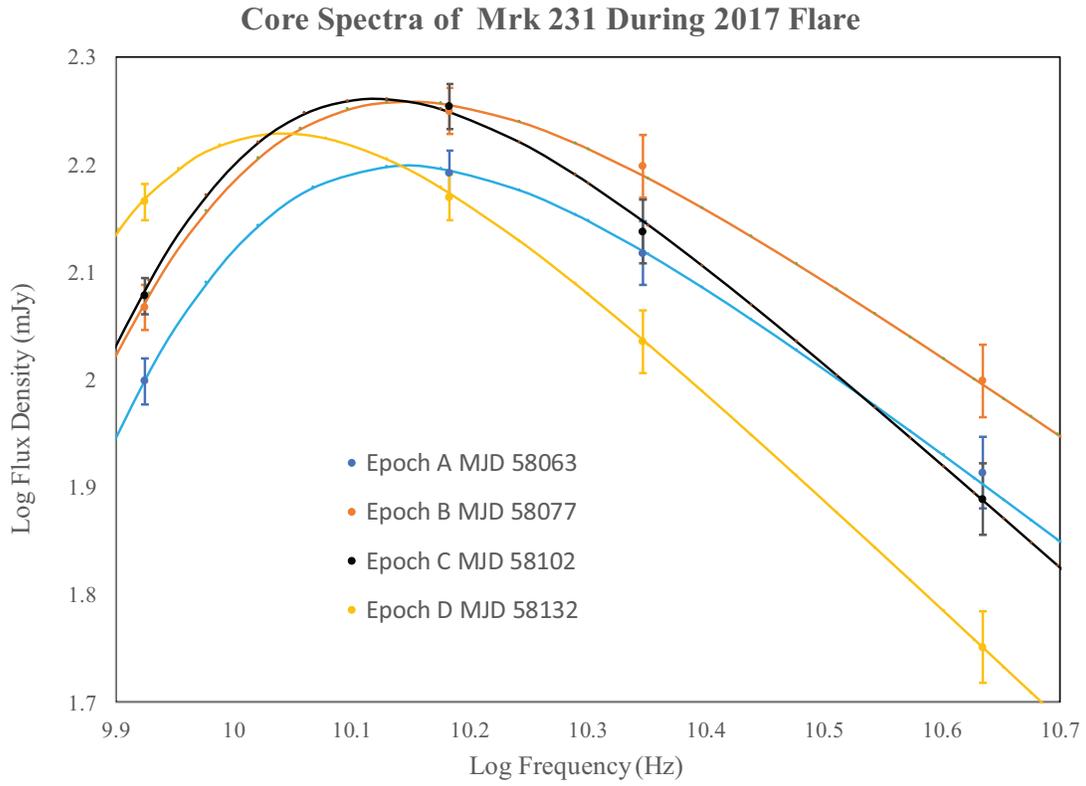}
\caption{\label{2017CS}The spectrum of the unresolved radio core in the four
    epochs. The data are from Table~\ref{modelfits}. The continuous curves are SSA
power-law fits to the data. These fits are the motivation for the physical models described in Sections 5--7.}
\end{center}
\end{figure*}

\begin{figure*}
\begin{center}
\includegraphics[width= 0.8\textwidth,angle =0]{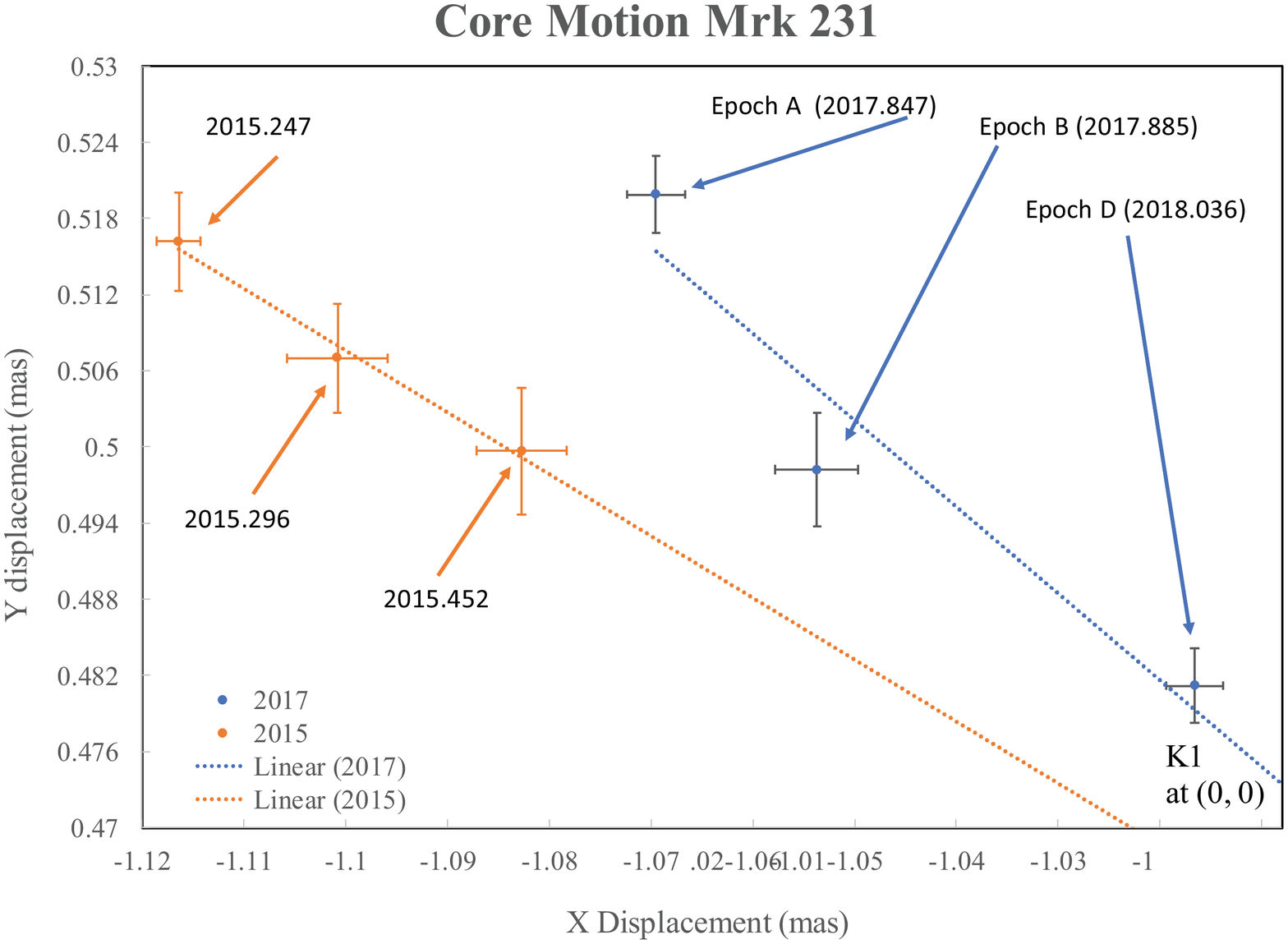}
    \caption{\label{2017CM}The position of the ``core" relative to the astrometric origin, K1, as a function of time. There was too much degradation to the epoch C observation for it to be useful in this sensitive measurement. The similar plot from \citet{rey17} was added for historical reference. The better fit to a linear trajectory might have been the consequence of having all 10 VLBA stations in the 2015 campaign.}
\end{center}
\end{figure*}

\subsection{Core Spectra} Even though a large amount of information was lost due to
poorer than expected resolution, we did anticipate the additional information that can be found in the core spectrum with a
multi-frequency observation plan.  The spectra of the
unresolved core in each epoch are shown in Figure~4. In \citet{rey09}, we used
the detailed nature of the spectral shape to argue that the spectrum of the
radio core was most likely a power-law synchrotron spectrum that was seen
through a SSA screen as opposed to a synchrotron power-law spectrum that was
seen through an opaque screen caused by free-free absorption. The plots in Figure~4 show synchrotron self-absorbed (SSA) power
law fits to the data. This is a simple spectral model that has been used to describe previous observations \citep{rey09,rey17}. In Section 6, we will interpret these fits in terms of the physical models that emit these spectra.
\subsection{Core Motion}
Figure~5 shows the other important piece of information that can be retrieved from Table~\ref{modelfits} about the core dynamics during the flare. It has been shown that K1 is stationary (with respect to the phase reference calibrator) within the uncertainties of the VLBA over two decades \citep{ulv99,ulv00,rey09,rey17}. Thus, as in \citet{rey17}, we use K1 as the origin for an astrometric description of core motion during the flare. There is  more scatter in the 2017 data then we see in the 2015 data due to the loss of observing stations in the current analysis. We do not consider the poor epoch C observation in this analysis, due to the problems discussed earlier in this section. The apparent velocity is $v_{\rm{app}} \approx 0.97 c $ based on the best fit line, directed towards K1. What we identify as the core appears to be a discrete ejection that dominates the total nuclear flux density at 43~GHz. The nuclear region and the ejected components are too small to be resolved based on the fits in Table~\ref{modelfits}. The displacements are about 1/3 of the beam width (Figure~2) in our two best observations epochs A and B, and less in later epochs. The next section will present models of the physical nature of the discrete high brightness feature moving towards K1.
\section{Synchrotron Self-Absorbed Homogeneous Plasmoids}
The core emission is unresolved at 43~GHz in all our campaigns including 2017/2018 (see Figure~3). Thus, there is no evidence of a preferred geometry. Consequently, there is no observational motivation to choose a model more complicated than a uniform spherical plasmoid. The simple homogeneous spherical volume model was shown by \citet{van66} to provide a very useful means of understanding the spectra and time evolution of astrophysical radio sources. Ostensibly, the invocation of a simple plasmoid seems less relevant than the notion of a continuous jet. A continuous jet has been inferred in extragalactic radio sources based on the need to replenish the distant radio lobes with energetic plasma in order to counteract synchrotron cooling \citep{kai97}. In a practical context blazar calculations typically reduce to plasmoid models. They are referred to as single zone homogeneous, spherical models \citep{ghi10}. But this is merely semantics. In order to appreciate the utility of these simple models, consider the following relevant circumstance. If there is a strong flare produced by a large explosive event in the jet (i.e., increase power injected at the base, a Parker instability at the point of origin or other instability) then the details of a much weaker background jet add no practical insight into the physical parameters responsible for the emission from an over-luminous finite region. This appears to be the case for Mrk\,231 in which a weak background jet feeds K1 as evidenced by the large effects of synchrotron cooling that results in a very steep spectrum. The plasmoid is more versatile than a continuous jet model for our purposes. Because the explosive event responsible for the flare is so intense, we cannot assume its origin or nature based on the weak, invisible background jet. The plasmoid need not be similar to the rest of the jet. We need not assume whether it is protonic or leptonic, nor do we need to assume a bulk velocity or whether it is magnetically dominated or dominated by bulk kinetic energy. Instead, we will determine this from the data.

\par The specific application of this model that is implemented in the following has been successful in describing a wide range of astrophysical phenomena. This spherical homogeneous discrete ejection model has been used to study the major flares in the Galactic black hole accretion system of GRS~1915+105 \citep{pun12}. It was applied to the neutron star binary merger that was the gravity wave source, GW170817, and associated gamma ray burst (GRB), GRB 170817A \citep{pun19}. It has been previously used in the study of quasar radio flares in Mrk\,231 \citep{rey09}. The primary advantage of the method is that the SSA turnover provides information on the size that cannot be obtained from unresolved VLBA images. For example, our best radio image, epoch A, can only resolve a region of $2.5 \times 10^{17}$~cm or larger. Assuming a size equal to the resolution limit of the telescope generally results in plasmoid energy estimates that are off by one or more orders of magnitude due to the exaggerated volume of plasma \citep{fen99,pun12}. The first subsection will describe the underlying physics and the next subsection describes physical quantities of interest in the spherical plasmoids.
\subsection{The Underlying Physical Equations}
It is important to distinguish between quantities measured in the plasmoid frame of reference and those measured in the Earth observer frame of reference. The strategy will be to evaluate the physics in the plasma rest frame then transform the results to the observer's frame for comparison with observation. The underlying power law for the flux density is defined as $S_{\nu}(\nu= \nu_{o}) = S\nu_{o}^{-\alpha}$, where $S$ is a constant. Observed quantities will
be designated with a subscript, ``o", in the following expressions. The observed frequency is related to the emitted frequency, $\nu$, by $\nu_{o} = \delta \nu$. The bulk flow Doppler factor of the plasmoid, $\delta$,
\begin{equation}
\delta = \frac{\gamma^{-1}}{1-\beta \cos{\theta}},\; \gamma^{-1} = 1- \beta^{2}\;,
\end{equation}
where $\beta$ is the normalized three-velocity of bulk motion and $\theta$ is the angle of the motion to the line of sight (LOS) to the observer.
The SSA attenuation coefficient in the plasma rest frame is given by \citep{gin69},
\begin{eqnarray}
&& \mu(\nu)=\overline{g(n)}\frac{e^{3}}{2\pi
m_{e}}N_{\Gamma}(m_{e}c^{2})^{2\alpha} \left(\frac{3e}{2\pi
m_{e}^{3} c^{5}}\right)^{\frac{1+2\alpha}{2}}\left(B\right)^{(1.5
+\alpha)}\left(\nu\right)^{-(2.5 + \alpha)}\;,\\
&& \overline{g(n)}= \frac{\sqrt{3\pi}}{8}\frac{\overline{\Gamma}[(3n
+ 22)/12]\overline{\Gamma}[(3n + 2)/12]\overline{\Gamma}[(n +
6)/4]}{\overline{\Gamma}[(n + 8)/4]}\;, \\
&& N=\int_{\Gamma_{min}}^{\Gamma_{max}}{N_{\Gamma}\Gamma^{-n}\,
d\Gamma}\;,\; n= 2\alpha +1 \;,
\end{eqnarray}
where $\Gamma$ is the ratio of lepton energy to rest mass energy, $m_{e}c^2$ and $\overline{\Gamma}$ is the gamma function. $B$ is the magnitude of the total
magnetic field. The power-law spectral index for the flux density is $\alpha=(n-1)/2$. The low energy cutoff, $E_{min} = \Gamma_{min}m_{e}c^2$,
is constrained loosely by the data in Table~\ref{core_spectra} and Figure~4. The fact that the core is very luminous at $\nu_{o}=8.4$~GHz, means that the lepton energy distribution is not cutoff near this frequency. This is not a very stringent bound since it is just a consequence of the fact that 8.4~GHz is our lowest observing frequency. It would be a major coincidence if this were not to provide a loose upper bound on $\Gamma_{min}$. Note that the SSA opacity in the observer's frame, $\mu(\nu_{o})$, is obtained by direct substitution of $\nu =\nu_{o} / \delta$ into Equation (2).
\begin{figure*}
\begin{center}
\includegraphics[width= 0.8\textwidth,angle =0]{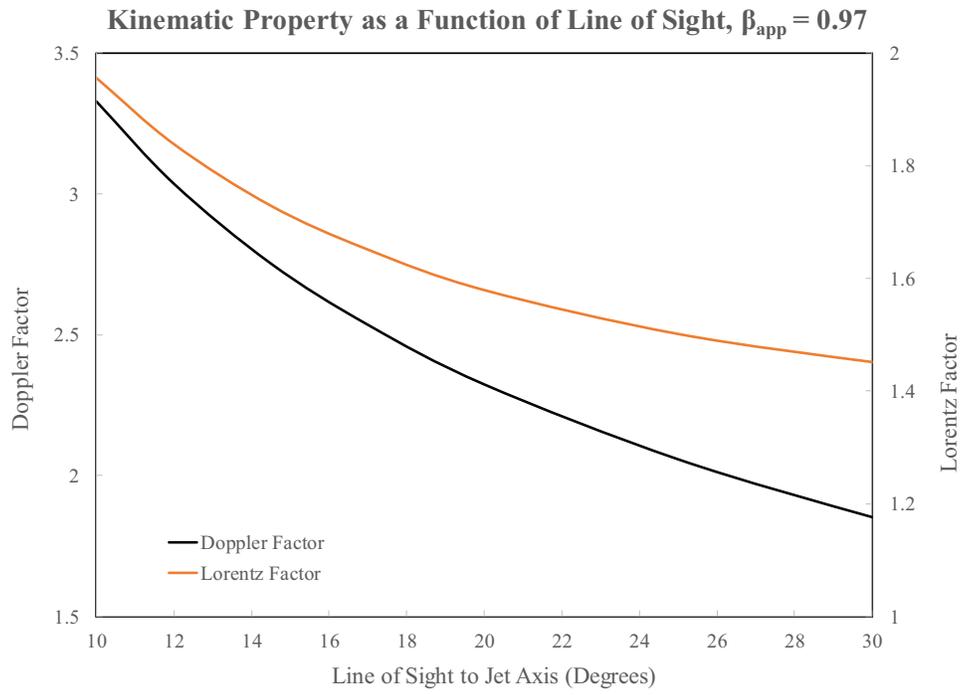}
    \caption{\label{2017.kin}The Doppler and Lorentz factors as a function of the line of sight to the direction of plasmoid motion. A value of $\beta_{\rm{app}}=0.97$ derived from Figure~5 is assumed in the calculations.}
\end{center}
\end{figure*}
\par The uniform, homogeneous approximation yields a simplified solution to the radiative transfer equation \citep{gin65,van66}
\begin{eqnarray}
&& S_{\nu_{\o}} = \frac{S_{o}\nu_{o}^{-\alpha}}{\tau(\nu_{o})} \times \left(1 -e^{-\tau(\nu_{o})}\right)\;, \; \tau(\nu_{o}) \equiv \mu(\nu_{o}) L\;, \; \tau(\nu_{o})=\overline{\tau}\nu_{o}^{(-2.5 +\alpha)}\;,
\end{eqnarray}
where $\tau(\nu)$ is the SSA opacity, $L$ is the path length in the rest frame of the plasma, $S_{o}$ is a normalization factor and $\overline{\tau}$ is a constant. There are three unknowns in Equation (5), $\overline{\tau}$, $\alpha$ and $S_{o}$. These are three constraints on the following theoretical model that are estimated from the observational data. These three constraints are used to establish the uniqueness and the existence of solutions in the following.
\par The theoretical spectrum is parameterized by Equations (2)--(5) and the synchrotron emissivity that is given in \citet{tuc75} as
\begin{eqnarray}
&& j_{\nu} = 1.7 \times 10^{-21} [4 \pi N_{\Gamma}]a(n)B^{(1
+\alpha)}\left(\frac{4
\times 10^{6}}{\nu}\right)^{\alpha}\;,\\
&& a(n)=\frac{\left(2^{\frac{n-1}{2}}\sqrt{3}\right)
\overline{\Gamma}\left(\frac{3n-1}{12}\right)\overline{\Gamma}\left(\frac{3n+19}{12}\right)
\overline{\Gamma}\left(\frac{n+5}{4}\right)}
       {8\sqrt\pi(n+1)\overline{\Gamma}\left(\frac{n+7}{4}\right)} \;.
\end{eqnarray}
One can transform this to the observed flux density, $S(\nu_{o})$, in the optically thin region of the spectrum using the relativistic transformation relations from
\citet{lin85},
\begin{eqnarray}
 && S(\nu_{o}) = \frac{\delta^{(3 + \alpha)}}{4\pi D_{L}^{2}}\int{j_{\nu}^{'} d V{'}}\;,
\end{eqnarray}
where $D_{L}$ is the luminosity distance and in this expression, the
primed frame is the rest frame of the plasma. These are the basic equations needed to fit the data in Figure~4.
\subsection{Physical Quantities that Characterize Spherical Models}
The apparent velocity constrains the kinematics of the plasmoid \citep{ree66}:
\begin{equation}
v_{app}/c=\beta_{app} = \beta \sin{\theta}/(1- \beta \cos{\theta})\approx 0.97\;,
\end{equation}
where we used the result from Figure~5. The apparent motion and the LOS upper limit $25^{\circ}.6^{+3^{\circ}.2}_{-2^{\circ}.6}$ noted in Section 2.1 constrain the dynamics. The validity of these numbers is a major assumption of this analysis.
\par The models that produce the fits in Figure~4 are of interest if we can deduce the physical parameters in Equations (1)--(9) that are responsible for the spectra. Furthermore, we want to know the kinematic consequences of these parameters within a specific model. There are two basic classes of models. First are discrete ejections of plasma, a ballistic ejection of a plasmoid. This is essentially a blob of magnetized gas. The gas is turbulent and the magnetic field is randomized and tangled \citep{mof75}. In this model, the magnetic field strength behaves as an extra component of the internal gas pressure and acts as an expansive force on the plasmoid \citep{pun08}. Secondly, the discrete unresolved emission might be ``knots" or regions of enhanced dissipation within a continuous jet. In the jet scenario, the magnetic field is predominantly an organized toroidal magnetic field, $B_{\phi}^{'}$, in the rest frame of the plasma in contrast to the turbulent magnetic field of the discrete ejections \citep{bla79}. The behavior of this magnetic field is much different than the turbulent magnetic field. It acts as a hoop stress that provides a confining force on the outflow \citep{pun08}.
\subsubsection{The Poynting Flux in a Strong Knot in the Jet}
We review the derivation of the dominance of the toroidal component of the magnetic field in order to derive an estimate of the poloidal Poynting flux. The dominance of the toroidal component of the magnetic field is a consequence of the perfect magnetohydrodynamic (MHD) assumption and approximate angular momentum conservation in the jet \citep{bla79}. The total angular momentum flux per unit magnetic flux in the observer's coordinate system is \citep{pun08},
\begin{equation}
L \approx k\mu r_{\perp}\gamma v^{\phi} - \frac{c}{4\pi}B^{\phi}r_{\perp}\;, \; k \equiv \frac{N\gamma v^{P}}{B^{P}};,
\end{equation}
where $k$ is the perfect MHD conserved mass flux per unit poloidal magnetic flux, $B^{P}$ and $v^{P}$ are the poloidal magnetic field and the poloidal velocity, respectively. The quantity $N$ is the number density in the plasma rest frame, $\mu$ is the specific enthalpy and $r_{\perp}$ is the cylindrical radius. The first term is the mechanical angular momentum flux per unit magnetic flux and the second term is the electromagnetic angular momentum flux per unit magnetic flux. The notion of describing the angular momentum per unit poloidal magnetic flux is physically motivated by angular momentum conservation and perfect MHD. In perfect MHD, angular momentum is transported conserved along each poloidal magnetic field line in a wind or jet \citep{pun08}. Jets expand as they propagate \citep{bla79}. Thus, $r_{\perp}$ is much larger in the knot than at the jet base. Consider a jet that is highly magnetic, either Poynting flux dominated or at a minimum the mechanical angular momentum flux per unit magnetic flux and the electromagnetic angular momentum flux per unit magnetic flux are comparable. Then, by Equation (10), if $L$ is approximately constant, in the limit of large $r_{\perp}$, the following must be true for highly magnetic jets (assuming angular momentum outflow, i.e. $L>0$, $B^{P}>0$, $B^{\phi}<0$)
\begin{equation}
 v^{\phi}< \frac{L}{k\mu r_{\perp}\gamma}\; ,\; 0.5\frac{4\pi}{c r_{\perp}}L< \mid B^{\phi}\mid < \frac{4\pi}{c r_{\perp}}L; .
\end{equation}
The lower bound in the second relationship merely expresses the fact that by assumption (magnetic jet), the electromagnetic angular momentum flux is at least equal to the mechanical angular momentum flux. By poloidal magnetic flux conservation $B^{P} \sim  r_{\perp}^{-2}$, so by Equation (11), $\mid B^{\phi} \mid \gg B^{P}$ at large $r_{\perp}$. To estimate the poloidal Poynting flux, $S^{P}$, in the plasmoid, first transform fields to the observer's frame
\begin{equation}
B^{\phi} =\gamma B_{\phi}^{'} \quad E^{\perp} = \frac{v^{P}}{c} \gamma B_{\phi}^{'} - \frac{v^{\phi}}{c} \gamma B^{P} \approx \frac{v^{P}}{c} \gamma B_{\phi}^{'} \;,
\end{equation}
where $ E^{\perp}$ is the poloidal electric field orthogonal to the magnetic field direction. The poloidal Poynting flux in the observer's frame, $S^{P}$, along the jet direction is \citep{pun08}:
\begin{equation}
S^{P}= \frac{c}{4\pi}E^{\perp} B^{\phi}  \approx \frac{c}{4\pi}\gamma^{2} \beta [B_{\phi}^{'}]^{2} \;,\; B_{\phi}^{'} \approx B\;.
\end{equation}
\begin{figure*}
\begin{center}
\includegraphics[width= 0.48\textwidth,angle =0]{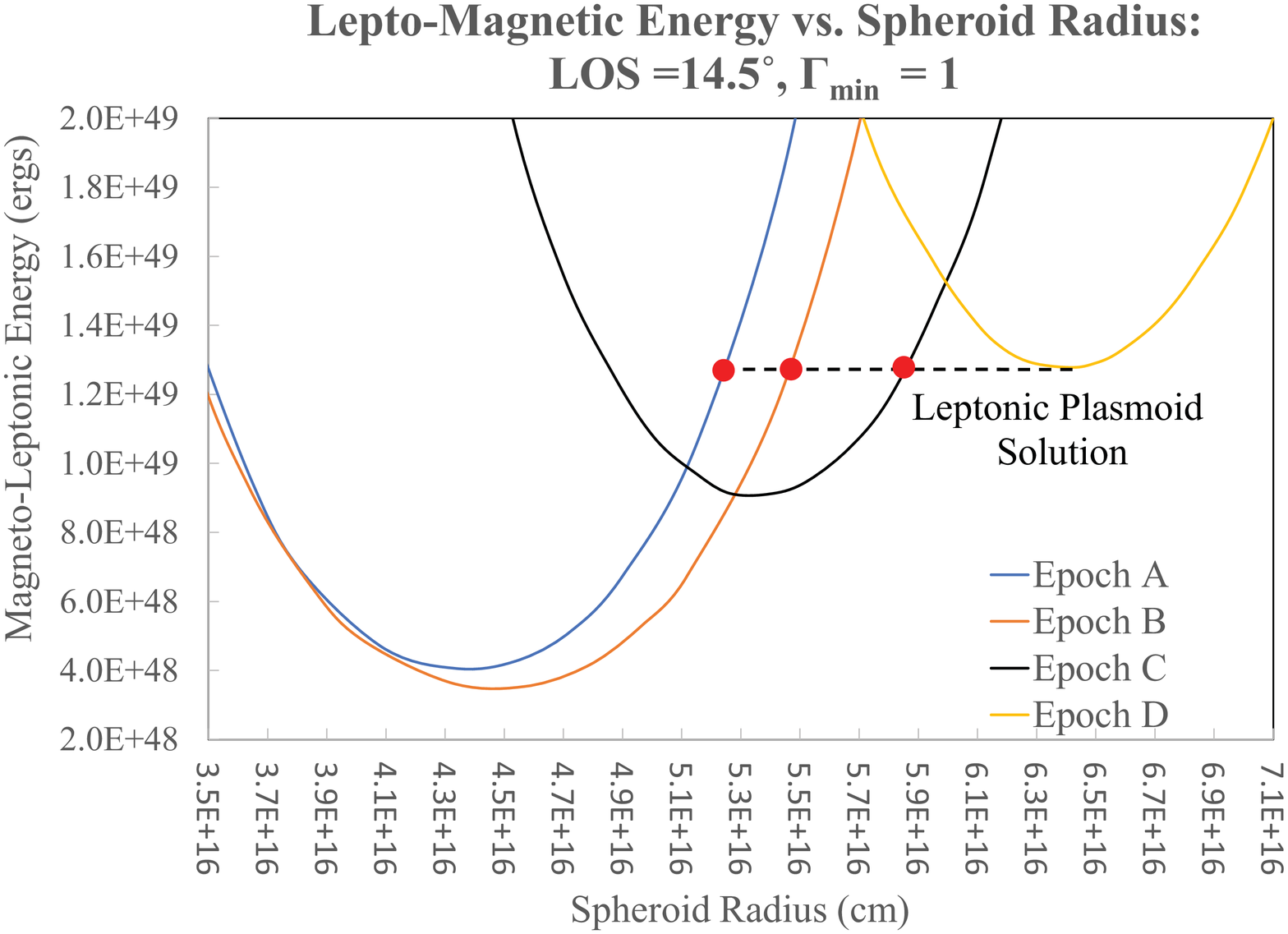}
\includegraphics[width= 0.48\textwidth,angle =0]{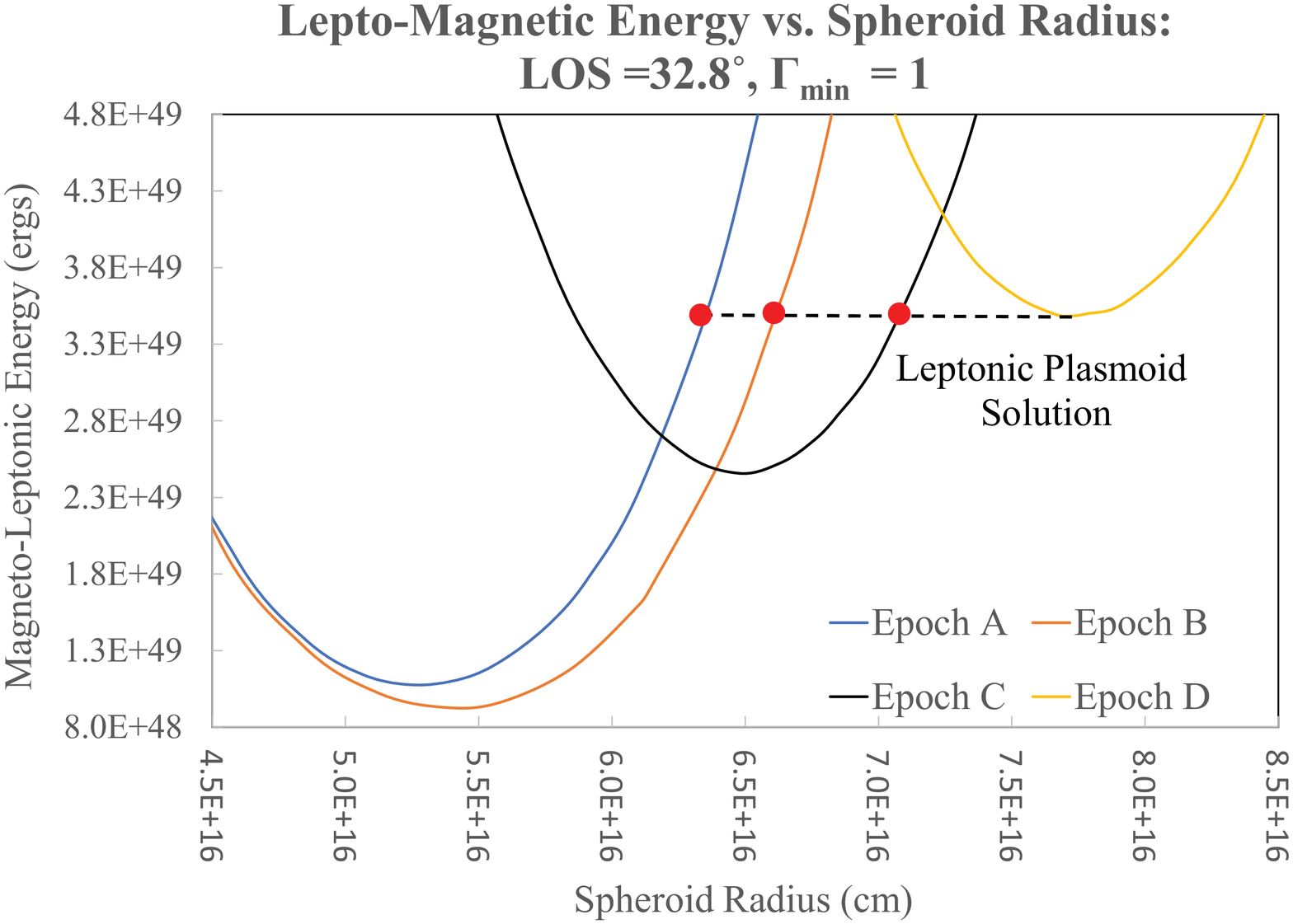}
\includegraphics[width= 0.48\textwidth,angle =0]{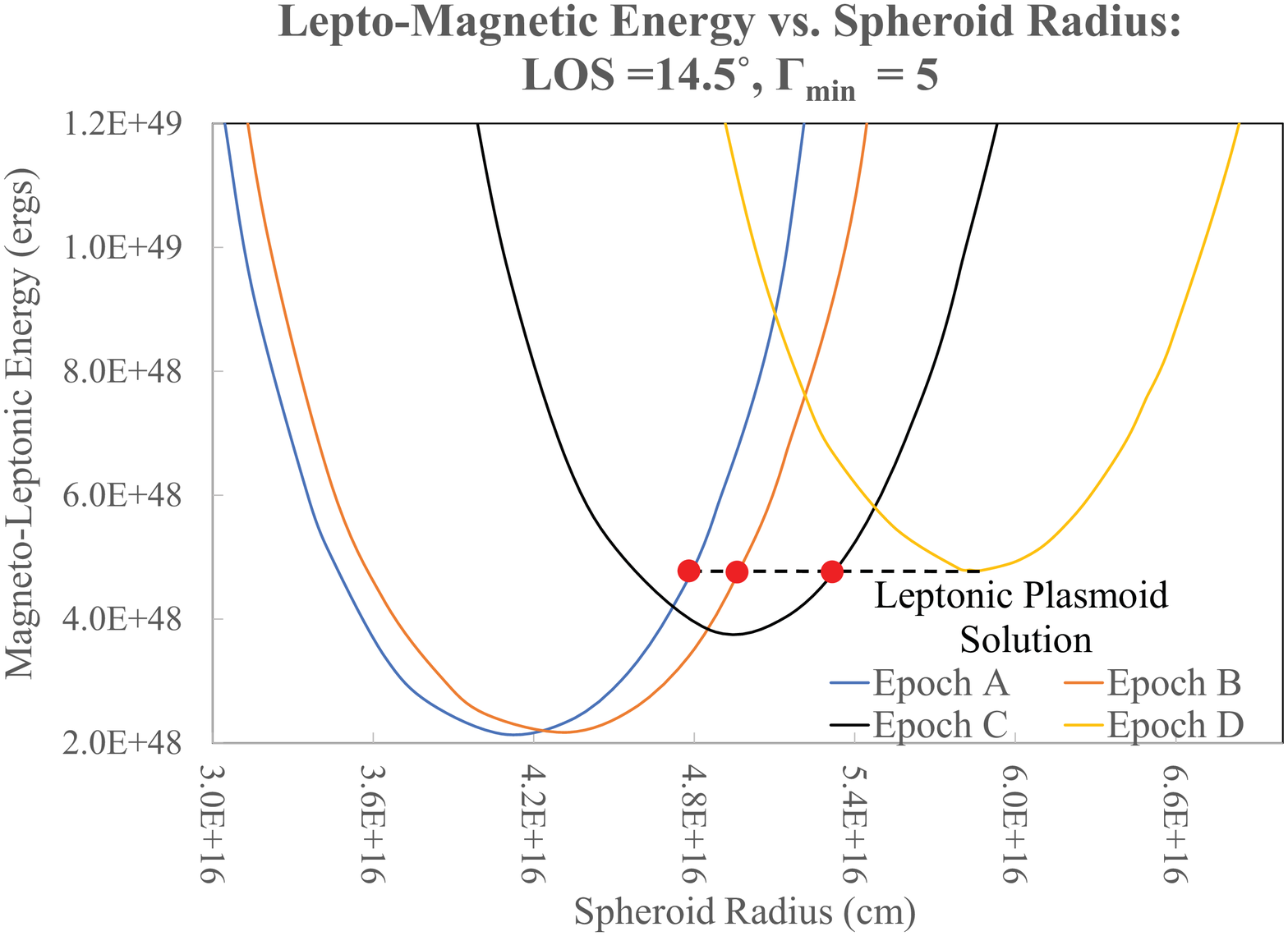}
\includegraphics[width= 0.48\textwidth,angle =0]{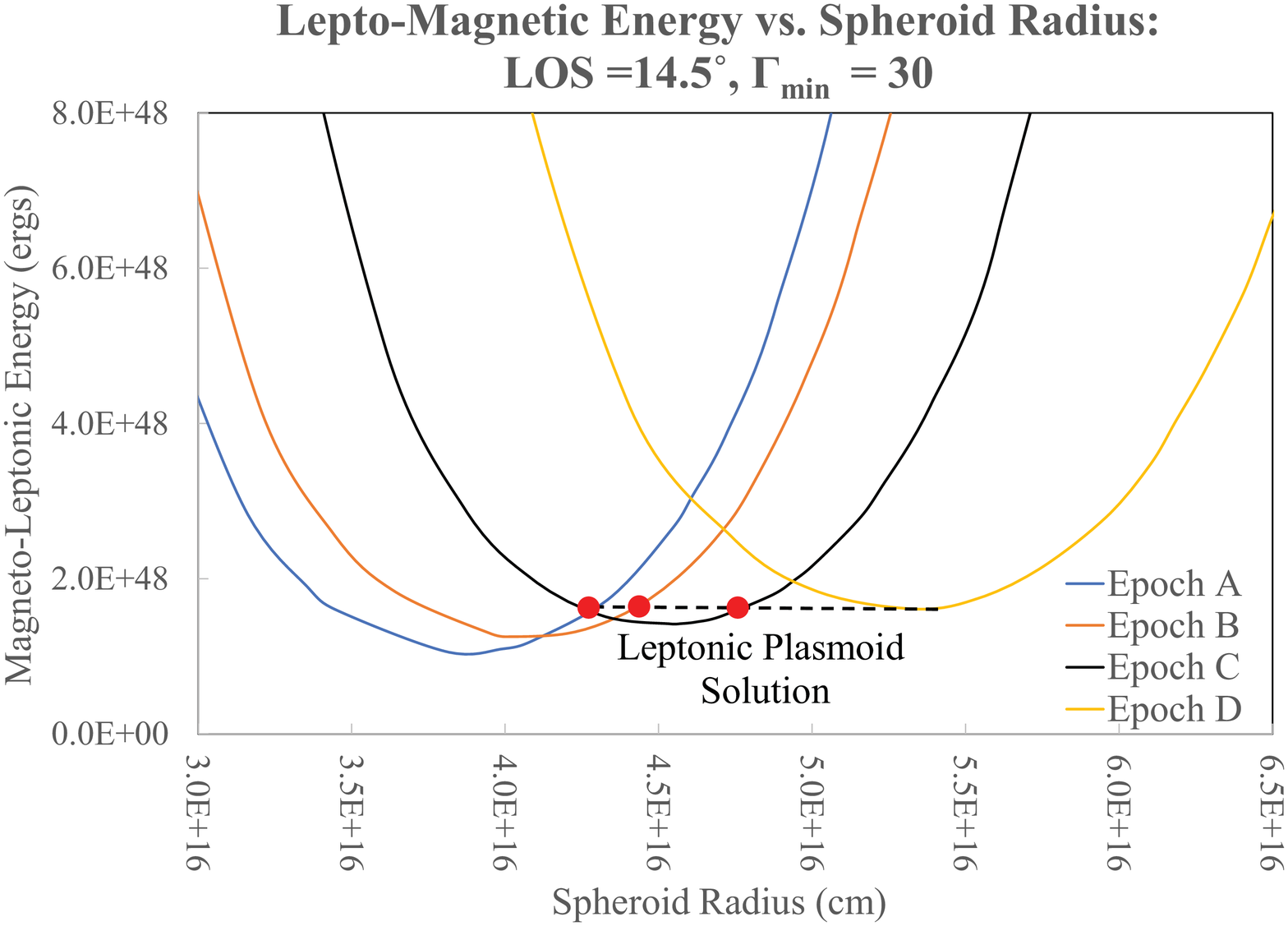}

    \caption{\label{2017E1}The time evolution of constant $E(\rm{lm})$ discrete electron-positron ejection solutions described in Section 7.1. Note that there could be some protons, but their effect is negligible in the energetics by assumption. All of these solutions generate the fitted spectra in Figure~4. The red dots show the location of the constant $E(\rm{lm})$ solution in each epoch on the infinite 1-D solution space plotted in the $R$-$E(\rm{lm})$ plane. Magnetic energy is converted to plasma internal energy throughout the time evolution. The top two frames show the effect of varying the LOS and the bottom two frames show the effect of varying $E_{\rm{min}}$.}
\end{center}
\end{figure*}
\subsubsection{Mechanical Contributions to the Energy Flux} The energy content is separated into two pieces. The first is the kinetic energy of the protons, $\mathcal{K}(\mathrm{protonic})$,
\begin{eqnarray}
 && \mathcal{K}(\mathrm{protonic}) = (\gamma - 1)Mc^{2}\;,
\end{eqnarray}
here $M$ is the mass of the plasmoid. The other piece is named the lepto-magnetic energy, $E(\mathrm{lm})$, and is composed of the volume integral of the leptonic internal energy density, $U_{e}$, and the magnetic field energy density, $U_{B}$. It is straightforward to compute the lepto-magnetic energy in a spherical volume,
\begin{eqnarray}
 && E(\mathrm{lm}) = \int{(U_{B}+ U_{e})}\, dV = \frac{4}{3}\pi R^{3}\left[\frac{B^{2}}{8\pi}
+ \int_{\Gamma_{min}}^{\Gamma_{max}}(m_{e}c^{2})(N_{\Gamma}E^{-n + 1})\, d\,E \right]\;.
\end{eqnarray}
The lepto-magnetic energy is often argued to be minimized in astrophysical sources \citep{har04,cro05,kat05}. The relevance of this will be discussed in terms of the time evolution of various models of the flare in the next section. The leptons also have a kinetic energy analogous to Equation (14),
\begin{eqnarray}
 && \mathcal{K}(\mathrm{leptonic}) = (\gamma - 1)\mathcal{N}_{e}m_{e}c^{2}\;,
\end{eqnarray}
where $\mathcal{N}_{e}$ is the total number of leptons in the plasmoid.
\par The other quantities of interest are the protonic and leptonic field aligned (jet axis aligned) poloidal energy fluxes in the frame of the observer. The protonic energy flux in the frame of the observer is approximately the kinetic energy flux density given
\begin{equation}
\mathcal{E}(\mathrm{proton}) = N(\gamma-1)\gamma v^{P}m_{p}c^{2}\;,
\end{equation}
where $m_{p}$ is the mass of the proton. The leptonic jet aligned poloidal energy flux density in the frame of the observer is
\begin{equation}
ke(\mathrm{leptonic}) = N \gamma v^{P}\left[\gamma\mu c^{2}\right]\;,
\end{equation}
where $k$ is the mass flux defined in Equation (10) and $e$ is the specific energy of a lepton \citep{pun08}.
From an energetics standpoint it is useful to subtract off the rest mass term which merely reflects the conservation of particle number from the source to the jet and cannot be converted into different forms of energy flux. This has been called the free thermo-kinetic energy flux and was defined in \citet{mck12} as
\begin{equation}
\mathcal{E}(\mathrm{leptonic}) = N\gamma v^{P}\left[\gamma\mu c^{2}-m_{e}c^{2}\right]\;.
\end{equation}
This would be appropriate for protonic outflows. However, for an electron-positron outflow, the mass is not conserved due to pair creation and in fact in the models considered the number of leptons in the plasmoid increases substantially over time. Thus, $ke(\mathrm{leptonic})$ is the more appropriate energy flux in this study. The specific enthalpy decomposes as
\begin{equation}
N\mu = U_{e} + P \;,
\end{equation}
where the relativistic pressure, $P \approx (1/3) (U_{e}-Nm_{e}c^{2})$ \citep{wil99}.
\begin{figure*}
\begin{center}
\includegraphics[width= 0.85\textwidth,angle =0]{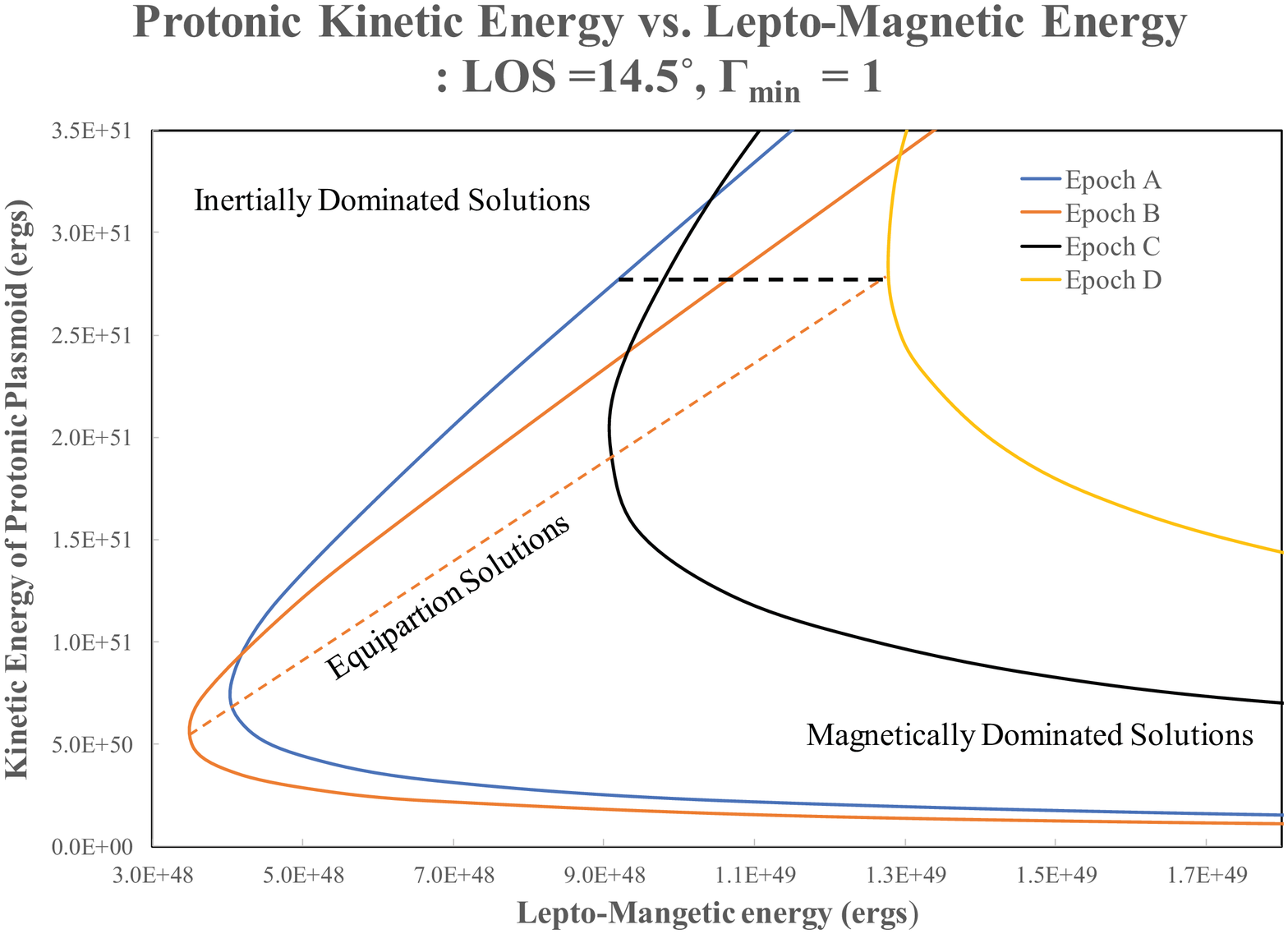}
    \caption{\label{2017KE}The discrete protonic plasmoid ejection model discussed in Section 7.2. It is a plot of the plasmoid kinetic energy as a function of $E(\rm{lm})$. The solution has a constant kinetic energy as indicated by the dashed black line. Note the dashed red line called ``equipartition". Above the line, $E(\rm{lm})$ is dominated by internal plasma energy. Below the line, $E(\rm{lm})$ is magnetic energy dominated. Thus, the solution converts mechanical energy to magnetic energy. This seems inconsistent with magnetic acceleration of the plasmoid to $\beta_{\rm{app}} =0.97$. Thus, the solution has no explanation of its existence.}
\end{center}
\end{figure*}
\begin{table}
    \caption{\label{core_spectra}Fits to the Core Spectra}
{\footnotesize\begin{tabular}{cccccc} \tableline\rule{0mm}{3mm}
Epoch &  Date & Peak Frequency & Peak Luminosity & Spectral Index & Number Index\\
  &  (MJD) & (GHz) & ergs/sec & $\alpha$ & $n$ \\
   &   & $\nu_{\rm{peak}}$ & $\nu L_{\nu}(\nu = \nu_{\rm{peak}})$ &  &  \\
\tableline \rule{0mm}{3mm}
A & 58063 & 13.9 & $8.08 \times 10^{40}$ & 0.83  & 2.66 \\
B & 58077 & 14.1 & $1.12 \times 10^{41}$ & 0.75 & 2.50  \\
C & 58102 & 13.2 & $1.05 \times 10^{41}$  & 0.97 & 2.94 \\
D & 58132 & 11.0 & $8.15 \times 10^{40}$ & 1.03 & 3.06
\end{tabular}}
\end{table}
\section{Fitting the Data with a Specific Spherical Model}
Mathematically, the theoretical determination of $S_{\nu}$ depends on 7 parameters in Equations (2)--(8),
$N_{\Gamma}$, $B$, $R$ (the radius of the sphere), $\alpha$, $\delta$, $E_{min}$ and $E_{max}$, yet as there are only 3 constraints from the observation $\overline{\tau}$, $\alpha$ and $S_{o}$, it is an under determined system of equations. Most of the particles are at low energy, so the solutions are insensitive to $E_{max}$. In order to study the solution space, $\delta$ and $E_{min}$ are pre-set to a 2-D array of trial values. The reason for separating these two variables out in this fashion is that there is information that constrains these variables, apriori. First of all, Equations (1) and (9) combined with the upper bound on the LOS (from time variability considerations in \citet{rey09} and the superluminal ejection in \citet{rey17}) and the $\beta_{\rm{app}} =0.97$ from Figure~5 greatly restricts the allowed values of $\delta$. The results are shown in Figure~6. There has been no evidence of extreme blazar properties so we do not pursue extreme blazar-like LOS values (i.e $<10^{\circ}$). Thus, we expect $\delta$ to be in a fairly narrow range of $\sim 1.8- 3$. Using the formula for the observed frequency at which the peak of the synchrotron emission occurs \citep{tuc75}
\begin{equation}
\nu_{\rm{peak}} \approx 3 \times 10^{6} \left[\frac{E_{\rm{min}}}{m_{e}c^{2}}\right]^{2} \delta B
\end{equation}
and noting from Table~\ref{core_spectra} there is strong core radiation at 8.4~GHz, the models quickly converge on an upper bound for $\Gamma_{\rm{min}}$ of $\approx 25 - 30$.
\par For each trial pair of values, $\delta$ and $E_{min}$, one has 4 unknowns, $ N_{\Gamma}$, $B$, $R$ and $\alpha$, but recall that there are only 3 constraints from the spectrum for each model. Thus, there is an infinite 1 dimensional set of solutions for each pre-assigned $\delta$ and $E_{min}$ that results in the same spectral output. First, a power-law fit to the high frequency optically thin synchrotron tail fixes $S_{o}$ and $\alpha$ in Equation (5). An arbitrary $B$ is chosen in the spheroid. Then $N_{\Gamma}$ and the spheroid radius, $R$, are iteratively varied to produce this fitted $S_{o}$ and a value of $\overline{\tau}$ that minimizes the least squares residuals of the SSA region at 15.2~GHz and 8.4~GHz. Another value of $B$ is chosen and the process repeated in order to generate two new values of $N_{\Gamma}$ and $R$ that reproduce the spectral fit. The process is repeated until the solution space of $B$, $N_{\Gamma}$ and $R$ is spanned for the pre-assigned $\delta$ and $E_{min}$.
\begin{figure*}
\begin{center}
\includegraphics[width= 0.85\textwidth,angle =0]{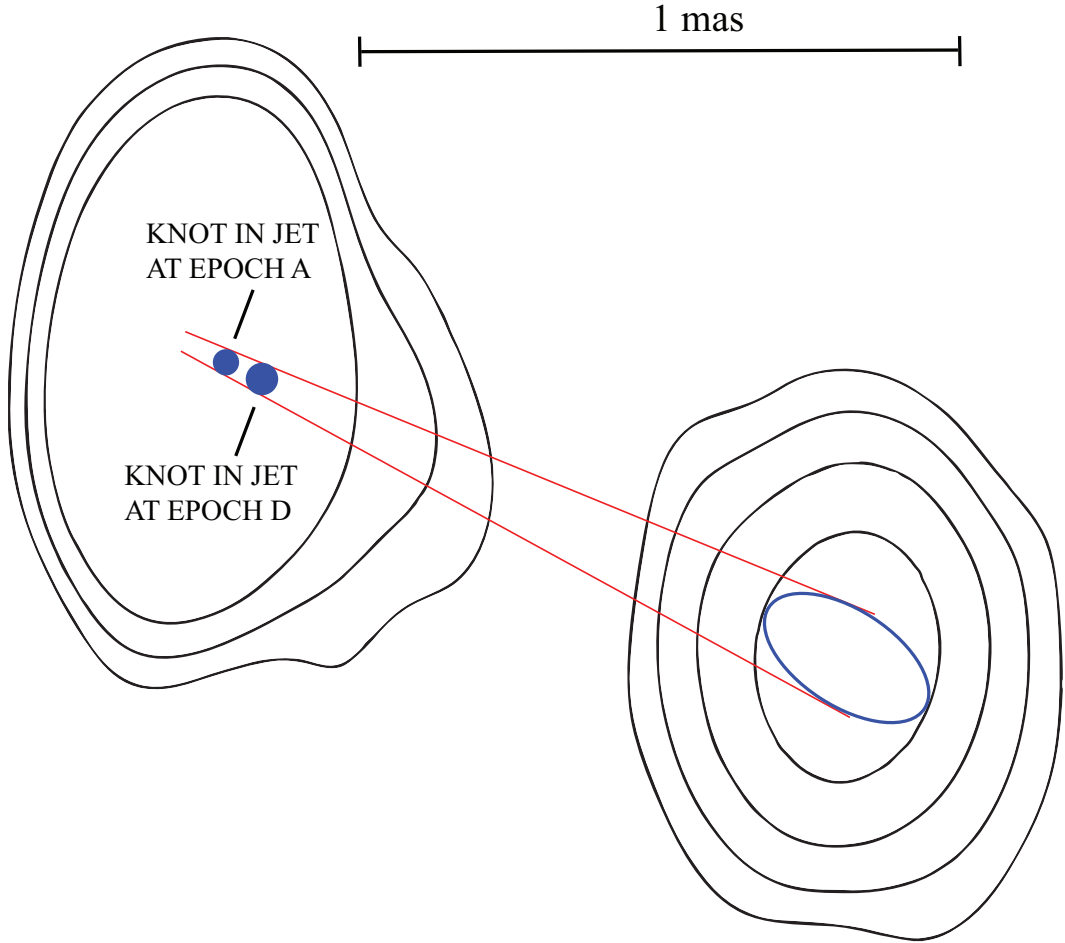}
    \caption{\label{2017jet4}This figure shows the geometry of the jet overlayed on our best (highest resolution and highest signal to noise) radio image from epoch 1 in 2015. The full image was published previously \citep{rey17}. In order to describe the details of our current campaign, the highest contour levels are removed from the core. The size of the plasmoids is approximated by the jet solution in the lower left hand frame of Figure~11 with $\Gamma_{min} = 1$, LOS = $14.5^{\circ}$. The separation from the ``nucleus" is somewhat arbitrary and is based on $\beta_{\rm{app}}=0.97$ and the time lapse to the beginning of the flare rise in Figure~1. The separation from epochs A to D is based on $\beta_{\rm{app}}=0.97$. The ellipse in K1 is the FWHM of the elliptical Gaussian fit to K1 \citep{rey17}. The overall geometry is very suggestive of a simple uniformly expanding conical jet with a half angle of $\sim 3^{\circ}$.}
\end{center}
\end{figure*}
\par To this point, we have described the mathematics of how to fit the spectra and how to convert this to the physical parameters of a spherical volume. It is not known ahead of time what region of solution space is relevant to a realistic physical solution. In order to determine this, we consider various circumstances related to the jet of Mrk\,231:
\begin{enumerate}
\item There needs to be a source for the long term (20 years) radio luminosity ($\sim 10^{41}$ ergs/sec) of the secondary source, K1, 2.5 lt-yrs away.
\item There needs to be a mechanism that accelerates the jet away from the gravitational attraction of the central black hole. The only known mechanisms for launching jets to relativistic speed involve strong magnetic acceleration and the conversion of magnetic energy to bulk plasma energy \citep{par58,lov76,bla76,bla82}.
\item At late times (i.e., epoch D), when the flare fades, the plasmoid likely tends toward the minimum energy configuration. This will be motivated more in the detailed discussions of the next section.
\end{enumerate}
These three constraints are essential for producing physically reasonable models in the following.
\section{Physical Models}
There are four types of emission regions that are possible.
\begin{itemize}
\item A discrete ballistic ejection of a turbulent magnetized plasmoid made of electrons and positrons. This was the preferred solution for the major flares in GRS~1915+105 \citep{fen99,pun12}. This will be referred to as a leptonic plasmoid in the following.
\item A discrete ballistic ejection of a turbulent magnetized plasmoid made of protons and electrons. This was a possible solution for the ejection from the neutron star merger and gravity wave source GW170817 \citep{pun19}. This will be referred to as a protonic plasmoid in the following.
\item A high dissipation region or knot in an organized magnetic jet of leptonic plasma. This was the most likely source of the radio emission in GW170817 \citep{pun19}. This will be referred to as a leptonic jet in the following.
\item Another possibility is a high dissipation region or knot in an organized magnetic jet of protonic plasma.
\end{itemize}
These types of physical models, all of which produce the fits in Figure~4, are described in this section.
\subsection{Leptonic Discrete Ejections} Figure~7 shows the dependence of $E(\mathrm{lm})$ in Equation (15) on $R$ for a wide span of pre-assigned values of $\delta$ and $E_{min}$. The figure allows one to see how $E(\mathrm{lm})(R)$ varies as one adjusts the pre-assigned $\delta$ and $E_{min}$. More importantly, it shows how the plasmoid changes, independent of the pre-assigned $\delta$ and $E_{min}$ form epoch A to epoch D. The dashed line and the red dots in each panel show the time evolution of a possible physical solution. It is the solution found for the major flares in GRS~1915+105 \citep{pun12}. A blob of turbulent electron-positron plasma is ejected from the black hole accretion system. It obeys energy conservation and the radiation losses are negligible in the 69 days of monitoring ($\sim 10^{47}$ ergs). To the right of the minimum of each $E(\mathrm{lm})$ curve, the stored plasmoid energy is more magnetic than mechanical and to the left of the minimum of $E(\mathrm{lm})$ of each curve in Figure~7, the energy is more mechanical than magnetic. The dashed curve indicates a scenario in which the flare evolves from magnetically dominated to near equipartition at late times. This satisfies physical requirement (3) from Section 6. Magnetic energy is converted to mechanical form. This is consistent with physical requirement (2) of Section 6, magnetic forces can launch the jet and this energy is converted to bulk motion and internal leptonic energy.
\par The main concern is physical requirement 1 from Section 6. We can estimate the radio luminosity of K1 from the VLBA observations. For example, in our best observation, epoch A, we note that the K1 luminosity is down about 10\% from its historic high in 2015. The data in Table~\ref{modelfits} is well fit by a power law with $\alpha=1.61$. Most of the radio luminosity is at low frequency, but our data only goes down to 8.4~GHz. A previous, VLBA campaign in 1996 covered frequencies of 1.5~GHz, 2.3~GHz, 4.9~GHz, 8.4~GHz and 15.4~GHz \citep{ulv99}. Even though the double is not resolved (even partially) below 8.4~GHz, the core was very weak at 8.4 and 17.6~GHz ($\sim 15\%$ of that in 2017). Thus, the low frequency emission can all be considered to be from K1 as a good approximation. It was argued that the power-law behavior began to be attenuated by free-free absorption below 8.4~GHz \citep{ulv99}. The flux density peaked at 5~GHz. Thus, we estimate the radio luminosity from K1 in epoch A by integrating the power-law fit from 5~GHz to 50~GHz , $\approx 8.4\times 10^{40}$ ergs/sec. Thus, from 2015 to 2018 this equates to $\sim 2.5\times 10^{48}$ ergs/yr. The total energy supplied every year must be $2.5\times 10^{48}/\epsilon_{R}$ ergs, where $\epsilon_{R}$ is the radiative efficiency of the plasma from 5~GHz to 50~GHz in K1. For the case, LOS = $14.5^{\circ}$ and $\Gamma_{min} =1$, in Figure~7, there needs to be $\approx [2.5\times 10^{48} \rm{ergs/yr}/1.25 \times 10^{49} \rm{ergs}]/\epsilon_{R}\approx 0.2 / \epsilon_{R}$ ejections per year. Since the light curve in Figure~1 shows at most 2 ejections per year. This implies a radiative efficiency of larger than 10\% in K1 which would be a very extreme requirement for the radio lobe, but such a high efficiency might be consistent with a compact steep spectrum object \citep{ode98}.
\begin{figure*}
\begin{center}
\includegraphics[width= 0.7\textwidth,angle =0]{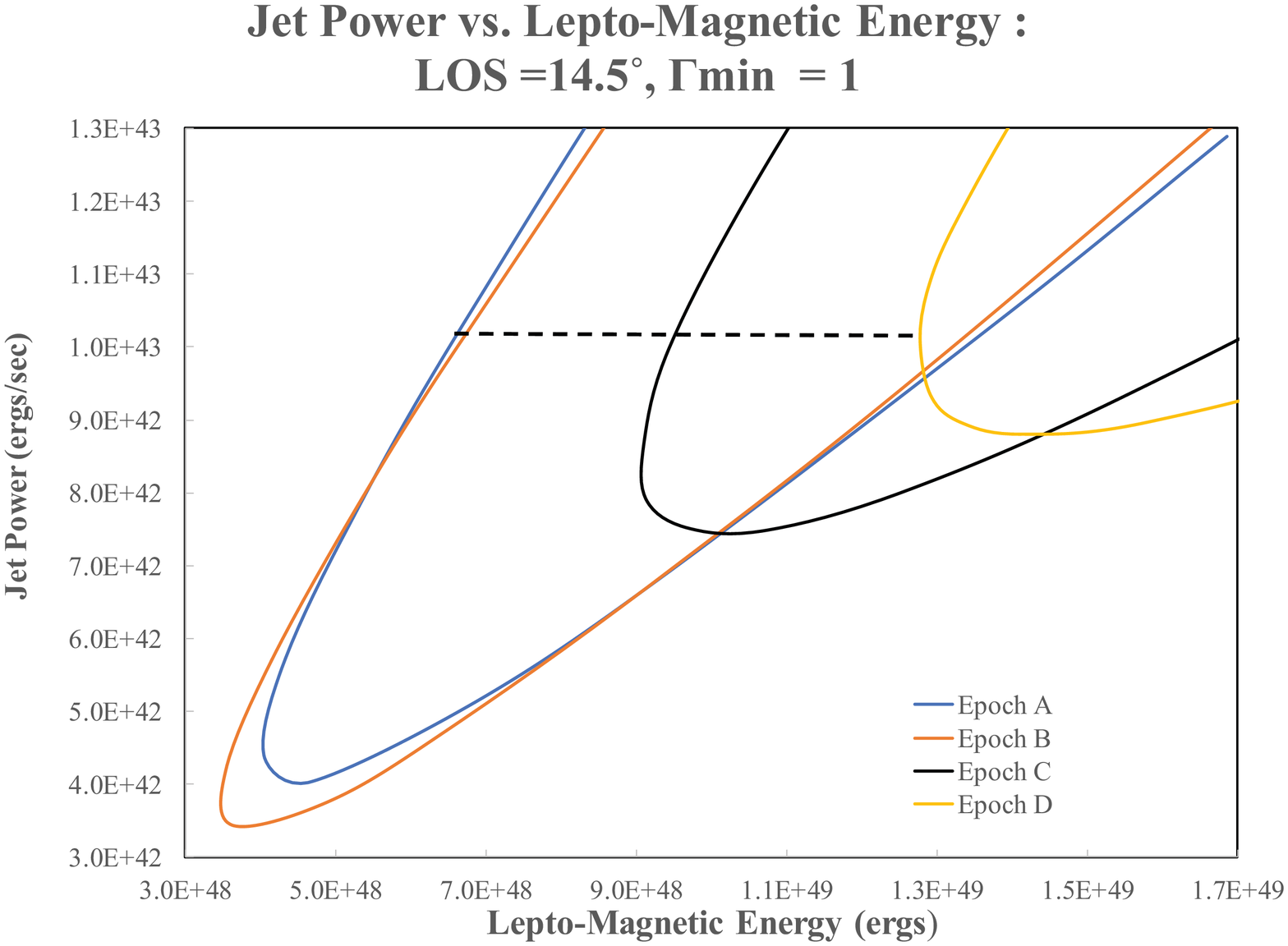}
\includegraphics[width= 0.7\textwidth,angle =0]{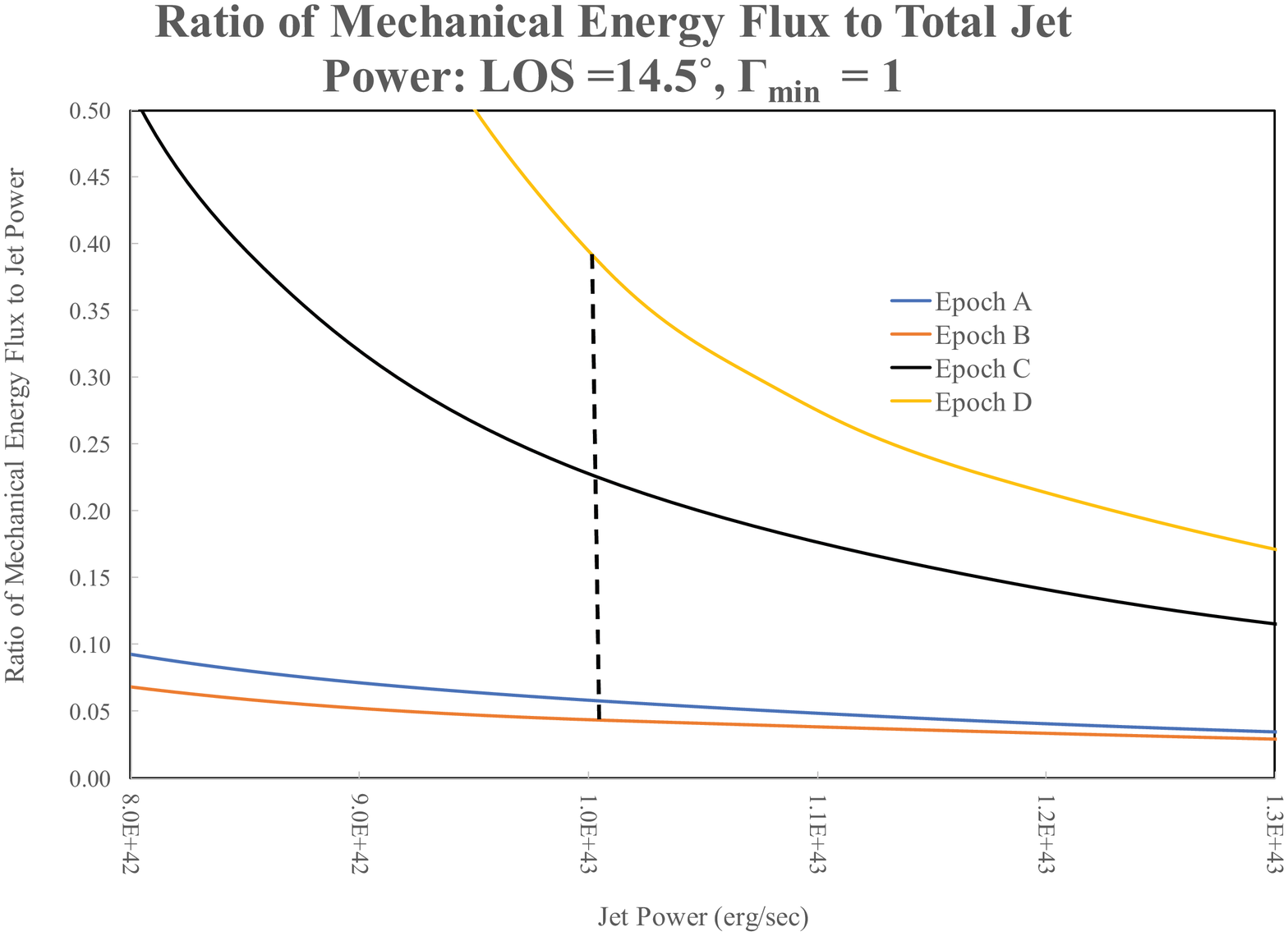}
    \caption{\label{2017S}The top frame shows the time evolution of a constant leptonic jet power, $Q$, solution for the flare from epoch A to epoch D in the $E(\rm{lm})$ - $Q$ plane. The conserved jet power solution is indicated by the dashed line. The bottom frame shows the rate at which Poynting flux is converted to mechanical energy flux as the propagating knot evolves from epoch A to epoch D.}
\end{center}
\end{figure*}
\subsection{Protonic Discrete Ejections} Another possibility is a ballistic ejection of a blob of protonic plasma. Figure~8 plots $\mathcal{K}(\mathrm{protonic})$ as a function of $E(\mathrm{lm})$ for the case  LOS = $14.5^{\circ}$ and $\Gamma_{min} =1$. This is useful for understanding the time evolution of the system. For a flare that decays, we posit that the final state is near minimum energy, minimum $E(\mathrm{lm})$, as in the discrete ejections in GRS~1915+105. If this is true then a solution associated with the dashed black line is indicated. The plasmoid is ejected with $\mathcal{K}(\mathrm{protonic})=2.7 \times 10^{51}$~ergs, a constant total energy from epoch A to epoch D. Epoch D is relaxed to the minimum energy configuration. Figure~8 shows that the time evolution is one in which plasma internal energy is converted to magnetic turbulent energy. Based on Figure~1, the longest time frame in which energy is pumped into the plasmoid is the time lapse, $T$, from epoch A to the start of the flare ($\sim 120$ days, see Section 8 for more details). The instantaneous jet power required to energize and eject the plasmoid is therefore $Q_{o}\approx \mathcal{K}(\mathrm{protonic})/T > 2.7\times 10^{44}$~ergs/sec. This is a large jet power by astrophysical standards. Theoretical efforts to explain the initiation of such powerful jets rely on some form of magnetic jet launching that requires the plasma dynamics to be dominated by magnetic forces \citep{lov76,bla76,bla82}. Thus, it is hard to understand a mechanism in which inertial energy is converted to magnetic energy during the course of the plasmoid evolution. One would expect there to be an excess of magnetic energy in the early stages as required for relativistic jet initiation. The indicated time evolution does not satisfy physical requirement (2) of Section 6.
\subsection{Leptonic Jet}Another possibility that is very common in extragalactic jet models is a magnetic jet filled with an electron-positron plasma \citep{bla79,wil99}. In this scenario, the spherical plasmoid is a knot in the jet as indicated in Figure~9. The knot represents a region of enhanced dissipation. It might be related to a wavefront or shock signaling an increase in jet power from the source that is propagated down the conduit of the jet. In this circumstance, the following estimates of jet power represent a transient state of enhanced jet power. Alternatively, the enhanced dissipation might have arisen from interactions of the jet with the ambient medium. For example, it is likely that the jet is interacting with the very dense BAL wind \citep{rey09}. This provides a piston for launching strong dissipative waves such as shocks as well as a source of instabilities created from the jet/wind interface \citep{bic90}. In this instance, the jet power estimate is likely more indicative of the jet over a long time frame, but the interaction has allowed us to sample the quasi-steady jet (by increasing the dissipation) for a short period of time.
\par The jet power in this model, $Q$, using Equations (13) and (18) is
\begin{equation}
Q = \int [S^{P} + ke(\mathrm{leptonic})] dA_{\perp} + L_{r}\;,
\end{equation}
where $dA_{\perp}$ is the cross sectional area element normal to the jet axis and $L_{r}$ is the energy flux lost to radiation. We constrain the solution space by assuming that the system approaches minimum $E(\rm{lm})$ at late times (epoch D). This is motivated by the discrete ejections in GRS~1915+105 that approach minimum energy at late times and it has been argued that the hot spots (the terminal jet knot) and the surrounding radio lobes are near minimum energy in extragalactic radio sources \citep{pun12,har04,cro05,kat05}. A common spheroid shape for the emission region is chosen to allow direct comparison between the four classes of models. However, it is not ideal for cross-sectional integration as in Equation (22) due to its nonuniform cross-sectional area. Thus, the evaluation of $Q$ in Equation (22) is the average over all the cross-sections
(i.e., $\pi R^{2} \rightarrow (2/3)\pi R^{2}$). Within the crude approximation of these simplified models this is not a huge source of uncertainty. The basic physics of the knot evolution is shown in Figure~10. The top frame plots $Q$ versus $E(\rm{lm})$. Epoch D is in the minimum energy configuration. The time evolution between epochs is indicated by the dashed line which is simply conservation of $Q$ (radiation losses are negligible). The evolution makes sense from jet/wind launching theory, requirement (2) of Section 6. Magnetic energy flux is converted to inertial energy flux. The bottom frame of Figure~10 is a plot of the fraction of the jet power that is inertial energy flux:
\begin{equation}
\rm{Inertial\, Fraction} = \left[\int ke(\mathrm{leptonic})) dA_{\perp} \right]/Q\;.
\end{equation}
The dashed line is the same solution as in the top frame. In epoch A, the inertial component of $Q$ is $\sim 5\%$ of $Q$ and increases steadily to epoch  D where it is $\sim 40\%$.

\begin{figure*}
\begin{center}
\includegraphics[width= 0.48\textwidth,angle =0]{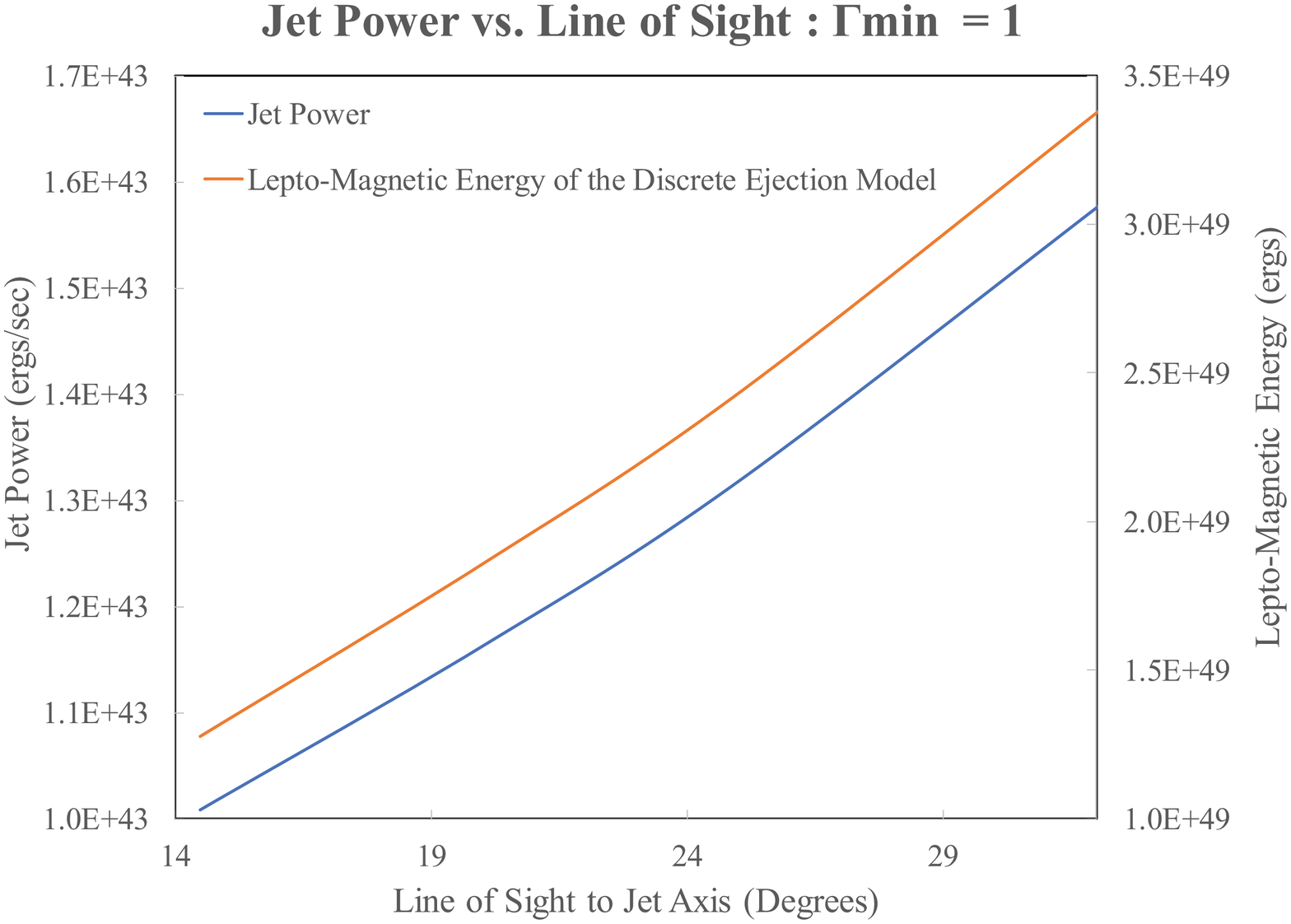}
\includegraphics[width= 0.48\textwidth,angle =0]{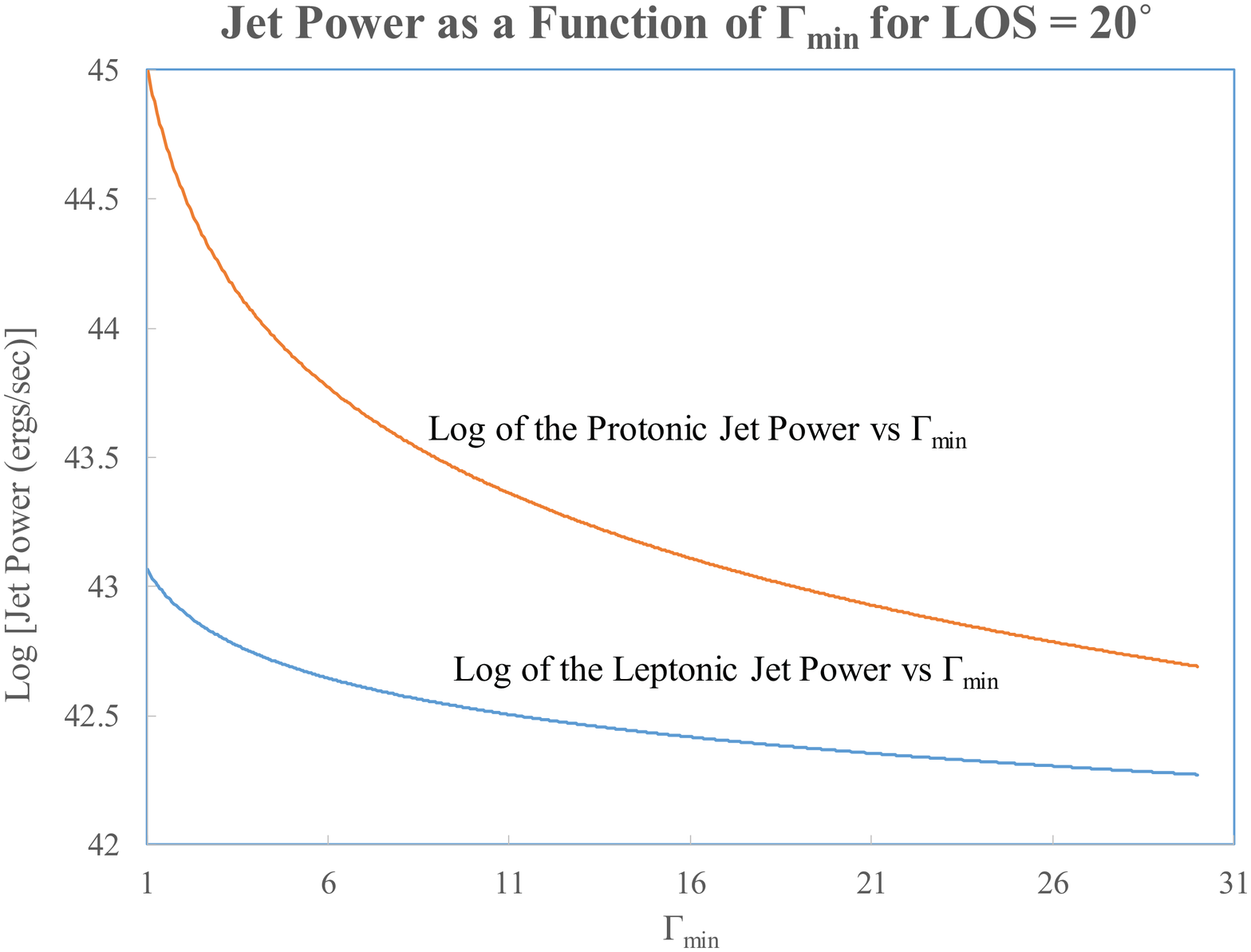}
\includegraphics[width= 0.48\textwidth,angle =0]{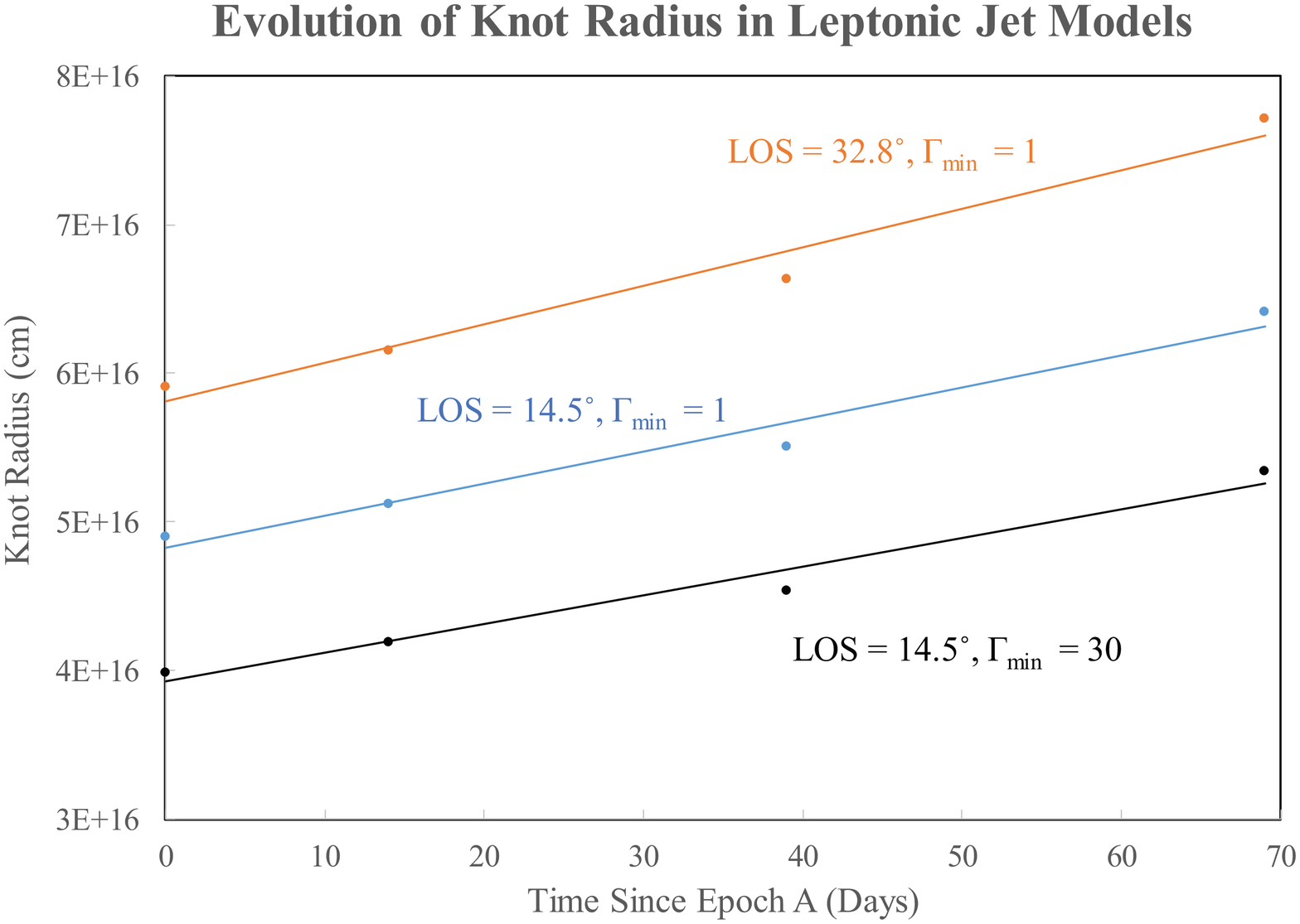}
\includegraphics[width= 0.48\textwidth,angle =0]{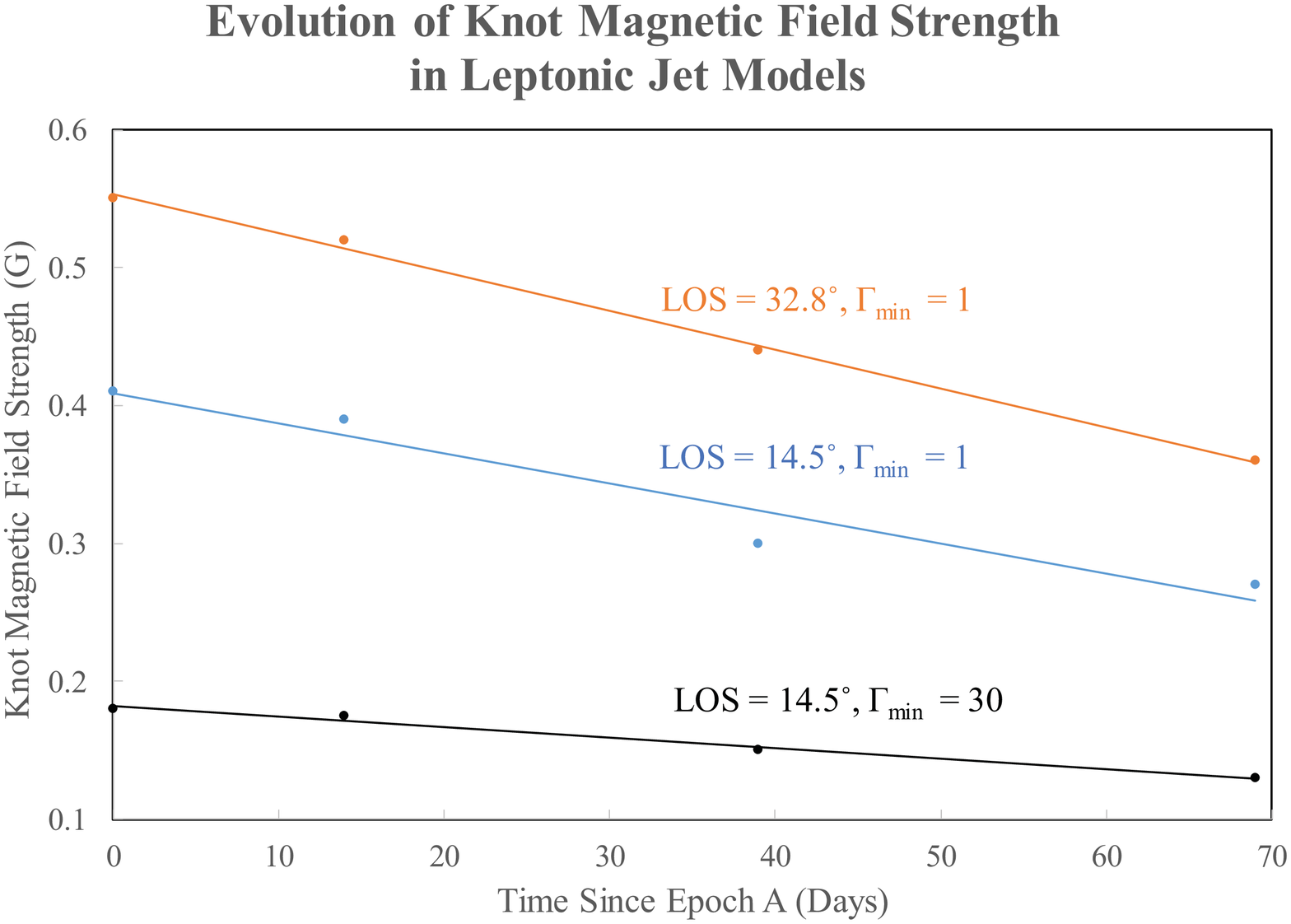}
    \caption{\label{2017Q}The top left frame shows the dependence of $Q$ on the assumed LOS for the leptonic jet model of Section 7.3. It also shows the dependence of $E(\rm{lm})$ on the LOS for the end state (epoch D) which is also the conserved energy in the discrete leptonic ejection model of Section 7.1. The top right hand frame shows the dependence of $Q$ on $\Gamma_{\rm{min}}$ for both the leptonic jet model of Section 7.3 and the protonic jet model of Section 7.4. The bottom left hand frame shows the time evolution of the spheroid radius in the leptonic jet model of Section 7.3 for various LOS and $\Gamma_{\rm{min}}$ choices. Notice that it is fairly steady. The bottom right hand frame shows the gradual decay of $B$ in these models. }
\end{center}
\end{figure*}
\par The dependence of jet power on the LOS is illustrated in the top left frame of Figure~11. A larger LOS equates to a more powerful jet as well as a stronger discrete ejection in the leptonic model of Section 7.1. The top right frame of Figure~11, shows that $Q$ decreases as $E_{\rm{min}}$ increases. It also shows that a protonic jet (see next section) is much more powerful. The lower left hand frame shows the time evolution of the spheroid radius. The solutions with $\Gamma_{\rm{min}}= 1$ were used to estimate the jet opening angle at K1 in Figure~9. The solution with $\Gamma_{\rm{min}}= 1$ and an LOS = $14.5^{\circ}$ has an expansion rate, $2(dR/dt)$ = 0.2c. The bottom right hand corner of Figure~11 shows that the magnetic field decreases as the knot propagates. This is expected from the conversion of magnetic to inertial energy indicated in Figure~10. Combining these results with Figure~6 and Equation (21) motivated our choice of $\Gamma_{\rm{min}}= 30$ as an upper limit for a viable lower energy cutoff. The solution has a smooth monotonic behavior after the peak of the flare at epoch B as might be expected for a plasmoid that relaxes to the minimum energy state.
\begin{figure*}
\begin{center}
\includegraphics[width= 0.6\textwidth,angle =0]{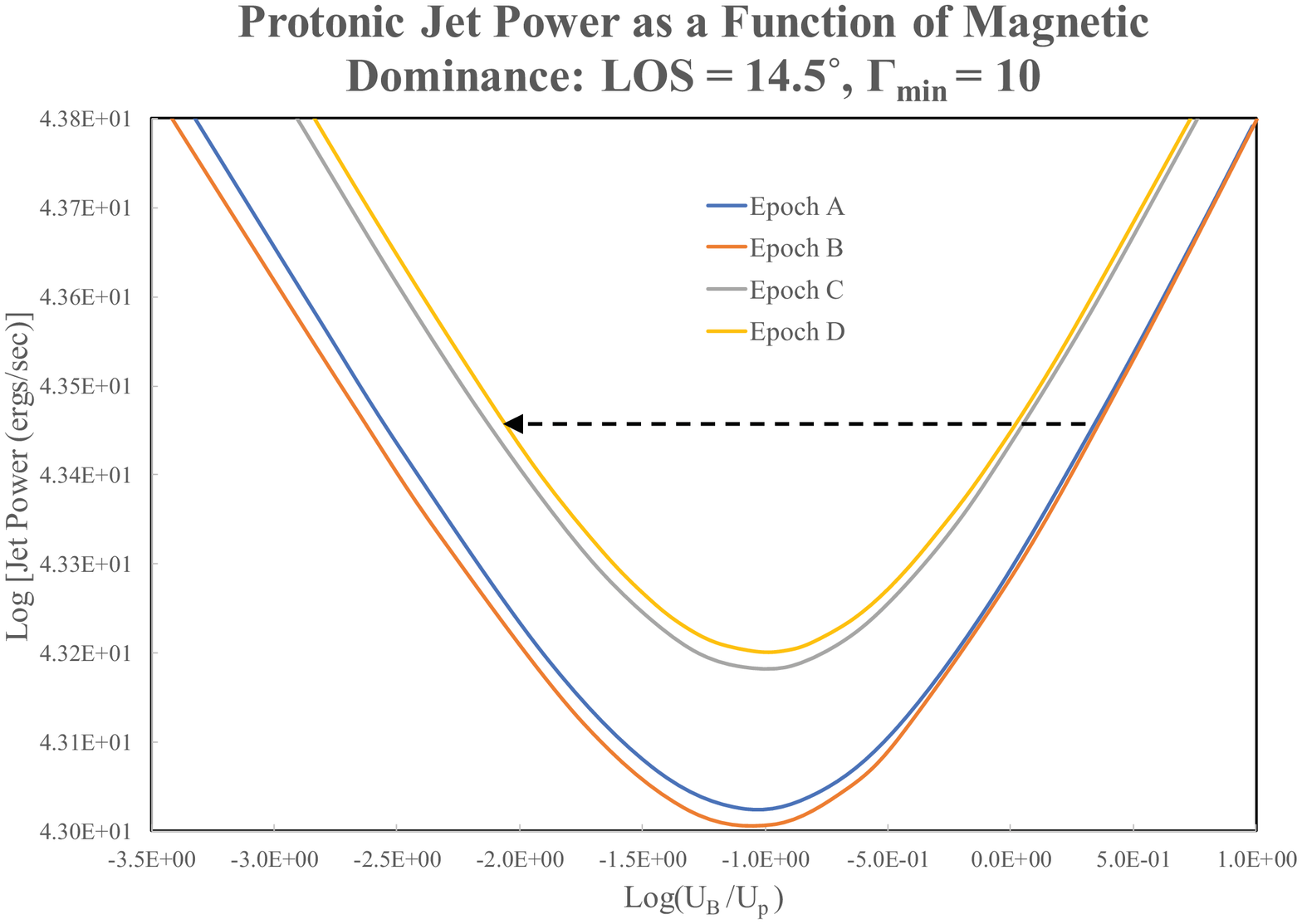}
\includegraphics[width= 0.65\textwidth,angle =0]{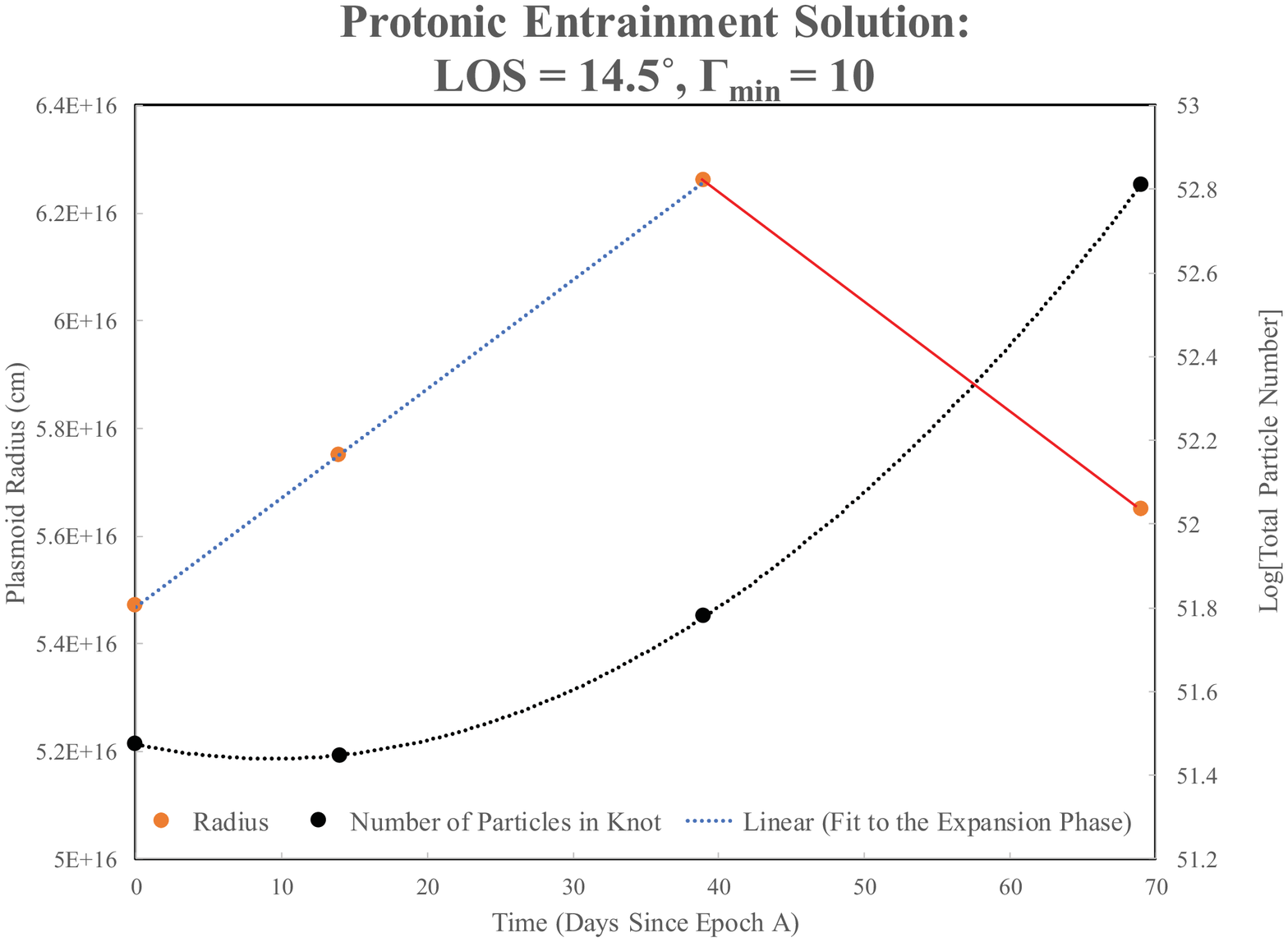}
    \caption{\label{2017UA}The dashed line in the top frame indicates a protonic jetted knot solution with the conversion of magnetic energy flux to protonic energy flux. The solution is smooth from epochs A to C and is consistent with our simple formalism. However, the knot needs a catastrophic change from epochs C to D in order to achieve conservation of energy and a relaxed state with $E(\rm{lm})$ near a minimum at the end of the flare. There is a two orders of magnitude change in $U_{B}/U_{P}$. The bottom frame highlights the implications of the large change. There is massive entrainment of plasma and an abrupt halt to the steady expansion of the knot from epochs A to C. Such dynamics goes far beyond our simple model, but the implication is clear. Strong dissipative physical process are required to execute this plasma state transition. However, there is no evidence of this in the spectra, see Table~\ref{core_spectra} and Figure~4.}
\end{center}
\end{figure*}
\subsection{Protonic Jet}
It is possible that the jet is made of primarily protonic not positronic plasma. Then there are two inertial contributions to $Q$, one from the energized leptons and a dominant component from the bulk motion of protons. Equation (22) needs to be modified accordingly with Equation (17),
\begin{equation}
Q_{\rm{protonic}} = \int [S^{P} + ke(\mathrm{leptonic})+\mathcal{E}(\mathrm{proton})] dA_{\perp} + L_{r}  \approx \int [S^{P}+ \mathcal{E}(\mathrm{proton})] dA_{\perp} \;.
\end{equation}
The solutions that are plotted in the upper right hand frame of Figure~11 are computed assuming a minimum $E(\rm{lm})$ state in epoch D, the same state as in all the other models. There is no concern with these solutions having enough power to energize K1 in accord with requirement (1) of Section 6. However, we are unable to find any solutions that satisfy physical requirements (2) and (3) that change monotonically and smoothly as the leptonic jet solutions of the previous subsection. Thus, our methods are not likely valid quantitatively, but might provide some qualitative insight into these solutions and will be discussed below. The biggest issue is the assumption of a constant velocity is likely grossly inaccurate.
\par The top frame of Figure~12, shows the solution for $\Gamma_{\rm{min}}=10$ and a LOS = $14.5^{\circ}$ that proceeds by converting magnetic energy to inertial energy. Based on the upper right hand frame in Figure~11, any value of $\Gamma_{\rm{min}}<10$ is likely too powerful (A Fanaroff-Riley~II, FR~II, extragalactic radio source level luminosity). In order to explore protonic jets, it is useful to introduce the protonic energy density,
\begin{equation}
U_{P} \approx Nm_{p}c^{2}\; .
\end{equation}
The horizontal axis in the top frame of Figure~12 plots $U_{P}/U_{B}$, the ratio of inertial energy density to magnetic energy density. Even though there are numerous changes from our original leptonic jet models inserted by the introduction of  protons, the leptons in the plasma still radiate the spectra in Figure~4 for all the solutions within Figures~11 and 12.
\par There are solutions that proceed towards more magnetic energy over time, but by physical requirement (2) of Section 6 this is untenable. The dashed arrow shows the direction of time evolution, assuming a constant jet power. The end of the arrowhead is the minimum energy solution in epoch D. This fixes the solution. Without this assumption, the solution is unconstrained and our analysis is not useful. The only solutions that satisfy physical requirements (1)--(3) of Section 6 require a dramatic change between epochs C and D. This is apparent from the large gap in $U_{P}/U_{B}$ from epoch C to epoch D in the top frame of Figure~12, two orders of magnitude. The bottom frame of Figure~12 indicates the required dynamics. From epochs A--C there is a smooth gradual expansion with modest entrainment of plasma. The jet power is so dominated by Poynting flux (97\%) that the jet can be either positronic or protonic in the early stages without affecting the solution. However, the bottom frame of Figure~12, indicates intense protonic entrainment between epochs C and D. This is not unreasonable as deduced in \citet{rey09}. At K1, there is evidence of overwhelming entrainment on parsec scales from the dense enveloping BAL wind. The solution described here is one of a powerful Poynting flux dominated jet in which all the power is converted into proton kinetic energy on the order of 1--2 light months from the source.
\par There are a couple of issues with this solution that violate our basic assumptions.
\begin{enumerate}
\item Clearly, large entrainment would slow down the jet. Our models are not sophisticated enough to capture this, so the models are only qualitative.
\item In order to keep $Q$ reasonable we had to choose an adhoc $\Gamma_{\rm{min}}\geq 10$ based on the upper right hand frame of Figure~11. The bottom frame of Figure~12 shows an entrainment solution in which over 90\% of the plasma enters the knot between epochs C and D. The plasma that surrounds the jet in the BAL wind is cold relative to the jet. This process should drastically lower $\Gamma_{\rm{min}}$. This complexity is far beyond our simple model and involves much uncertain physics and geometry.
\end{enumerate}
These shortcomings raise the question if our model is of any value in the analysis of this scenario. In order to explore this, we note that the model of a Poynting flux dominated jet with a steadily increasing radius and a slow and steady conversion of magnetic energy to mechanical energy seems qualitatively reasonable from a wind theory point of view \citep{bla82,pun08}. But, can we say anything about the notion of intense entrainment between epochs C and D that is implied by the prediction of the model? The model predicts its own failure in this time frame. First, we note that the knot in the Poynting jet is a very powerful dynamic object from epochs A--C. It propagates with $\beta_{app}=0.97$c, $Q > 4\times 10^{43}$ ergs/sec of pure Poynting flux and it is expanding (based on the top frame of Figure~12) at $2(dR/dt)=0.16$c. In a one month period, $>95\%$ of the Poynting jet power is converted to protonic kinetic energy flux. This process also introduces 90\% of the jet plasma. The Poynting flux fraction of $Q$ changes from $97\%$ to $3\%$. The radius stops expanding at 0.08c and contracts rapidly. This seems to be a very violent process involving dissipative instabilities and fast shocks \citep{ken84}. The introduction of new plasma and strong dissipative process should change the energy spectrum of the electrons. This should be independent of the entrainment details. Now, we look outside of the simple, spherical (likely invalid due to the strong entrainment induced changes) model for empirical data, in particular Table~\ref{core_spectra} and Figure~4. Note that $\alpha$ and $n$ in the last two columns are very similar in epochs C and D, almost the same considering the crudeness of the model. The entrainment scenario should have produced a significant change unless there is a major coincidence that the process recreates the previous particle spectrum. By contrast, note the shift of the spectral peak and the decrease of peak flux intensity implied by Table~\ref{core_spectra} and Figure~4. This is the defining characteristic of cooling by adiabatic expansion: the same spectral shape with a shift of the spectral peak to lower frequency and lower intensity \citep{mof75}. This empirical (model independent) spectral change is consistent with the steady expansion of the leptonic jet model of the last subsection and not the violent contraction indicated for an the intense entrainment scenario of the protonic jet. For this reason, we consider a knot in a leptonic jet to be more likely to represent the physical solution than a knot in a protonic jet. We also note that the leptonic jet favors $\Gamma_{\rm{min}}\approx 1$ and does not require the adhoc condition $\Gamma_{\rm{min}}\geq 10$ needed for the protonic jet models.
\section{High Energy Observations}
The energetic flare at 7mm can manifest itself as enhanced X-ray emission. If so, this can provide a valuable tool for analyzing the possible physical nature of the flare. We performed a deep 80 ks NuSTAR observation during the flare rise on October 19, 2017 for this purpose. We were searching for evidence of a pronounced change from the NuSTAR observation on August 27, 2012, the only NuSTAR observation that occurred during a state of low radio jet activity. The older observation was for 41 ks and showed evidence of X-ray absorption indicative of an out-flowing wind with a velocity $17000 \pm 4500$ km/s \citep{fer15,rey17}. Thus, we proposed a deeper 80ks observation to clearly distinguish the high radio state X-ray spectrum from the low radio sate X-ray spectrum. This was demonstrated only marginally with the four $\sim 30$ks NuSTAR observations during high radio states that were analyzed previously \citep{rey17}.
\begin{figure*}
\begin{center}
\includegraphics[width= 0.45\textwidth,angle =-90]{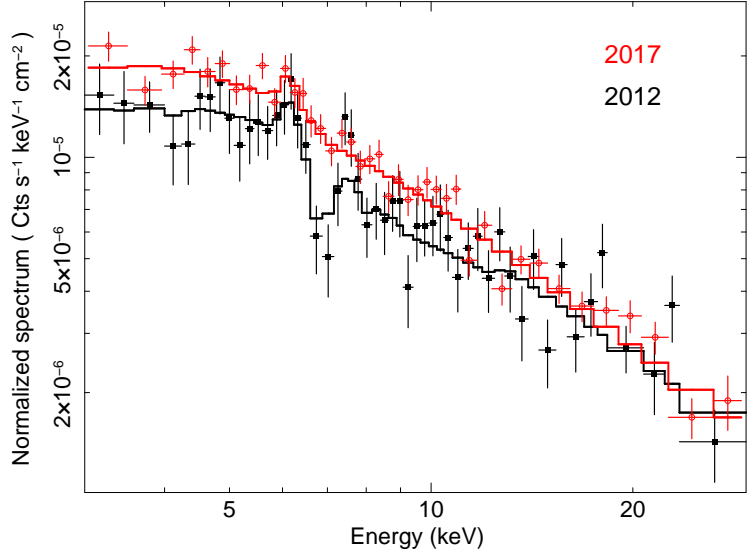}
    \caption{\label{2012_2017}The high radio state NuSTAR spectrum from October 2017 (red empty circles) is compared to the low radio state spectrum from August 2012 (black filled squares), with solid lines representing  the respective best-fitting models. The X-ray continuum level is unchanged, but the X-ray ultra-fast outflow is suppressed in the high radio state. Although, in our analysis, we fit the NuSTAR data from the FPMA and FPMB detectors jointly, we show here the co-added spectra at each epoch for visual clarity; each spectrum is divided by the respective effective area to allow direct comparison.}
\end{center}
\end{figure*}
\begin{figure}
\begin{center}
\includegraphics[width= 0.75\textwidth,angle =0]{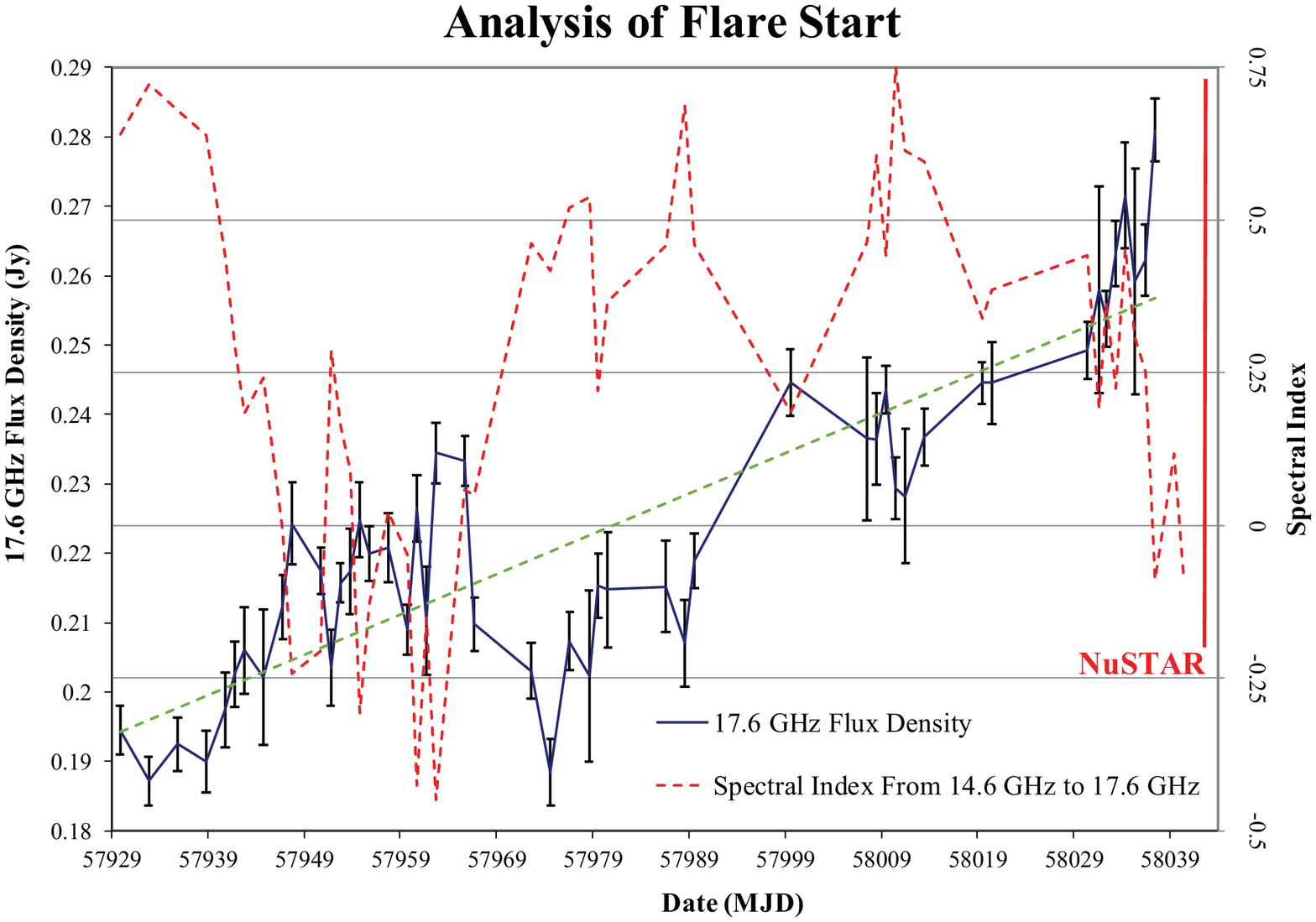}
    \caption{\label{2017start}The plasmoid ejection time is estimated from this plot of the spectral index and the 17.6~GHz flux density near the start of the flare. Since the spectrum flattens rapidly and eventually inverts near MJD 57940 this is a more viable starting point for the flare than the other relative minimum near MJD 57974.}
\end{center}
\end{figure}
Figure~13 clearly shows that the X-ray absorbing wind is suppressed during the high jet state. This corroborates the claim that the X-ray absorbing wind was suppressed during previous states of high jet activity based on shorter exposures \citep{rey17}. We note the related phenomenon, the disappearance of the high ionization ultraviolet broad absorption line wind during high radio states that was discussed in that study as well.

\begin{table}
    \caption{\label{nustar}Long Term Evolution of NuSTAR Flux}
\begin{tabular}{ccc} \tableline\rule{0mm}{3mm}
Date &  3--10 keV Flux ($10^{-12}$ erg/sec-$\rm{cm}^{2}$) & 10--30 keV Flux ($10^{-12}$ erg/sec-$\rm{cm}^{2}$)\\
 & 90\% Confidence Interval   & 90\% Confidence Interval   \\
\tableline \rule{0mm}{3mm}
2012 August 26  & 0.65-0.77&  1.68-2.09 \\
2013 May 09 &  0.76-0.89 &  1.36-1.75 \\
2015 April 02  & 0.74-0.86 & 1.34-1.67 \\
2015 April 19  & 0.61-0.74 & 1.25-1.66 \\
2015 May 28  & 0.67-0.79 &  1.23-1.58 \\
2017 October 19  &  0.90-0.98 &  1.77-2.00
\end{tabular}
\tablenotetext{}{Confidence intervals of X-ray fluxes obtained by fitting jointly the NuSTAR FPMA and FPMB spectra at each epoch with a cross-normalization constant accounting for the different detector responses.}
\end{table}

\par As with the other high radio states, there is no pronounced increase in the X-ray luminosity. Similar to the other 5 NuSTAR epochs in \citet{rey17}, the spectrum is fit with a photon number spectral index $1.50\pm0.1$ and a neutral hydrogen column density of $10\pm 5 \times 10^{22}\rm{cm}^{-2}$. Table~\ref{nustar} shows the observed flux for the 6 epochs. The flux is constant to within 20\% over 5 years. The 10--30 keV column is the most robust for tracking flux variation, since it is not subject to changes in the intrinsic observing column. The observed 3--30 keV X-ray luminosity in the high radio state (2017) is  $\sim 1.1 \times 10^{43}$ ergs/sec to $1.2 \times 10^{43}$ ergs/sec.

\par In the remainder of this section, we estimate the expected X-ray flux in our plasmoid model and compare it to the continuum flux level. The epoch A observation/plasmoid model is the most suitable in time, being just 17 days after the NuSTAR observation. The energetic plasmoid is not only a source of synchrotron emission, but the energetic leptons can also up-scatter soft background photons to the X-ray energies detectable with NuSTAR. There are a few possibilities relevant to blazars. There is synchrotron self-Compton scattering (SSC) of the synchrotron photons in the plasmoid as well as the external inverse Compton (EIC) scattering of the accretion disk photons, broad emission line photons and infrared photons from the dusty torus. In quasars, the EIC process dominates the SSC mechanism because the quasar is a very luminous source of soft photons. For highly relativistic jets, EIC from the photon fields of the broad emission lines or the torus tend to be dominant since they are extremely enhanced in the plasma frame of reference due to Doppler blue-shifting and conversely, the more luminous accretion disk photon field is significantly Doppler redshifted in the frame of reference of the plasma \citep{der93}. However, in the plasmoid models of epoch A, the bulk Lorentz factor $\sim 1.6$, so the dynamics are only trans-relativistic and the Doppler effects are modest. Thus motivated, we cannot neglect the EIC luminosity from the copious accretion disk photons apriori in our analysis as shown below.

\subsection{Inverse Compton Scattering of Accretion Disk Photons}
\par It is not trivial to estimate the accretion disk luminosity. Mrk\,231 has significant intrinsic extinction in the nuclear region \citep{smi95,lip94,vei13,vei16}. This not only makes the direct observation of the accretion disk spectrum impossible, but it also makes the origin of the weak X-ray emission from October 2012 in Figure~13 very uncertain \citep{rey17,vei13}. An estimator of the accretion disk luminosity is the H$\alpha$ broad emission line and this is very prominent in Mrk\,231. The intrinsic extinction has been corrected with a LMC extinction law with E(B-V)=0.63, yielding a plausible intrinsic quasar spectrum in the optical band \citep{lip94,bok77}. This same extinction law does not work in the ultraviolet \citep{smi95,vei16}. This extinction correction for H$\alpha$ is a factor of $\sim 4.5$. There is so much intrinsic absorption (including that of Fe~II) that H$\beta$ is very weak and is difficult to use for estimating the bolometric luminosity of the accretion flow, $L_{\rm{bol}}$. The situation is even worse for the ultraviolet broad lines \citep{smi95}. The H$\alpha$ line is a bonafide quasar broad line with an extinction corrected luminosity of $L_{H\alpha}=4.7 \times 10^{43}$ ergs/sec and a full width half maximum (FWHM) of 2800 km/sec \citep{smi95}. $L_{\rm{bol}}$ can be estimated with the formula \citep{gre07},
\begin{equation}
L_{\rm{bol}} \approx 2.34\times 10^{44} \left[\frac{L_{H\alpha}}{10^{42} \rm{erg/sec}}\right]^{0.86} \, \rm{ergs/sec}\approx 6.45 \times 10^{45} \rm{erg/sec}\;.
\end{equation}
\par Secondly, we can also estimate $L_{\rm{bol}}$ from the reprocessed photons in the infrared from the dusty torus, but this is not simple in Mrk\,231 either. The infrared band is very luminous and is actually dominated by starburst emission. The breakdown of the different components with detailed modeling indicates that the AGN luminosity at wavelengths longer than $3\mu$m is $L_{IR}(\rm{AGN})\approx 4 \times 10^{45} \rm{ergs/sec}$ and the star burst luminosity is $L_{IR}(\rm{SB})\approx 10^{46} \rm{ergs/sec}$ \citep{far03}. The AGN component is complicated by the existence of two strong components, a hotter component with the SED peaked at $\approx 4\mu$m and a stronger cooler component peaked at $\approx 60\mu$m. The second component is likely associated with the molecular disk imaged in the hydroxyl maser line with VLBI with an inner radius of 30~pc and an outer radius of 100~pc \citep{klo03}. Regardless of this complex nature we use the IR based bolometric estimators \citep{spi95},
\begin{eqnarray}
&& \log[L_{\rm{bol}}] = 0.942\log[\lambda L_{\lambda}(\lambda=
    12\mu\mathrm{m})] + 3.642 \approx 45.82 \;,\\
&& \log[L_{\rm{bol}}] = 0.837\log[\lambda L_{\lambda}(\lambda=
    25\mu\mathrm{m})] + 8.263 \approx 45.75\;,
\end{eqnarray}
where $\lambda L_{\lambda}(\lambda=12\mu\mathrm{m})$ and $\lambda L_{\lambda}(\lambda=25\mu\mathrm{m})$ are from the AGN component of the fit to the IR SED \citep{far03}. The three estimation methods in Equations (26)--(28) closely agree with each giving us confidence in our estimate of $L_{\rm{bol}}$.

\par The EIC luminosity, $L_{\rm{EIC}}$, and frequency, $\nu_{\rm{EIC}}$, are related to the synchrotron luminosity, $L_{\rm{synch}}$ and frequency, $\nu_{\rm{synch}}$ by \citep{tuc75}
\begin{equation}
\frac{L_{\rm{EIC}}}{L_{\rm{synch}}}\approx \frac{U_{ph}}{U_{B}}\;, \quad \frac{\nu_{\rm{EIC}}}{\nu_{\rm{synch}}}\sim \Gamma^{2}\;,
\end{equation}
where $U_{ph}$ is the energy density of the accretion disk photon field in the frame of reference of the plasmoid. The flux of the photon field of the accretion disk is diluted in the plasmoid in epoch A, by geometric dilution and also experiences Doppler redshifting. For the dissipative knot in a leptonic jet model (our preferred model in Section 7.3 and Figure~10), assuming a LOS = $14.5^{\circ}$ and $\Gamma_{min}=1$ for illustrative purposes, the relevant parameters are
\begin{equation}
B=0.41\rm{G}\quad U_{B} = 6.69 \times 10^{-3} \rm{ergs/sec} \quad \beta =0.816\;.
\end{equation}
In order to compute $U_{ph}$, taking into account geometric dilation, we need to know the distance the plasmoid was from the accretion disk during the NuSTAR observation. Based on the light curve in Figure~1, the plasmoid seems to have been ejected $\sim 100$ days earlier. We illustrate this in Figure~14 which plots both the 17.6~GHz flux density and the spectral index from 14.6~GHz to 17.6~GHz at the beginning of the flare. The spectral index is important since as a plasmoid is ejected at the beginning of the flare it begins very compact and has large SSA at 17.6~GHz. Therefore, the flare begins at high frequency first and evolves to lower frequency \citep{van66,mof75}. For strong flares, even in the presence of the $\sim 120$~mJy steep background flux density, the spectrum flattens as the new ejection emerges and can even invert (as in Figure~14) between 14.6~GHz and 17.6~GHz \citep{rey17}. These conditions occur near MJD 57940, thus this is a more likely flare ejection time than the other relative minimum near MJD 57974. We estimate $\sim 100$ days interval between the flare ejection time and the NuSTAR observation.

Assuming the velocity of Equation (30) is constant, it would have propagated $D_{d-p}\sim 2.12\times 10^{17}$ cm from the disk.  In order to compute the relevant Doppler redshift, note that as the plasmoid moves away from the accretion disk, the distance from the disk become much larger than the radius of the bulk of most of the disk luminosity and the photons approach from behind to first approximation \citep{der93}. Thus, the Doppler factor of the plasmoid relative to the disk would be computed from Equation (1) with $\theta=180^{\circ}$ and $\beta$ from Equation (30),
$\delta_{d-p} = 0.318$. We can then compute the photon energy density in the plasmoid from Equation (8) and \citep{lig75,tuc75}
\begin{equation}
U_{ph}\approx \delta_{d-p}^{4}\frac{L_{bol}}{4\pi \rm{c} D_{d-p}^{2}} =3.93 \times 10^{-3}\;.
\end{equation}
From Equation (21), (29) and (30), the range of $\Gamma$ that results in EIC photons in the NuSTAR 2-30 keV band is $ 594 < \Gamma < 1170$. These same leptons radiate synchrotron photons in the observer's frame in the far infrared, $1.36 \times 10^{12} \rm{Hz} < \nu_{o} < 5.31 \times 10^{12} \rm{Hz}$. Based on the Epoch A fit in Table~\ref{core_spectra}, these leptons radiate a synchrotron luminosity of $L_{\rm{synch}} = 3.34 \times 10^{41} \rm{ergs/sec}$ as observed at earth. From Equations (29) - (31), the expected EIC luminosity observed at earth in the NuSTAR band from these leptons is $L_{\rm{EIC}} = 1.96 \times 10^{41} \rm{ergs/sec}$ or a flux of $F_{\rm{EIC}} = 4.48\times 10^{-14} \rm{ergs/sec-\rm{cm}^{2}}$. This value is negligible compared to the fluxes in Table~\ref{nustar}.
\par Note that given the large distance from the accretion disk to the plasmoid, $D_{d-p}\sim 2.12\times 10^{17}$ cm, the broad line region is likely an order of magnitude closer to the black hole, and also illuminates the plasmoid primarily from behind \citep{gre07}. Thus, the EIC flux is also weak, perhaps an order of magnitude less than  that induced by the accretion disk photon field.
\subsection{Inverse Compton Emission of Infrared Photons} The primary controlling factor in the determination of the EIC from the dusty torus is the geometric dilution. Thus, the closest significant component in the complex IR spectrum is the most relevant. This is the hot dust on the inner face of the torus. The hot dust is typically considered to be region in which graphite and silicate grains are sublimated with a temperature between $\sim 1500$~K and $\sim 1800$~K \citep{mor12}. It creates significant flux at wavelengths $1.6 \mu$m - $2.2 \mu$m. There is also ``hot dust" that radiates a similar luminosity at $3 \mu$m--$4 \mu$m in Mrk\,231 with a temperature $\sim 750$~K \citep{lop17}. These components have the least geometric dilution of any of the IR components and contribute about $>1/3$ of the total AGN IR luminosity \citep{far03}.
\par Beside geometric dilution, one needs to know the relevant Doppler factors and that is based on the distribution of the hot dust. The primary piece of evidence is from the 3CRR catalog in which radio galaxy and radio loud quasar IR spectra are compared. The radio galaxy IR spectra are heavily attenuated at wavelengths short of $5 \mu$m. This implies that the hot dust is viewed through the optically thick torus \citep{hon11}. The hot dust lies on the inner face of the torus and is restricted to a solid angle less than that subtended by the main body of the torus, less than $\approx 45^{\circ}$ \citep{bar89}. The covering factor, CF, of the hot dust has been estimated at 0.15-0.35 \citep{mor12}. We assume $0.15<CF<0.35$ and use this to compute an upper limit on the EIC flux in the calculation below. We note that the distribution of hot dust at the inner edge of the torus is restricted to within $45^{\circ}$ of the equatorial plane.
\par The IR spectrum from data in \citet{lop17} at wavelengths shorter than $4\mu$m was fit with two blackbody components, $T= 750$~K as in \citet{lop17} and a hotter component $T= 1650$~K, \citet{mor12}, with a luminosity of $1.06 \times 10^{45}$ ergs/sec and $1.65 \times 10^{45}$ ergs/sec, respectively. The radius at which these hot dust distributions reside are $\approx \sqrt{0.35/CF}6.5 \times 10^{18}$ cm and $\approx \sqrt{0.35/CF}1.7 \times 10^{18}$ cm, respectively. We assume that the dust is isotropically distributed relative to the equator in the range $-45^{\circ}< \theta< +45^{\circ}$. If the hot dust is concentrated closer to the equatorial plane than $45^{\circ}$ then the Doppler factors are smaller and this calculation is an upper limit. For the hottest component,
\begin{equation}
    U_{ph}\approx\frac{1}{c}\int_{0}^{2\pi} \int_{-\theta_{o}}^{\theta_{o}}\delta(\beta=0.816, \, \theta)^{4} I_{BB}(T= 1650\mathrm{K})d\phi\, d\theta\ = \frac{CF}{0.35}(1.16 \times 10^{-2})\;,
\end{equation}
where $\theta_{o}= 45^{\circ}$.
From Equations (29), (30) and (32), the EIC flux  in the NuSTAR band from the $1650$~K dust is (for $0.15<CF<0.35$)
\begin{equation}
5.68\times 10^{-14} < F_{EIC}(T =1650 K) < 1.33\times 10^{-13}\;.
\end{equation}
Similarly, for the other hot dust component
\begin{equation}
2.42\times 10^{-15}<F_{EIC}(T =750 K) < 5.55\times 10^{-15}\;.
\end{equation}
\par Note how much more inefficient the $T =750 K$ component is at inducing EIC emission in the plasmoid, since it is farther out. For completeness, we consider the remaining $\sim 3 \times 10^{45}$ ergs/sec of AGN IR emission from the SED that is peaked at 60$\mu$m \citep{far03}. Assuming a blackbody this would require a radius of 60~pc at $T \lesssim 40 K$ \citep{far03}. The rotating molecular disk resolved with VLBI is roughly this size, 30--100~pc \citep{klo03}. It seems reasonable to assume that these are the same regions. The VLBI image reveals gas that is distributed in a region that is approximately orthogonal to the plasmoid velocity in Figure~5 as was the case for the hot dust \citep{lon03,klo03}. There would be no strong Doppler enhancement for such a distribution. The geometric dilution is $10^{4}$ times that of
$T =1650 K$ dust with only twice the luminosity. Thus, there is no evidence of a pole on distribution of luminous dust associated with the AGN. VLBI observations revealed that the AGN dominates the starburst emission within $\sim 100$~pc of the quasar \citep{lon03}. If we assume the most extreme upper bound of placing all the starburst emission along the jet axis 100~pc away and recompute Equation (31) with $\theta =0 $, we get
$U_{ph}< 3.93\times 10^{-3} (10^{4}) \left[\frac{2\times 10^{17}\rm{cm}}{3\times 10^{20}\rm{cm}}\right]^{2}< 3 \times 10^{-5}$. This is $\sim10^{-3}$ of the hot dust value in Equation (32). Of course, the starburst regions are probably more isotropically distributed and the contribution would be significantly less than this upper limit. Therefore, we do not consider starburst emission as a viable seed field for X-ray emission. In summation, the combined EIC flux from Sections 8.1 and 8.2 is $< 8\%$ of the observed continuum flux in Table~\ref{nustar}.
\section{Summary and Concluding Remarks}
In this article we used a multi-frequency, multi-epoch VLBA campaign that suffered from significant degradation in resolution and sensitivity at 43~GHz in order to explore a luminous 17.6~GHz flare detected with AMI. The bad luck with lost stations was offset by a fortuitous time sampling that caught the rise, the peak and the decay of the flare. Our primary conclusion is that the central engine of Mrk\,231, at the end of 2017, produced a strong moving knot in a Poynting flux dominated region of the jet. We determined that the knot was filled primarily with an electron-positron plasma that transports a power of $Q\sim 10^{43}$ ergs/sec along the conduit of the jet.
\begin{table}
    \caption{\label{conformance}Conformance of Models to Observation}
{\tiny\begin{tabular}{cccccc} \tableline\rule{0mm}{3mm}
Ejection &  Requirement 1 & Requirement 2 & Requirement 3 & Comments & Assessment\\
Type  &  $Q$ Sufficient  & Magnetic & Evolves to &  &  \\
      & to Power K1  & Acceleration & Minimum $E(\rm{lm})$ &  &  \\
\tableline \rule{0mm}{3mm}
Discrete Leptonic & Marginal & C & C & Requires $>10\%$   &  \\
  &          &   &  & Radiative Efficiency at K1  & Marginal \\
\tableline \rule{0mm}{3mm}
Discrete Protonic & C & NC & C & No Driving Mechanism & Excluded  \\
\tableline \rule{0mm}{3mm}
Leptonic Knot in Jet & C & C & C &  & Conforms to Observation \\
\tableline \rule{0mm}{3mm}
Protonic Knot in Jet & C & C & C & Requires Dissipative & \\
                     &   &   &   & Entrainment Between & \\
       &   &   &   & Epochs C and D & \\
              &   &   &   & Conflicts with Constant $\alpha$ & Excluded
\end{tabular}}
\tablenotetext{}{C means conformant and NC means nonconformant}
\end{table}
\par The existence of the simple SSA power-law spectrum of the unresolved radio core from 8.4~GHz to 43.1~GHz motivated a simple model, an optically thick uniform spherical volume for the emission region. There were four categories of models. If the spheroid is a discrete ejection, it can be positronic or protonic plasma. If the spheriod approximates a knot in a continuous jet it can be positronic or protonic plasma as well. This paper considered physical constraints that could narrow down the possible models. First, there were some hard constraints from observation. The LOS to the velocity of the ejected plasmoid was bounded to $<25-30^{\circ}$ in \citet{rey09,rey17} based on light curve variability and a superluminal ejection, respectively. In this paper, we found that the apparent core (the spheroid) was moving towards K1 in Figure~5 at $\beta_{app}\approx 0.97$. This constrained the kinematics, but did not favor one of the four categories over another. In order to differentiate amongst these 4 possibilities we established three more constraints in Section 6. The implications of the constraints introduced in Section 6 for the four classes of models were demonstrated in Section 7 and are summarized in Table~\ref{conformance}. Clearly, the most robust model is a knot in a magnetic positron-electron jet. Figure~11 shows a gradual temporal evolution approaching adiabatic expansion at late times. The jet power based on Figure~11 is $Q\sim 10^{43}$ ergs/sec. We also note that this estimate is more reliable (based on more data) than the \citet{rey09} estimate of a flare $Q \sim 3\times 10^{43}$ ergs/sec.
\par The discrete leptonic plasmoid ejections model is not ruled out. It would imply a long term time averaged power output, an order of magnitude less, with the impulsive power or launch power of $Q_{\rm{impulsive}} \approx E(\rm{lm})/T$, with $T$ loosely constrained by the data as less than the flare rise time of $\sim 120$ days,
$Q_{\rm{impulsive}} > 1.25 \times10^{49}\rm{ergs}/1\times 10^{7} sec > 1.25 \times 10^{42} \rm{erg/sec}$. In this estimate, it is assumed that $\Gamma_{\rm{min}}=1$ and we used the top left hand frame of Figure~7 and Figure~1. The advantage of this model is that it is the same class of solution as was found for the major flares in GRS~1915+105 \citep{fen99,pun12}. Secondly, the source of the positronic plasma might explain the weak X-ray luminosity, an order of magnitude less than expected from a quasar of similar bolometric luminosity \citep{lao97,rey17}. The corona above the accretion disk that normally produces the X-ray luminosity is filled with pair plasma, but for some reason it is being episodically ejected as these radio plasmoids. This phenomenon has theoretical origins as a Parker instability or a twisted magnetic tower that acts as a piston or propellor, ejecting the plasmoid as it expands vertically relieving the magnetic stress \citep{lyn03}.
\begin{figure*}
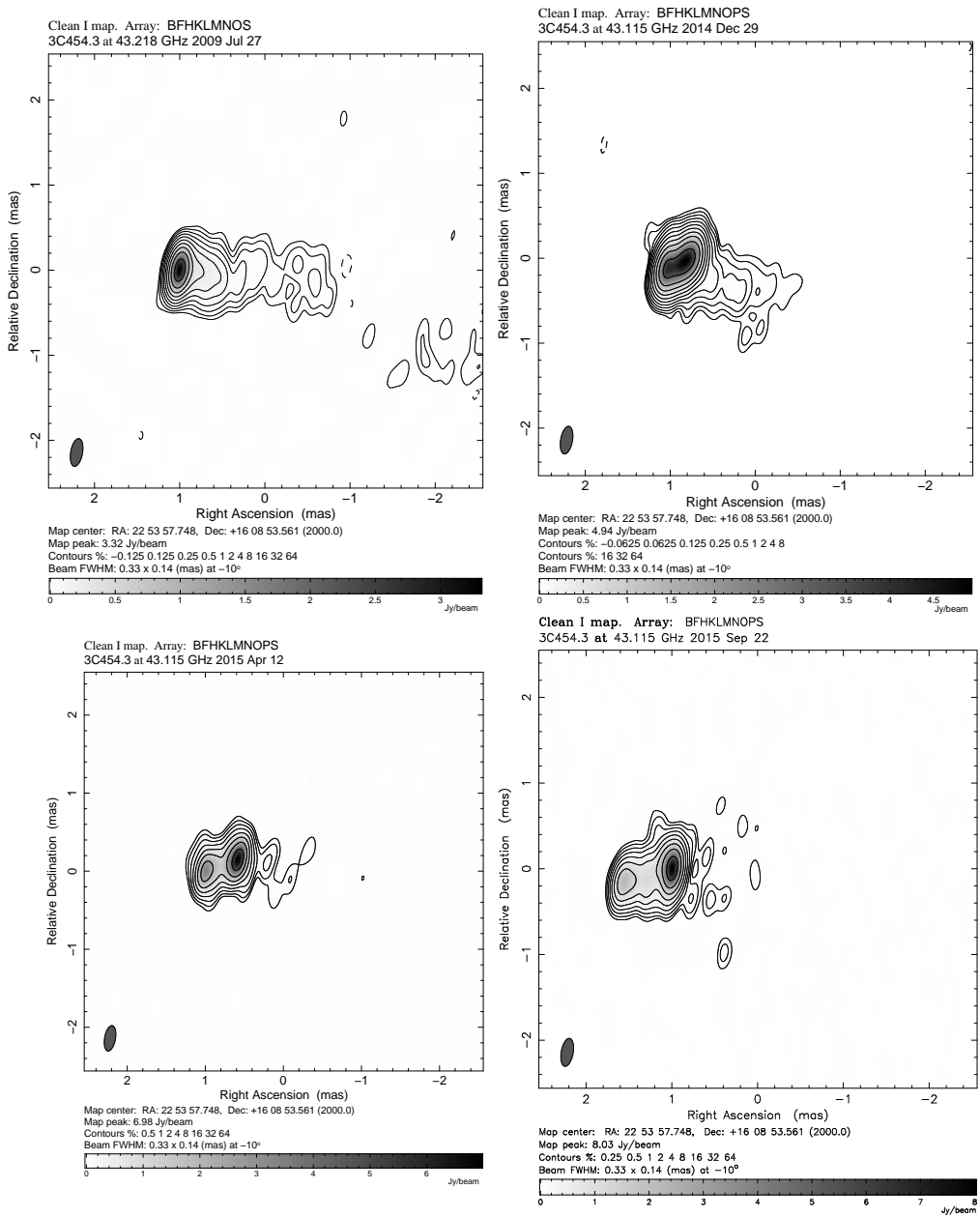

\begin{center}
\includegraphics[width= 0.4\textwidth,angle =0]{f29.ps}
\includegraphics[width= 0.4\textwidth,angle =0]{f30.PS}
\includegraphics[width= 0.4\textwidth,angle =0]{f31.PS}
\includegraphics[width= 0.4\textwidth,angle =0]{f32.PS}
    \caption{\label{454_7.09}\footnotesize{An ejected peak surface brightness feature in the blazar 3C454.3 as seen with 43~GHz VLBA. The top left hand panel shows the peak surface brightness as a compact core at the east end of a jet $>2$mas. The top right panel shows a very bright knot ejected from the west end of the jet in December 2014, $~0.25$ mas from the feature that was previously defined as the core. The bottom left hand frame shows that in April 2015 the feature is 0.5 mas from the feature previously identified as the core, but now $\sim$ twice the brightness of the core. In September 2015, the ejected knot is $>0.7$ mas from the core and more than 4 times as bright. This is a similar dynamic to what we see in Mrk\,231 Figure~5, but on a scale two orders of magnitude larger. It would be very difficult or impossible estimate the point of origin with lower resolution.}}
\end{center}
\end{figure*}
\par The continuous jet model, on the other hand, needs to be understood in terms of other extragalactic jets. First, we consider the ``radio loudness" of Mrk 231. This determination is not a trivial or unambiguous calculation because of intrinsic extinction of the optical flux and the significant SSA of the radio emission. Radio loudness is typically defined by the ratio of the 5 GHz flux density, f(5), to the blue band flux density, f(B), $R\equiv f(5)/f(B)$ \citep{kel89,sto92}. A value of $R>10$ is considered radio loud. Taking the values from the NASA Extragalactic Database, $R= 280\, \rm{mJy}/13\,\rm{mJy}\approx 22$ which ostensibly seems to be radio loud. However, recall from Section 8 that there is strong attenuation of the continuum in this source. Based on the extinction law discussed in Section 8, we can compute an extinction corrected $R$, $R^{*}$. The intrinsic extinction in B-band is a factor of $\approx 10$. Thus, $R^{*}\approx 280\, \rm{mJy}/[(10)(13\,\rm{mJy})] \approx 2$ which is formally radio quiet.
\par The jet in Mrk 231 feeds plasma to the luminous secondary, K1. In multiple campaigns, we have looked at wide field images at 8.4~GHz to see if there is any sign of emission emerging from the approximately stationary, K1. We have never detected any evidence of such a flow. The question is whether the jet is short because it is young or because its propagation is thwarted. The time it takes the jet to propagate to K1, 0.8 pc on the sky plane from the point of origin, is approximately 2.5 years based on our estimated $v_{app}=0.97$c, The core has been active for much longer than this, so this is not the origin of the short jet length. We conclude that the jet terminates at K1. As such, it should be considered a failed jet. We note that there is low frequency (1.4 GHz - 5 GHz) emission on scales of 10-100 pc in Mrk 231. But, it is not aligned with the jet, nor does it connect with the jet termination point at K1, as noted above. This has been discussed in detail elsewhere \citep{ulv99}.
\par In spite of its radio quiet classification, Mrk 231 has an intrinsically powerful jet, $\sim 10^{43}$ ergs/sec. It is comparable to some of the most powerful nearby, jets in active galactic nuclei such as M87 and 3C120 that propagate a few hundred kpc from the source (see below for a quantitative discussion). There are even radio quiet quasars with large scale radio structures hundreds of kpc in extent \citep{blu03}. Yet, paradoxically, there is a failed jet in Mrk 231. To explore this, we quantify the magnitude of the jet truncation compared to extragalactic jets with similar power. In the extragalactic radio source lexicon, the Fanaroff-Riley classification of radio source morphology separates large radio sources into two classes FRI and FRII \citep{fr74}. The FRII sources are edge brightened with hot spots in their radio lobes and FRI are edge darkened and the lobes resembles plumes. Empirically, there is an FRI/FRII luminosity divide as well, the FRIIs more luminous. In terms of long term time averaged jet power this divide occurs at $Q\sim 5 \times 10^{43}$ ergs/sec \citep{wil99}. Thus, we have estimated that Mrk\,231 has the jet power of a strong FRI radio source. At redshifts similar to Mrk\,231, FRI radio sources are usually narrow line radio galaxies (NLRGs) with a radio structure that extends well beyond galactic dimensions, with a linear size of  $\sim 100$~kpc or more \citep{wil99}. The jet in Mrk\,231 $\sim 10^{-5}$ the size of the FRI radio sources at similar redshift with similar jet power.

\par In order to understand the origin of the termination of the powerful jet in Mrk 231, consider the very low accretion rates in these FRI NLRGs, 3-4 orders of magnitude less than a quasar \citep{chi99,chi02,har09}. However, there are the occasional FRI broad line galaxies such as the Seyfert 1 galaxy, 3C 120 that has a very prominent FRI morphology on a 400~kpc scale \citep{wal87}. Thus, the low accretion rate of FRI NLRGs is not the full explanation of the discrepancy with Mrk\,231, but is likely related. One difference between 3C 120 and Mrk\,231 is that Mrk\,231 has a low ionization BAL wind \citep{lip94,smi95}. In low radio states, it has also displayed evidence of a high ionization X-ray absorbing wind \citep{fer15,rey17}. There is also extreme amounts of intrinsic optical absorption in the galaxy itself from dusty gas \citep{lip94,smi95}. All three circumstances point to a very dense nuclear environment through which the jet must propagate. Furthermore, in \citet{rey09}, it was argued that the density of the free-free absorbing screen at K1 is consistent with the jet being stopped by the BAL wind at K1. By contrast, the extremely low, undetectable accretion of gas in FRI NLRGs is consistent with a very low density nuclear environment. The launching of the jet in Mrk\,231 does not seem to be stopped by the dense nuclear environment, but its propagation does seem to be thwarted. The case of 3C 120 does not seem to be accommodated by this discussion. It is noted that there is no evidence of either a BAL wind or a high ionization X-ray absorbing wind in 3C 120 \citep{oke79,bal04}. Thus, it might be the case that there is not an extremely dense nuclear environment near the source of the jet in 3C 120, hence jet propagation is not hindered. We cannot rule out the possibility that the BAL wind in Mrk\,231 also diminishes the power of the jet launching mechanism as well as a providing a drag on its propagation.
\par This conclusion does not exist in isolation. We apply the original definition of
BAL quasars (BALQSOs) as quasars with UV absorbing gas that is blue
shifted at least 5,000 km/s relative to the QSO rest frame and
displaying a spread in velocity of at least 2,000~km~s$^{-1}$,
\citep{wey91}. This definition was designed to exclude the ``mini-BALQSOs," with the BALNicity index = 0 \citep{wey97}. This
definition is preferred here since mini-BALQSOs tend to resemble
non-BALQSOs more than BALQSOs in many spectral and broadband
properties \citep{pun06,zha10, bru13, hay13}. It was found that the large scale radio emission ($> 20$~kpc in linear extent)
in BALQSOs is strongly anti-correlated with the BALnicity index \citep{bec00,bec01}. Dense outflows seem to prevent large scale jets from forming in BALQSOs.
\par We note that Section 8 provided further support of the details of our model of a dissipative moving knot in a leptonic jet. We found that the plasmoid model predicts an X-ray flux $<8\%$ of that of the historical levels detected by NuSTAR. Thus, no elevated NuSTAR flux levels are expected during the flare and none were observed. This shows consistency between the models and the high energy observations.
\par The astrometry of the SSA core in Figure~5 provides a valuable laboratory for exploring the notion of a ``core" in blazars. Mrk\,231 has the advantage of the nearby stationary secondary for astrometric measurements, unparalleled in other blazars. In Mrk\,231, the ``apparent core" moves outward then eventually fades to the background surface brightness of the faint jet. This discovery might provide insight into more luminous blazar jets. Figure~15 shows an extreme example of ``apparent core" motion in the blazar, 3C454.3, (z=0.859) observed with 43~GHz VLBA \footnote{The data were obtained with permission of the Boston University Blazar Monitoring program and was downloaded from http://www.bu.edu/blazars/VLBAproject.html}. The ejected knot becomes much brighter than the feature previously identified as the core in earlier epochs. It appears to move $\sim 10.5$ light years in 9 months, an apparent velocity of $\sim 13$c. By contrast, the core in Mrk\,231 moves only 0.2 light years in our observing campaign (see Figure~5) and even less in 2015. Such ``small" displacements would be very difficult to detect without the low redshift of Mrk\,231 and the precision astrometry made possible by the nearby stationary secondary, K1. Evidence of this phenomenon appears in simulations of blazar jets \citep{gom97}. More importantly, there is evidence of this in the component motion of blazar jets. Some highly relativistic components appear to suddenly decelerate \citep{jor17}. This can be explained by a ``core" that is actually a strong knot that moves downstream slightly (below the resolution limits of VLBA), before fading (A. Marscher private communication 2019). The lesson of Mrk\,231 and 3C 454.3 in Figure~15 is that the ability of continuous jet models to predict the point of jet origin and the width of the jet at the point of origin is limited. They are too simplistic. The notion that the shift in the core position with frequency as a consequence of the change in the SSA opacity, Equation (2), can be extrapolated to infinite frequency to find the point of jet origin depends on the assumption that the low resolution observations are actually observing the core region \citep{bla79}. Namely, the unresolved peak surface brightness is assumed to be the continuation of the jet to the frequency dependent photosphere of the launch region. However, the procedure can very likely be estimating the SSA shift of a bright (possibly moving) knot ``near" the launch point and therefore estimates the location of an ``arbitrary" point upstream of the launch point. In Mrk\,231, it is 0.2 light years away from the point of origin, which is very large if this is the observational data that is input into a model of jet launching. The situation can be much more severe in a bright blazar like 3C 454.3 where the error in this method of estimating the point of jet origin can be 10 light years. In highly relativistic blazars, Doppler aberration can cause large swings in the apparent jet direction \citep{lin85}. Evidence of this wild jet bending near the base of the jet is the Event Horizon telescope image of 3C 279 \citep{eht19}. In 3C 279, the brightest features are not likely to be the core, but knots in an apparently bent jet. In such circumstances, the accuracy of a ``core shift" extrapolation to the point of jet origin is even more likely to be misleading.
\begin{acknowledgements}
This research has made use of data obtained with NuSTAR, a project
led by Caltech, funded by NASA and managed by NASA/JPL, and has
utilized the NUSTARDAS software package, jointly developed by the
ASDC (Italy) and Caltech (USA). This study makes use of 43~GHz VLBA data from
the VLBA-BU Blazar Monitoring Program VLBA-BU-BLAZAR funded by NASA
through the Fermi Guest Investigator Program. The National Radio Astronomy Observatory is a facility of the National Science Foundation operated under cooperative agreement by Associated Universities, Inc. This research has made use of NASA's Astrophysics
Data System Bibliographic Services.
This work made use of the Swinburne University of Technology software correlator, developed as part of the Australian Major National Research Facilities Programme and operated under licence.
GM acknowledges funding by the Spanish State Research Agency (AEI) Projects No. ESP2017-86582-C4-1-R and No. MDM-2017-0737 Unidad de Excelencia ``María de Maeztu”- Centro de Astrobiología (CSIC-INTA).
C.O. was supported by  the Natural Sciences and Engineering Research Council of Canada (NSERC).
\end{acknowledgements}


\begin{thebibliography}{}
\bibitem[Ballentyne et al.(2004)]{bal04}Ballentyne, D., Fabian, A., \& Iwasawa, K.\ 2004, MNRAS,354, 839.
\bibitem[Barthel(1989)]{bar89}Barthel, P. D. 1989, ApJ, 336, 606
\bibitem[Becker et al.(2000)]{bec00}Becker., R.,  White, R. L., Gregg, M. D.,  et al. 2000, ApJ, 538, 72
\bibitem[Becker et al.(2001)]{bec01}Becker., R., White, R. L., Gregg, M. D., et al. 2001, ApJS, 135, 227
\bibitem[Bicknell et al.(1990)]{bic90}Bicknell, G., DeRuiter. H., Fanti, R., Morganti, R., \& Parma, P.\ 1990, ApJ, 354, 98
\bibitem[Blandford(1976)]{bla76}Blandford, R. D. 1976, MNRAS 176 465
\bibitem[Blandford and K{\"o}ingl(1979)]{bla79}Blandford, R. \& K{\"o}nigl, A. 1979, ApJ 232 34
\bibitem[Blandford and Payne(1982)]{bla82}Blandford, R. D., \& Payne D. 1982, MNRAS 199 883
\bibitem[Blundell et al.(2003)]{blu03}Blundell, K., Beasley, A., \& Bicknell, G. 2003, ApJL, 591, 103
\bibitem[Bokensberg et al.(1977)]{bok77}Bokensberg, A., Carswell, R., Allen, D., et al. 1977, MNRAS 178 451
\bibitem[Braito et al.(2004)]{bra04}Braito, V., Della Ceca, R., Piconcelli, E.,  et al. 2004, A\&A 420, 79
\bibitem[Briggs (1995)]{briggs95}Briggs, D., 1995, PhD Thesis, New Mexico Institute of Mining and Technology
\bibitem[Bruni et al.(2013)]{bru13}Bruni, G, Dallacasa, D., Mack, K.-H., et al. 2013 A\&A 554 94.
\bibitem[Brunthaler et al.(2000)]{bru00}Brunthaler, A., Falcke, H., Bower, G. C., et al. 2000, A\&A, 357, L45
\bibitem[Chiaberge et al.(1999)]{chi99}Chiaberge, M., Capetti, A., \& Celotti, A. 1999, A \& A, 349, 77
\bibitem[Chiaberge et al.(2002)]{chi02}Chiaberge, M., Macchetto, F.D., Sparks, W.B., et al. 2002, ApJ, 571, 247
\bibitem[Croston et al.(2005)]{cro05}Croston J. H., Hardcastle M. J., \& Birkinshaw M. 2005, MNRAS, 357, 279
\bibitem[Deller et al.(2011)]{deller2011}Deller, A., Brisken, W. F., Phillips, C. J., et al. 2011, PASP, 123, 275
\bibitem[Dermer and Schlickeiser (1993)]{der93}Dermer C. D., \& Schlickeiser R., 1993, ApJ, 416, 458
\bibitem[Edelson (1987)]{ede87}Edelson, R., 1987, ApJ, 313, 651
\bibitem[Evans \& Koratkar (2004)]{eva04}Evans, I., \& Koratkar, A. 2004, ApJS 150, 73
\bibitem[Event Horizon Telescope Collaboration (2019)]{eht19}Event Horizon Telescope Collaboration, Akiyama, K., Alberdi, A., et al. 2019, ApJ, 875, L4
\bibitem[Fanaroff \& Riley (1974)]{fr74}Fanaroff, B.~L., \& Riley, J.~M., 1974, MNRAS, 167, 31P
\bibitem[Farrah et al.(2003)]{far03}Farrah, D., Afonso, J., Efstathiou, M., et al. 2003, MNRAS 343, 585
\bibitem[Fender et al.(1999)]{fen99}Fender, R., Garrington, S. T., McKay, D. J., et al. 1999, MNRAS 304, 865
\bibitem[Feruglio et al.(2015)]{fer15}Feruglio, C.; Fiore, F.; Carniani, S., et al. 2015 A\&A, 583 99
\bibitem[Fritz et al.(2006)]{fri06}Fritz, J., Franceschini, A.,  \& Hatziminaoglou, E. 2006, MNRAS 366, 767
\bibitem[Gallagher et al.(2002)]{gal02}Gallagher, S.,  Brandt, W. N., Chartas, G., Garmire, G. P., \& Sambruna, R. M. 2002 ApJ, 569, 655
\bibitem[Ghisellini et al.(2010)]{ghi10}Ghisellini, G, Tavecchio, F. \& Foschini, L., et al. 2010, MNRAS 402, 497
\bibitem[Ghosh \& Punsly (2007)]{gho07}Ghosh, K., \& Punsly, B. 2007 ApJL 661 139
\bibitem[Ginzburg \& Syrovatskii (1965)]{gin65}Ginzburg, V., \& Syrovatskii, S. 1965, Annu. Rev. Astron. Astrophys. 3 297
\bibitem[Ginzburg \& Syrovatskii (1969)]{gin69}Ginzburg, V., \& Syrovatskii, S. 1969, Annu. Rev. Astron. Astrophys. 7, 375
\bibitem[Gomez et al.(1997)]{gom97}Gomez, J., Marti, J., Marcher, A., Ibanez, J., \& Alberdi, A. 1997, ApJL, 482, 33
\bibitem[Greene \& Ho (2007)]{gre07}Greene, J. E., \& Ho, L. C. 2007, ApJ, 670, 92
\bibitem[Hamilton \& Keel(1987)]{ham87}Hamilton, D. \& Keel, W. 1987, ApJ, 321, 211
\bibitem[Hardcastle \& Worrall(2000)]{har00}Hardcastle, M. \& Worrall, D. 2000, MNRAS, 314, 359
\bibitem[Hardcastle et al.(2004)]{har04}Hardcastle, M. J., Harris, D. E., Worrall, D. M., \& Birkinshaw, M. 2004, ApJ, 612, 729
\bibitem[Hardcastle et al.(2009)]{har09}Hardcastle, M., Evans, D. \& Croston, J. 2009, MNRAS, 396, 1929
\bibitem[Hayashi et al.(2013)]{hay13}Hayashi, T., Doi, A., \& Nagai, H. 2013 ApJ 772 4
\bibitem[Homan et al.(2002)]{hom02}Homan, D. C., Ojha, R., Wardle, J. F. C., et al. 2002, ApJ, 568, 99
\bibitem[H{\"o}nig et al.(2011)]{hon11}H{\"o}nig, S. Leipski, C., Antonucci, R., \& Haas, M. 2011, ApJ, 736, 26
\bibitem [Jorstad et al.(2017)]{jor17}Jorstad, S., Marscher, A., Morozova, D., et al. 2007, ApJ, 846, 98
\bibitem[Kalberla et al.(2005)]{kal05}Kalberla, P., Burton, W., Hartmann, D., et al. 2005, A\&A, 440, 775
\bibitem[Kaiser \& Alexander (1997)]{kai97}Kaiser C. R., \& Alexander P.  1997, MNRAS, 286, 215
\bibitem[Kataoka \& Stawarz (2005)]{kat05}Kataoka, J. \& Stawarz, L. 2005, ApJ, 622, 797
\bibitem[Kellermann et al.(1989)]{kel89}Kellermann, K. I.; Sramek, R. A.; Schmidt, M.;Shaffer, D.; BGreen, R. F. 1989, AJ, 98, 1195
\bibitem[Kennel \& Coroniti(1984)]{ken84}Kennel, C., \& Coroniti, F. 1984 ApJ 283 694
\bibitem[Kettenis et al.(2006)]{kettenis2006}Kettenis,~M., van Langevelde,~H.~J., Reynolds,~C. \& Cotton,~B., 2006, Astron. Data Anal. Software Syst. XV, 351, 497
\bibitem[Klein-Wolt et al.(2002)]{kle02}Klein-Wolt, M.  Fender, R. P., Pooley, G. G., et al.  2002, MNRAS 331, 745
\bibitem[Kl{\"o}ckner et al.(2003)]{klo03}Kl{\"o}ckner, H-R, Baan, W. A., \& Garrett, M. A.  2003, Nature 421, 821
\bibitem[Knigge et al.(2008)]{kni08}Knigge, C., Scaringi, S., Goad, M., \& Cottis, C. 2008, MNRAS 386, 1426
\bibitem[Krongold et al.(2003)]{kro03}Krongold, Y., Nicastro, F., Brickhouse, N. S., Elvis, M., Liedahl, D., \& Mathur, S. 2003 ApJ, 597, 832
\bibitem[Laor and Davis(2014)]{lao14}Laor, A., \& Davis, S. 2014 ApJ 428 3024
\bibitem[Laor et al.(1997)]{lao97}Laor, A. Fiore, F., Elvis, M., Wilkes, B., \& McDowell, J. 1997, ApJ 477, 93
\bibitem[Lightman et al.(1975)]{lig75}Lightman, A., Press, W., Price, R. \& Teukolsky, S. 1975, \emph{Problem Book in Relativity and Gravitation}(Princeton University Press, Princeton)
\bibitem[Lind and Blandford(1985)]{lin85}Lind, K., \& Blandford, R. 1985, ApJ 295, 358
\bibitem[Lipari et al.(1994)]{lip94}Lipari, S., Colina, L., \& Macchetto, F. 1994, ApJ 427, 174
\bibitem[Leipski et al.(2009)]{lei09}Leipski, C., Antonucci, R., Ogle, P., \& Whysong, D. 2009 ApJ, 701, 891
\bibitem[Lonsdale et al.(2003)]{lon03}Lonsdale, C., Lonsdale, C., Smith, H., \& Diamond, P. 2003, ApJ 592 804
\bibitem[Lopez-Rodriguez et al.(2017)]{lop17}Lopez-Rodriguez, E., Packham, C., Jones, T., et al. 2017, MNRAS 464, 1762
\bibitem[Lovelace (1976)]{lov76}Lovelace, R. V. E. 1976, Nature 262 649
\bibitem[Lynden-Bell (2003)]{lyn03}Lynden-Bell, D. 2003, MNRAS, 341, 1360
\bibitem[Macleod et al. (2012)]{mac12}MacLeod, C. L., Ivezić, Ž., Sesar, B., et al. 2012, ApJ 753, 106
\bibitem[McCutcheon and Gregory (1978)]{mcc78}McCutcheon, W., \& Gregory, P. 1978, AJ, 83, 566
\bibitem[McKinney et al.(2012)]{mck12}McKinney, J., Tchekhovskoy, A., \& Blandford, R. 2012, MNRAS 423, 3083
\bibitem[McNamara et al.(2011)]{mcn11}McNamara, B., Rohanizadegan, M, \& Nulsen, P. 2011 ApJ, 727, 39
\bibitem[Moffet(1975)]{mof75}Moffet, A. 1975 in \emph{Stars and Stellar Systems, IX: Galaxies and the Universe}, eds. A. Sandage, M. Sandage \& J. Kristan (Chicago University Press, Chicago), 211.
\bibitem[Mor and Netzer (2012)]{mor12}Mor R., \& Netzer H.  2012, MNRAS 420, 526
\bibitem[Neilsen and Lee (2009)]{nei09}Neilsen, J. \& Lee, J. 2009, Nature 458, 481
\bibitem[O'Dea(1998)]{ode98}O'Dea, C. 1998, PASP, 110, 493
\bibitem[Oke \& Zimmerman(1979)]{oke79}Oke, B. \& Zimmerman, B. 1979, ApJL, 231, 13
\bibitem[Owen et al.(2000)]{owe00}Owen F. N., Eilek J. A., \& Kassim N. E. 2000, ApJ, 543, 611
\bibitem[Parker(1958)]{par58}Parker E. N., 1958, ApJ, 128, 664
\bibitem[Piconcelli et al.(2013)]{pic13}Piconcelli, E., Miniutti, G., Ranalli, P., et al. 2013, MNRAS 428, 1185
\bibitem[Prat et al.(2010)]{pra10}Prat, L., Rodriguez, J, \& Pooley, G. 2010, ApJ 717, 1222
\bibitem[Prieto et al.(2016)]{pri16}Prieto, M. A., Fernandez-Ontiveros, J. A., Markoff, S., Espada, D., \& Gonzalez-Martin, O. 2016, MNRAS 457, 3801
\bibitem[Punsly(1999a)]{pun99}Punsly, B. 1999a, ApJ 527, 609
\bibitem[Punsly(1999b)]{pun00}Punsly, B. 1999b, ApJ 527, 624
\bibitem[Punsly(2006)]{pun06}Punsly, B. 2006, ApJ, 647, 886
\bibitem[Punsly(2008)]{pun08}Punsly, B. 2008, \emph{Black Hole Gravitohydromagnetics}, second edition (Springer-Verlag, New York)
\bibitem[Punsly(2012)]{pun12}Punsly, B. 2012, ApJ 746, 91
\bibitem[Punsly(2014)]{pun14}Punsly, B. 2014, ApJL 797, 33
\bibitem[Punsly(2015)]{pun15}Punsly, B. 2015, ApJ 806, 47
\bibitem[Punslyet al.(2016)]{pun16}Punsly, B., Marziani, P., Kharb, P., O'Dea, C., \& Vestergaard, M. 2016, ApJ 812, 79
\bibitem[Punsly (2019)]{pun19}Punsly, B. 2019 ApJL 871, 34
\bibitem[Rees (1966)]{ree66}Rees, M. J. 1966, Nature 211: 468-70
\bibitem[Reynolds et al.(2009)]{rey09}Reynolds, C., Punsly, B. Kharb, P., O'Dea, C. \& Wrobel, J. 2009, ApJ, 706, 851
\bibitem[Reynolds et al.(2013a)]{rey12}Reynolds, C., Punsly, B. \& O'Dea, C. P. 2013a, ApJL, 773, 10
\bibitem[Reynolds et al.(2013)]{rey13}Reynolds, C., Punsly, B., O'Dea, C. P., \& Hurley-Walker, N. 2013b, ApJL, 776, 21
\bibitem[Reynolds et al.(2017)]{rey17}Reynolds, C., Punsly, B., Miniutti, G., O'Dea, C., \& Hurley-Walker, N. 2017, ApJ, 836, 155
\bibitem[Shankar et al.(2008)]{sha08}Shankar, F., Dai, X., \& Sivakoff, G. 2008 ApJ 687 859
\bibitem[Shepherd et al.(1994)]{shepherd94}Shepherd, M., Pearson, T., \& Taylor, G. 1994, BAAS, 26, 987
\bibitem[Shepherd et al.(1995)]{shepherd95}Shepherd, M., Pearson, T., \& Taylor, G. 1995, BAAS, 27, 903
\bibitem[Smith et al.(1995)]{smi95}Smith, P., Schmidt, G., Allen ,R., \& Angel, J.R.P. 1995, ApJ 444, 146
\bibitem[Spinoglio et al.(1995)]{spi95}Spinoglio, L., Malkan, M. A., Rush, B., Carrasco, L., \& Recillas-Cruz, E. 1995, ApJ 453, 616
\bibitem[Stocke et al.(1992)]{sto92}Stocke, J. T., Morris, S. L., Weymann, J. T., \& Foltz, C. B. 1992, ApJ, 396, 487
\bibitem[Sun and Malkan(1989)]{sun89}Sun, W.-H., \& Malkan, M. A 1989, ApJ 346 68
\bibitem[Szuszkiewicz et al.(1996)]{szu96}Szuszkiewicz, E., Malkan, A., \& Abramowicz, M. A. 1996, ApJ 458, 474
\bibitem[Teng et al.(2014)]{ten14}Teng, S. H., Brandt, W.N., Harrison, F. A. et al. 2014, ApJ, 758, 19
\bibitem[Telfer et al.(2002)]{tel02}Telfer, R., Zheng, W., Kriss, G., \& Davidsen, A. 2002 ApJ 565, 773
\bibitem[Tucker(1975)]{tuc75}Tucker, W. 1975, \emph{Radiation Processes in Astrophysics}(MIT Press, Cambridge).
\bibitem[Ulvestad et al.(1999a)]{ulv99}Ulvestad, J., Wrobel, J. \& Carilli, C. 1999, ApJ 516, 134
\bibitem[Ulvestad et al.(1999b)]{ulv00}Ulvestad, J. et al. 1999, ApJL 517, L81
\bibitem[van der Laan (1966)]{van66}van der Laan, H. 1966, Nature 211, 1131
\bibitem[Veilleux et al.(2013)]{vei13}Veilleux, S., Trippe, M., Hamann, F., et al. 2013, ApJ, 764, 15
\bibitem[Veilleux et al.(2016)]{vei16}Veilleux, S., Melendez, M., Tripp, T., Hamann, F., \& Rupke, D. 2016, ApJ, in press
\bibitem[Walker (2014)]{walker2014}Walker, R. 2014 VLBA Scientific Memo \#37.
% \\ \url{https://library.nrao.edu/public/memos/vlba/sci/VLBAS\_37.pdf}
\bibitem[Walker et al.(1987)]{wal87}Walker, R., Benson, J., \& Unwin, S. 1987, ApJ 316, 546
\bibitem[Welsh et al.(2011)]{wel11}Welsh, B., Wheatley, J., \& Neil, J. 2011, A\&A 527, A15
\bibitem[Wilhite et al.(2005)]{wil05}Wilhite, B. et al. 2005, ApJ  633, 638
\bibitem[Weymann et al.(1991)]{wey91}Weymann, R.J., Morris, S.L., Foltz, C.B., \& Hewett, P.C. 1991, ApJ 373, 23
\bibitem[Weymann(1997)]{wey97}Weymann, R. 1997 in ASP Conf. Ser. 128, Mass Ejection from Active Nuclei ed, N.Arav, I. Shlosman and R.J. Weymann (San Francisco: ASP), 3
\bibitem[Willott et al.(1999)]{wil99}Willott, C., Rawlings, S., Blundell, K., \& Lacy, M. 1999, MNRAS 309, 1017
\bibitem[Zhang et al.(2010)]{zha10}Zhang, S., Wang, T-G., Wang, H., et al. 2010, ApJ, 714, 367
\bibitem[Zheng et al.(1997)]{zhe97}Zheng, W.,  Kriss, G. A., Telfer, R. C., Grimes, J. P., \& Davidsen, A. F. 1997, ApJ 475, 469
\bibitem[Zwart et al.(2008)]{zwa08}Zwart, J., Barker, R. W., Biddulph, P., et al. 2008, MNRAS 391, 1545
\end{thebibliography}
\end{document}